\documentclass{aa}  

\usepackage{graphicx}
\usepackage{orcidlink}
\usepackage{txfonts}
\usepackage{float}
\usepackage{caption}
\usepackage{xcolor}
\usepackage[utf8]{inputenc} 
\usepackage[T1]{fontenc}    
\usepackage{orcidlink}
\begin{document}

   \title{From compact jets to extended lobes: radio morphologies of distant quasars at $z>4$}

   \author{K. \'E. Gab\'anyi
   \inst{\ref{inst1},\ref{inst2},\ref{inst3}, \ref{inst4}}\orcidlink{0000-0003-1020-1597}
          \and
          M. Kunert-Bajraszewska\inst{\ref{inst4}}\orcidlink{0000-0002-6741-9856}
          \and
          A. Krauze\inst{\ref{inst4}}\orcidlink{0009-0002-5915-4592}
          \and
          S. Frey\inst{\ref{inst2},\ref{inst1}}\orcidlink{0000-0003-3079-1889}
          \and
          M. Krezinger\inst{\ref{inst2}}\orcidlink{0000-0002-8813-4884}
          \and
          Z. Paragi\inst{\ref{inst5}}\orcidlink{0000-0002-5195-335X}
          \and
          H.~ Cao\inst{\ref{inst6}}\orcidlink{0000-0003-1514-881X}
          \and
          T.~ An\inst{\ref{inst7}}\orcidlink{0000-0003-4341-0029}
          \and
          L.~I.~ Gurvits\inst{\ref{inst5},\ref{inst8}, \ref{inst11}}\orcidlink{0000-0002-0694-2459} 
          \and 
          K.~ Perger\inst{\ref{inst2}}\orcidlink{0000-0002-6044-6069}
          \and
          T. Sbarrato\inst{\ref{inst9}}\orcidlink{0000-0002-3069-9399}
          \and K. Rozgonyi\inst{\ref{inst10}}\orcidlink{0000-0002-5611-9292}
          }
   \institute{Department of  Astronomy, Institute of Physics and Astronomy, ELTE E\"otv\"os Lor\'and University,
              P\'azm\'any P\'eter s\'et\'any 1/A, 1117 Budapest, Hungary\\
              \email{k.gabanyi@astro.elte.hu} \label{inst1}
         \and
         Konkoly Observatory, HUN-REN Research Centre for Astronomy and Earth Sciences, MTA Centre of Excellence, Konkoly-Thege Mikl\'os \'ut 15-17, 1121 Budapest, Hungary \label{inst2}
         \and
         HUN-REN--ELTE Extragalactic Astrophysics Research Group, ELTE E\"otv\"os Lor\'and University,
              P\'azm\'any P\'eter s\'et\'any 1/A, 1117 Budapest, Hungary
              \label{inst3}
        \and
       Institute of Astronomy, Faculty of Physics, Astronomy and Informatics, NCU, Grudzi\k{a}dzka 5/7, 87-100, Toru\'n, Poland
       \label{inst4}
       \and
       Joint Institute for VLBI ERIC, Oude Hoogeveensedijk 4, 7991~PD Dwingeloo, The Netherlands
       \label{inst5}
       \and
       School of Electronic and Electrical Engineering, Shangqiu Normal University, 298 Wenhua Road, Shangqiu, Henan 476000, People's Republic of China
       \label{inst6}
       \and
       Department of Astronomy, University of Science and Technology of China, 96 Jinzhai Road, Hefei, Anhui 230026, People's Republic of China
           \label{inst7}
       \and
       Faculty of Aerospace Engineering, Delft University of Technology, Kluyverweg 1, 2629\,HS Delft, The Netherlands
       \label{inst8}
         \and Shanghai Astronomical Observatory, Chinese Academy of Sciences, 80 Nandan Road, 200030, People's Republic of China
       \label{inst11}
       \and
       INAF - Osservatorio Astronomico di Brera, via E. Bianchi 46, 23807, Merate (LC), Italy 
\label{inst9}
\and
       C3S LLC, K\"{o}nyves K\'{a}lm\'{a}n krt. 12-14, 1097 Budapest, Hungary
       \label{inst10}
             }

   \date{Received ... 2026; accepted ...}

 
  \abstract
   {High-redshift quasars play an essential role in studying the growth and evolution of supermassive black holes and active galactic nuclei (AGN). Radio-loud quasars additionally enable us to investigate the interactions between the jets and their environment.}
   {We aimed to reveal the radio morphology of three radio quasars at redshifts $z>4$ that contain milliarcsecond (mas) scale compact radio features according to previous very long baseline interferometry (VLBI) observations, but show significant flux density at arcsecond scales, indicating the presence of extended radio structure that cannot be sampled by the highest-resolution observations.}
   {We analysed radio interferometric data obtained at various angular resolutions and multiple frequencies, including observations made by the international Low-Frequency Array and the enhanced Multi-Element Remotely-Linked Interferometer Network. We also re-imaged archival European VLBI Network observations of our targets.}
   {Two quasars (J0813+3508 and J1231$+$3816) exhibit complex radio structures with hotspots, extended to tens of kpc. They resemble Fanaroff--Riley II-type radio galaxies with extremely bent jet morphology. The third object (J1548$+$3335) shows a one-sided structure of $\lesssim 10$\,kpc size. There was no sign of relativistic boosting at its previous mas-scale resolution radio observations. Its radio power and (inferred) linear size derived from the lower-resolution observations are similar to the known high-power compact steep-spectrum sources. We found that the previously detected mas-scale compact radio features are related to the centres of the AGN or in one case to one of the hotspots in an extended lobe.
   }
   {}


   \maketitle
%

\section{Introduction}

High-redshift ($z>4$) quasars are key to understanding the growth of supermassive black holes and the evolution of active galactic nuclei (AGN) in the early Universe. The very existence of black holes with masses of several billion solar masses at $z\gtrsim6$ places strong constraints on black hole formation and accretion processes (e.g., \citealt{Volonteri2011}). Within this population, jetted \citep{padovani} or traditionally named as radio-loud quasars\footnote{The radio-loudness factor was originally defined as the ratio of the $5$-GHz luminosity to the $4400\AA$ optical luminosity by \citet{Kellermann1989,Kellermann2016}. Its value is larger than $10$ for radio-loud quasars.} are particularly valuable because they can be probed with the highest angular resolution achievable through radio interferometric techniques. Among them, high-redshift radio galaxies provide crucial insight into the evolution of large-scale structure, as they are frequently located at the centers of clusters and proto-clusters. At early cosmological epochs, such radio galaxies are expected to be young, compact systems.

At present, only a handful of high-redshift radio galaxies are known \citep[e.g.][]{drouart2020,vanBreugel1999}. One of the most powerful examples is RC\,J0311$+$0507, a young FR~II-type \citep{FR} radio galaxy at $z=4.514$, whose Multi-Element Remotely-Linked Interferometer Network (MERLIN) and European Very Long Baseline Interferometry Network (EVN) observations revealed a complex morphology \citep{Parijskij2014}. More recently, $144$-MHz Low-Frequency Array (LOFAR) observations with international baselines uncovered the extended radio jet of the highly radio-loud quasar, J1601$+$3102 at $z=4.9$ \citep{Gloudemans2025}. With a projected size of $\sim66$\,kpc, it is currently the largest known radio jet at $z>4$. No radio emission was detected connecting the two extended lobes, which may be due to the suppression of the radio emission by the inverse-Compton scattering against the cosmic microwave background (CMB) photons. Its discovery also highlights the importance of low-frequency observations in revealing the true sizes of high-redshift radio galaxies, which would otherwise be underestimated from GHz-only measurements \citep{Gloudemans2025}.

In contrast to radio galaxies, blazars are radio-loud AGN with relativistic jets aligned close to the line of sight. They are therefore subject to strong relativistic effects, such as Doppler boosting of the approaching jet, de-boosting of the receding jet, and apparent superluminal motion of jet components (e.g., \citealt{Kravchenko2025}). Their radio structures at milliarcsecond (mas) angular scales are dominated by compact, flat-spectrum radio cores.  
However, their structures at larger scales are often resolved out with mas-resolution GHz-frequency very long baseline interferometry (VLBI) imaging observations and can only be studied with lower-resolution data. The importance of such observations is demonstrated by e.g. \citet{Kharb2010}. Based on a Very Large Array (VLA) survey of $27$ low-redshift ($z<2.7$) blazars at $1.4$\,GHz, they found arcsecond-scale extended emission in $93$\% of the sample. While these objects showed compact appearance on pc scales with VLBI observations. 

However, observing the radio emission from radio jets in blazars and extended jet-related features in radio galaxies in high-redshift sources is difficult because of the cosmological redshift of the electromagnetic radiation. For sources at higher and higher redshifts, lower and lower observing frequencies have to be employed to recover emission from these steep-spectrum features. The low-frequency observations provided at an angular resolution of few hundred mas by the international LOFAR are well-suited to study the jets of high-redshift radio quasars.

Throughout this paper, we adopt a $\Lambda$CDM cosmological model with $H_0=70$\,km\,s$^{-1}$\,Mpc$^{-1}$, $\Omega_\mathrm{m}=0.27$, and $\Omega_\Lambda=0.73$. The power-law radio spectral index $\alpha$ is defined as $S\propto\nu^\alpha$, where $S$ is the flux density and $\nu$ the observing frequency. The emitted (rest-frame) frequency is $\nu_\mathrm{em}=(1+z)\,\nu$.

\section{Sample selection}

The aim of this work is to investigate the radio structure of high-redshift quasars by combining high-resolution VLBI data with low-frequency LOFAR observations. In particular, we focus on identifying sources with extended emission that may remain undetected at GHz frequencies. 

To construct our sample, we first selected radio quasars at $z>4$ that \textit{(i)} have existing high-resolution VLBI observations \citep[for that we used the list in][]{Krezinger2022}, and \textit{(ii)} were detected in the LOFAR Two-meter Sky Survey (LoTSS; \citealt{lotssdr2}). From this initial list, we retained only those sources that are classified as having multiple radio components in LoTSS DR2 and have a total flux density of at least 100\,mJy. This flux density threshold was adopted to maximize the likelihood of obtaining high-quality imaging on the international LOFAR baselines.

\begin{table*}[ht]
\begin{minipage}{\textwidth}
\centering
\caption{\label{tab:data} Basic properties of the high-redshift sources discussed. }
\begin{tabular}{llllcccc}
\hline 
\hline
\noalign{\smallskip}
Name & RA & Dec & $z$ & $S_\mathrm{LoTSS}$ & $P_\mathrm{LoTSS}$ & VLBI reference & Scale \\
 & (h m s) &  ($\degr$ $\arcmin$ $\arcsec$) & & (mJy) & (mJy\,beam$^{-1}$) & & (kpc\,arcsec$^{-1}$) \\
\noalign{\smallskip}
\hline
\noalign{\smallskip}
J0813$+$3508 & 08 13 33.32789 & +35 08 10.7698 & $4.920$  & $311.2 \pm 2.3 $ & $119.6\pm0.3$ & \cite{Frey2010} & $6.55$ \\ 
J1231$+$3816 & 12 31 42.17206 &+38 16 59.0429 & $4.131$ & $146.3 \pm 0.7$ & $74.6 \pm 0.1$ & \cite{Krezinger2022} & $7.09$\\ 
J1548$+$3335 & 15 48 24.01400 & +33 35 00.0862 & 4.678 & $176 \pm 1$ & $138.8 \pm 0.2$ & \cite{Coppejans2016} & $6.71$ \\ 
\noalign{\smallskip}
\hline
\end{tabular}
\tablefoot{Col.~1 -- source name, Cols.~2--3 -- right ascension and declination coordinates measured with VLBI, Col.~4 -- redshift from \citet{SDSS_DR16}, Cols.~5--6 -- flux density and peak intensity from the LoTSS DR2 \citep{lotssdr2} measured in the frequency range of $120-168$\,MHz, Col.~7 -- reference for the VLBI coordinates, Col.~8 -- angular scale \citep{Wright_2006}.
}
\end{minipage}
\end{table*}

Applying these criteria resulted in four candidate quasars. One of them J1430$+$4204 ($z=4.705$) is known to host a one-sided extended jet with $24$\,kpc projected linear size detected at GHz frequencies with the VLA, and in X-rays with the Chandra satellite \citep{J1430_largejet}. Unfortunately, it was observed with only seven international LOFAR stations, providing insufficient $(u,v)$ coverage for reliable calibration and imaging. We therefore excluded it from the analysis. Our final sample thus consists of three quasars whose basic properties are summarized in Table~\ref{tab:data}.

\section{Observations and data reduction}

\subsection{LOFAR observations}

All data were obtained from the LOFAR Long-Term Archive.
J0813$+$3508 was observed on 2016 September 27 in project LC6\_015 (PI: T.W.~Shimwell) with $46$ stations in total, including $9$ international stations.
J1231$+$3816 was observed on 2019 March 22 as part of project LT10\_010 (PI: T.W.~Shimwell) with $49$ stations, including $11$ international stations.
J1548$+$3335 was observed on 2019 January 5 in project LC11\_019 (PI: H.T.~Intema) with $50$ stations, including $12$ international stations.

All sources were observed in the frequency range $120.3$–$187.3$\,MHz, providing a total bandwidth of $67$\,MHz, with standard $8$-hour on-source integrations. The raw data were recorded at $1$\,s time resolution and $3.051$\,kHz spectral resolution, and subsequently averaged to $2$\,s and $12.207$\,kHz, respectively. Flux density calibration was performed using 3C196 for J0813$+$3508 and J1231$+$3816, and 3C295 for J1548$+$3335.

To produce $0\farcs5$ resolution images, the data were processed with the LOFAR Initial Calibration (LINC) pipeline\footnote{\url{https://git.astron.nl/RD/LINC}} 
and the Pipeline for the International LOFAR Telescope (PILOT)\footnote{ \url{https://git.astron.nl/RD/pilot}}, respectively. First, the data from the Dutch array (short baselines: core and remote stations) were calibrated with LINC, which corrects for phase irregularities due to the ionospheric effects, corrects clock offsets, and performs phase-only calibration using the Tata Institute of Fundamental Research (TIFR) Giant Metrewave Radio Telescope (GMRT) Sky Survey \citep[TGSS;][]{tgss} as a sky model. 
These calibration solutions were then supplied to the LOFAR-VLBI pipeline \citep{LOFAR-VLBI_pipeline}, which combines the subbands, corrects phases and delays for the international stations, and applies delay calibrator solutions.
Final imaging was carried out using the \textsc{facetselfcal} software \citep{lofar_facet_selfcal}, performing $15$ cycles of phase self-calibration with Briggs weighting (using a robust parameter of $-0.5$). For each source, either an e-MERLIN or an EVN image was adopted as the initial sky model.

The resulting images were analyzed with the task \textsc{imfit} in the NRAO Astronomical Image Processing System \citep[{\sc aips};][]{aips}, in order to measure the flux densities of individual components.

\subsection{e-MERLIN observations}

J1548$+$3335 was observed with the e-MERLIN at $1.5$\,GHz and $5$\,GHz on 2017 May 15, June 30 and on 2017 May 13, June 26, respectively, in project CY5211 (PI: K. Gab\'anyi). The observing array included the following telescopes: Jodrell Bank Mk2, Pickmere, Darnhall, Knockin, Cambridge, and Defford. At $1.5$\,GHz and $5$\,GHz, $8$ and $4$ spectral windows were used, respectively. The total bandwidth in both bands was $512$\,MHz. The observations were done in phase-reference mode \citep{p-ref}, using the calibrator source ICRF\,J154405.6$+$324048. The on-source integration time was $16.2$\,h and $17.9$\,h at $1.5$\,GHz and $5$\,GHz, respectively. 

The data reduction was done with the e-MERLIN pipeline\footnote{ \url{ http://www.e-merlin.ac.uk/data_red/tools/eMCP.pdf}} (version v0.7.1) with extensive help from the e-MERLIN science team (J. Moldon). For the initial flagging to remove data affected by radio frequency interference (RFI), the \textsc{AOflagger} software \citep{aoflagger} was used. Subsequent calibration and editing steps were performed within the Common Astronomy Software Application \citep[\textsc{casa},][]{casa}. These involved complex gain and phase calibrations, bandpass and delay calibrations, and additional flagging. The absolute flux density scale was tied to the primary flux density calibrator 3C286. 

We did not use the images of the target sources created by the automatic pipeline. Instead, we imaged the sources using the hybrid mapping method, iteratively improving the \textsc{clean} model \citep{Hogbom_clean} with subsequent phase self-calibration steps \citep{selfcal}. We used the
\textsc{clean} method implemented in \textsc{casa} with multi-frequency synthesis imaging and second-order Taylor term. Self-calibration was performed for the phases with first for a time interval of $195$\,s afterwards decreased to $95$\,s. Amplitude self-calibration was attempted after two or three runs of phase-only self-calibration (and subsequent improvement of the \textsc{clean} models). The solution interval of amplitude self-calibration was first set to $2$\,h or $1$\,h and was decreased gradually. 

The self-calibrated data of the target source were exported from \textsc{casa} to \textsc{uvfits} format with each spectral window separately. Then the \textsc{difmap} software \citep{difmap} was used to fit the calibrated visibilities with Gaussian brightness distribution model components. We also explored whether similar results could be obtained by fitting the images. To this end, we used the \textsc{imstat} task in \textsc{aips} and found that the obtained flux densities agree within the uncertainties.

\subsection{Archival VLBI data} \label{sec:past_obs}

J0813$+$3508 was observed with the EVN at $1.6$\,GHz and $5$\,GHz on 2008 October 29 and 22, respectively. The details of the observations and data reduction were presented in \citet{Frey2010}. We re-imaged the calibrated datasets with using only the European antennas (thus excluding the longest baselines to China) in {\sc difmap}. At $1.6$\,GHz, we imaged the source with and without phase-only self-calibration. Only Effelsberg, Westerbork, Toru\'n, and Medicina were used to correct the phases. When using phase-only self-calibration, we found that the peak intensity of the image and the total summed flux density in the \textsc{clean} components were $\sim 10$\% larger compared to the ones obtained without self-calibration \citep{Frey2010}, in general agreement with the expected coherence loss \citep[e.g.,][]{coherence_loss1, coherence_loss2}. At $5$\,GHz, no self-calibration could be attempted due to the faintness of the source.

J1231$+$3816 was observed with the e-EVN + e-MERLIN array at $1.7$\,GHz on 2020 June 23. The details of the observation and the data reduction were presented in \citet{Krezinger2022}. We re-imaged the calibrated dataset with using only the e-MERLIN data in {\sc difmap}. (Self-calibration was not performed due to the faintness of the source.) In both cases, to describe the brightness distributions, we used {\sc difmap} to fit Gaussian models to the visibilities. J1231$+$3816 was also observed at $5$\,GHz on 2019 March 19, however, only with the EVN array\footnote{The observing array in \cite{Krezinger2022} erroneously lists the e-MERLIN antennas.}.

J1548$+$3335 was observed with the EVN in 2014 and 2016. The details and results of the observations are given in \citep{Coppejans2016}. 

\subsection{Low-resolution radio surveys}

For all three sources, we collected flux density measurements from various radio surveys, the TGSS at $150$\,MHz \citep{tgss}, the Westerbork Northern Sky Survey \citep[WENSS,][]{wenss} at $330$\,MHz, the Rapid ASKAP Continuum Survey \citep[RACS,][]{racs_1.37, racs_1.66} at $1.37$\,GHz and $1.66$\,GHz, the NRAO VLA Sky Survey \citep[NVSS,][]{nvss}, the Faint Images of the Radio Sky at Twenty-Centimeters survey \citep[FIRST,][]{first} at $1.4$\,GHz, and the Very Large Array Sky Survey \citep[VLASS,][]{lacy_vlass} at $3$\,GHz. We also included the $144$-MHz flux density value from LoTSS DR2 \citep{lotssdr2}, and the detection or an upper limit at $74$\,MHz from the Very Large Array Low-frequency Sky Survey Redux dataset \citep[VLSSr, ][]{VLSSr}. J1548$+$3335 has an additional GMRT $610$-MHz observation done by \citet{Coppejans2017} which we included as well.

In the case of the VLASS data, we downloaded the quick-look images of all three available epochs (additionally, a `single-epoch', SE, image in the case of J1548$+$3335 was also available) from the Canadian Astronomy Data Centre (CADC\footnote{\url{https://www.cadc-ccda.hia-iha. nrc-cnrc.gc.ca/en/vlass/}, accessed on 27 May 2025}) and fitted the images with the {\sc aips} task \textsc{imfit}. When it was appropriate (i.e., at the third epoch for J1231$+$3816 and at all epochs for J0813$+$3508), we used more than one component to fit the data. For the first epoch of the VLASS observations, \citet{VLASS-memo13} discuss the flux density scale accuracies. They found that the flux densities obtained from image-fitting the quick-look images obtained between 2017 September and 2018 February are underestimated by $10$\%. Flux densities for the other observing times are accurate within $3$\%. Therefore, we increased the flux density given by the image fitting for VLASS first-epoch observation of J1548$+$3335, which took place on 2017 December 2. In the case of J1231$+$3816, we only used the third-epoch VLASS observation as it had lower noise level, thus all three components seen in the high-resolution LOFAR image can be recovered and fitted. The first-epoch VLASS observation of J0813$+$3508 was obtained on 2019 May 6. Therefore, flux density scaling was not needed.

We did not detect significant epoch-to-epoch differences in the $3$-GHz flux densities of the VLASS observations of J0813$+$3508 and J1548$+$3335. Therefore, we use the 3-epoch average flux density values in the following. We list the VLASS flux densities in Table\,\ref{tab:radioflux}. For J0813$+$3508 and J1231$+$3816, where multiple components were fitted, we give the sum of those.

We also downloaded the LoTSS DR2 images of our sources and used image fitting to quantify the brightness distribution. In the case of J0813$+$3508, we used the results of the image fitting instead of the automated Gaussian decomposition given in the LoTSS DR2\footnote{\url{https://vo.astron.nl/lotss_dr2/q/gaus_cone/info}}. The flux densities of these components are given in Table\,\ref{tab:low-res}.

\section{Results}

\begin{figure*}
    \centering
    \includegraphics[width=17cm, bb=0 140 690 370, clip]{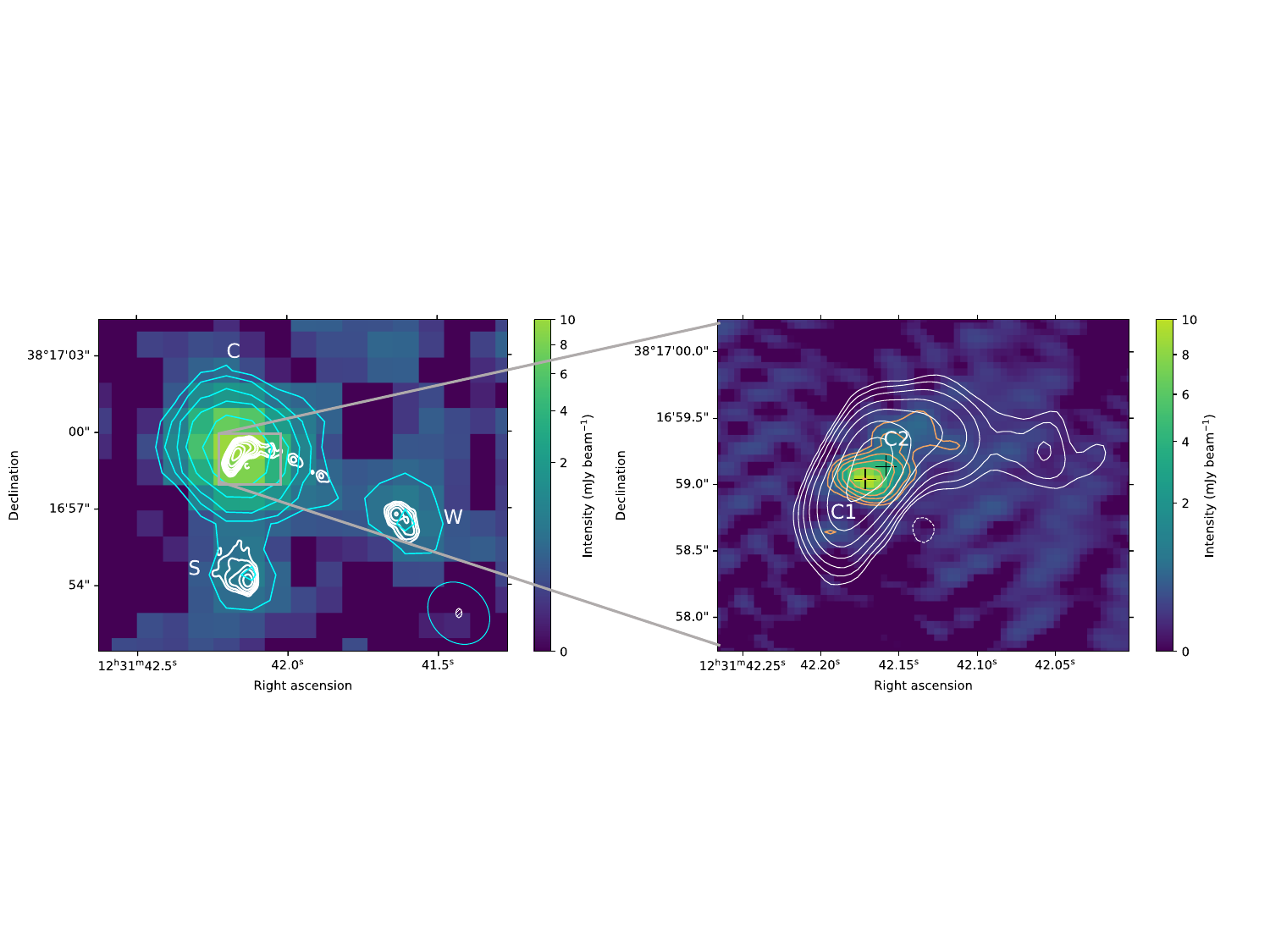}
    \caption{Radio images of J1231$+$3816. White contours show the high-resolution LOFAR image, with contours starting at $\pm 0.3$\,mJy\,beam$^{-1}$ ($\pm 5\sigma$ of the off-source image noise) and increasing by a factor of $2$. The peak intensity is $28.6$\,mJy\,beam$^{-1}$, and the restoring beam (white hatched ellipse shown in the lower right corner of the left panel) is $367\,\mathrm{mas} \times 242\,\mathrm{mas}$ with $\mathrm{PA}=-1\degr$. {\it Left panel:} The colour scale and cyan contours show the $3$-GHz VLASS third-epoch image. Contours start at $0.36$\,mJy\,beam$^{-1}$ ($3 \sigma$ of the off-source image noise) and increase by factors of $2$; the peak brightness is $11.9$\,mJy\,beam$^{-1}$. The VLASS restoring beam (cyan line, lower right corner of the image) is $2\farcs65 \times 2\farcs20$ and the major axis position angle is $\mathrm{PA}=44\degr$. {\it Right panel:} Zoom-in on the LOFAR feature C. The colour scale and the orange contours show the $1.7$-GHz e-MERLIN image, with a peak brightness of $11.8$\,mJy\,beam$^{-1}$. The orange contours start at $0.4$\,mJy\,beam$^{-1}$ (at $3\sigma$ image noise level) and increase by a factor of $2$. The two black crosses and white labels mark the positions of the two components fitted to the e-MERLIN visibilities.}
    \label{fig:J1231_big}
\end{figure*}

The radio images obtained for the three quasars are shown in Figs.\,\ref{fig:J1231_big}, \ref{fig:J0813_maps}, \ref{fig:J1548_lofar_emerlin}, and \ref{fig:J1548_lofar_emerlinC}. The image noise levels are given in Table\,\ref{tab:image_noise}.

\subsection{J1231$+$3816}

\subsubsection{Comprehensive radio morphology}

The high-resolution LOFAR image of J1231$+$3816 reveals a complex morphology with three bright radio-emitting regions, labeled as C, S, and W in Fig.~\ref{fig:J1231_big}. A faint ``bridge'' consisting of a few knots appears to (partially) connect features C and W. The angular separations of S and W from the brightest feature (C) are $\sim 5\arcsec$ and $\sim 7\arcsec$, corresponding to projected linear sizes of $\sim 35.5$\,kpc and $\sim 49.6$\,kpc, respectively. The overall radio structure is bent into a U-shaped form, with a jet bending angle (S--C--W) visually estimated to be less than $90\degr$. On the other hand, the structure of region C implies a larger bending in the inner region (see right panel of Fig.\,\ref{fig:J1231_big}).

The $1.7$-GHz e-MERLIN-only data show a compact component with an extension to the north-northwest, aligned with the extension of feature C in the 144-MHz high-resolution LOFAR image (right panel of Fig.~\ref{fig:J1231_big}). No radio emission is detected at the locations of features S and W down to the $3\sigma$ image noise level of 0.36\,mJy\,beam$^{-1}$. The e-MERLIN visibilities are well described by one circular (C1) and one elliptical (C2) Gaussian components. Their positions are shown in Fig.\,\ref{fig:J1231_big}, the fitted model parameters are given in Table~\ref{table:J1231_eMERLIN}.

\subsubsection{Radio spectrum}

The integrated radio spectrum of J1231$+$3816 using archival low-resolution data (Table~\ref{tab:radioflux}) is shown in  Fig.~\ref{fig:J1231_spectrum}. From these measurements, we obtain a spectral index of $\alpha_\mathrm{J1231}=-0.79 \pm 0.01$. The expected flux density of $\sim 248$\,mJy at $74$\,MHz, extrapolated using this spectral index, is consistent with the upper limit from VLSSr, providing no evidence for a low-frequency spectral turnover.

Spectral indices for the three regions detected in the high-resolution LOFAR image can also be derived separately. All three are detected in the third-epoch VLASS image (Fig.\,\ref{fig:J1231_big}), and for region C, we additionally include the summed flux density from the e-MERLIN components C1 and C2. The resulting spectral indices are $\alpha^\mathrm{J1231}_\mathrm{C}=-0.58 \pm 0.01$, $\alpha^\mathrm{J1231}_\mathrm{W}=-0.93 \pm 0.08$, and $\alpha^\mathrm{J1231}_\mathrm{S}=-1.12 \pm 0.08$. Region C has the flattest spectrum, while S and W show steeper spectra, as expected for extended jet-related structures. 

The spectral index of region C predicts a 5-GHz flux density of $\sim 10.6$\,mJy, slightly higher than the $(6.8 \pm 0.7)$\,mJy recovered in the 5-GHz EVN observations \citep{Krezinger2022}. This discrepancy may indicate the presence of a few mJy extended emission at 5\,GHz, or a spectral break at higher frequencies. Since 5\,GHz corresponds to $\sim 27$\,GHz in the rest frame of J1231$+$3816, the latter explanation is more plausible.

\subsection{J0813$+$3508}

\begin{figure*}
    \centering
    \includegraphics[width=17cm, bb = 0 0 710 540, clip]{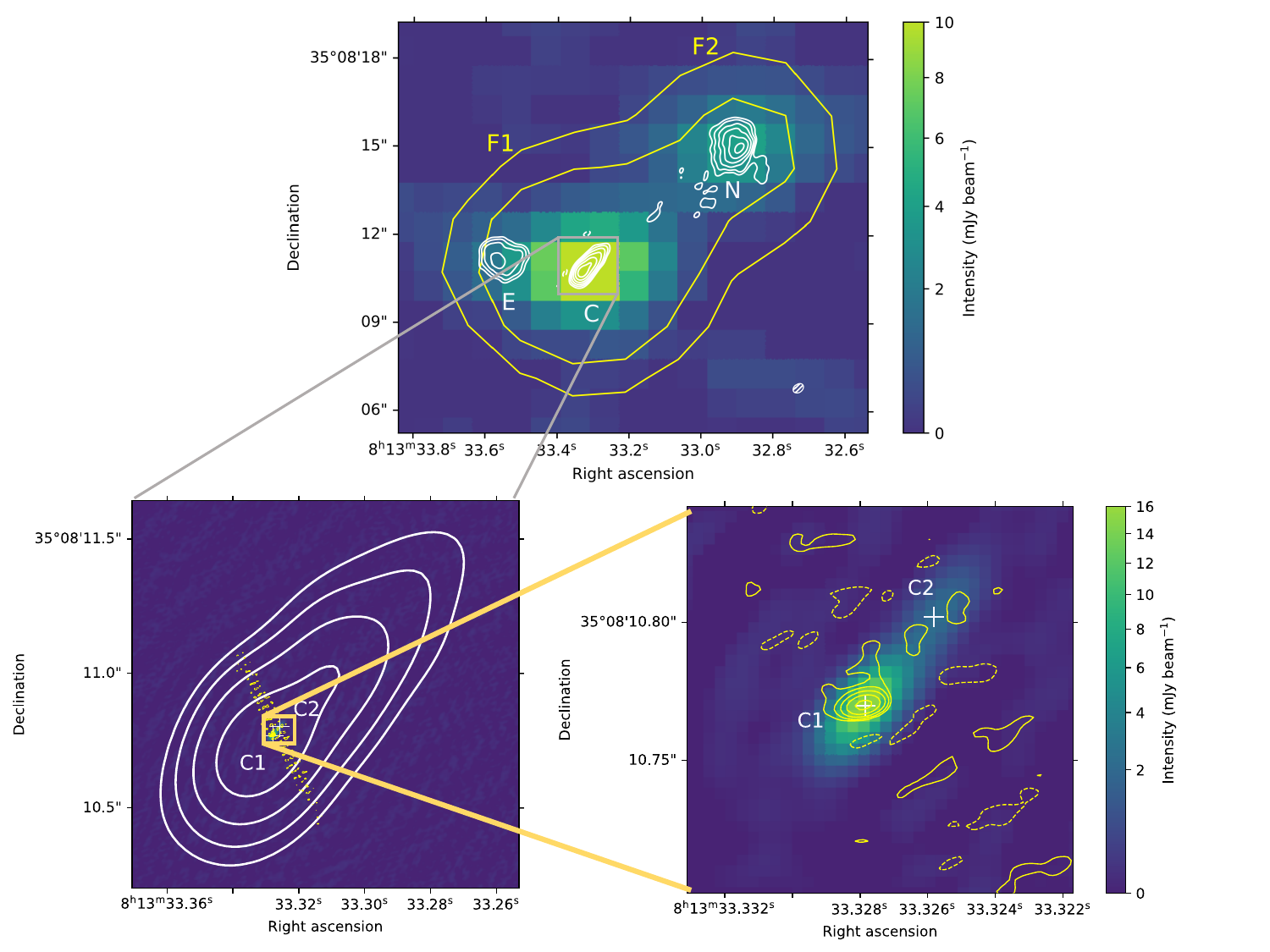}
    \caption{Radio images of J0813$+$3508. Top: White contours show the high-resolution LOFAR image, with the lowest levels at $\pm 0.8$\,mJy\,beam$^{-1}$ (corresponding to $\pm 4\sigma$ off-source noise level); higher contours increase by a factor of $2$. The peak brightness is $42.7$\,mJy\,beam$^{-1}$. The restoring beam, shown as a hatched white ellipse in the lower right corner, is $368\,\mathrm{mas} \times 272\,\mathrm{mas}$ with $\mathrm{PA}=130\degr$. The third-epoch VLASS image is displayed in colour scale. Yellow contours show the $1.4$-GHz emission from the FIRST image, drawn at $4.3$\,mJy\,beam$^{-1}$ and at $8.6$\,mJy\,beam$^{-1}$. The emitting regions identified in the international LOFAR image are shown by white letters. The two FIRST components are indicated by yellow symbols. Bottom: Zoom-in on feature C. The colour scale shows the $1.6$-GHz EVN image (using only European antennas) obtained on 2008 October 29 \citep{Frey2010}, with a peak brightness of $17.2$\,mJy\,beam$^{-1}$. The restoring beam is $26.1\,\mathrm{mas} \times 13.1\,\mathrm{mas}$ with $\mathrm{PA}=-37\degr$. White symbols and labels mark the positions of the Gaussian components fitted to the $1.6$-GHz EVN visibilities. Yellow contours show the $5$-GHz EVN image obtained using the European antennas only, and with only one Gaussian component fitted to the visibilities. The high residual noise level indicates the presence of additional emission which could not be adequately fitted. The peak intensity is $5.6$\,mJy\,beam$^{-1}$. The lowest contours are drawn at $\pm 4\sigma$ image noise level corresponding to $\pm 0.3$\,mJy\,beam$^{-1}$. The restoring beam size is $10.6\,\mathrm{mas} \times 5.1\,\mathrm{mas}$ with $\mathrm{PA}=-80\degr$. White contours show the high-resolution LOFAR image in the left panel; contour levels increase by a factor of $2$, starting at $3.2$\,mJy\,beam$^{-1}$.}
    \label{fig:J0813_maps}
\end{figure*}

\subsubsection{Comprehensive radio morphology}

The high-resolution LOFAR image of J0813$+$3508 reveals three main emission regions, labeled C, N, and E (Fig.~\ref{fig:J0813_maps} top panel). 
Regions N and E are located $\sim 6\farcs8$ to the north (corresponding to $\sim 44.5$\,kpc projected linear size) and $\sim 4\farcs6$ to the east (corresponding to $\sim 30.3$\,kpc) of the brightest part of region C, respectively. 
The overall radio structure is bent into a C-shape, with a visually estimated jet bending angle slightly greater than $90\degr$.

We used the \textsc{aips} task \textsc{imfit} to model the three regions with Gaussian components. In each of the regions C and N, two elliptical Gaussians were fitted, while in region E, one elliptical Gaussian could adequately describe the radio brightness distribution. The summed flux densities of these fitted features are given in Table~\ref{tab:lofar}.

Re-imaging the archival EVN data, but considering the intra-European baselines only, yielded results similar to those of \citet{Frey2010}. At $1.6$\,GHz, the source shows an elongated structure aligned with the extension of region C seen in the LOFAR image (Fig.~\ref{fig:J0813_maps}). 
The visibilities can be fitted with two elliptical Gaussian components (C1 and C2, in the bottom panels of Fig.\,\ref{fig:J0813_maps}, and in Table\,\ref{table:J0813_EVN}), although their sizes must be held fixed to avoid unrealistically small values. At $5$\,GHz, a single compact feature is recovered, but residual emission remains in the central part, inconsistent with a simple Gaussian model (see bottom right panel of Fig.\,\ref{fig:J0813_maps}, and Table\,\ref{table:J0813_EVN}). The compact component has a $5$-GHz flux density of $(7.0 \pm 0.4)$\,mJy after correcting for a $10$\% coherence loss (see Sect.~\ref{sec:past_obs}). From the residual map, we estimate there is additional $\sim 1$\,mJy extended emission not accounted for by the model. At both frequencies, the flux densities are slightly higher and component sizes are significantly larger than those obtained by \citet{Frey2010} with the inclusion of the longest intercontinental baselines, as expected.

\subsubsection{Radio spectrum}

Generally, at the lower-resolution images -- at $1.4$\,GHz with FIRST, at $3$\,GHz with VLASS, and at $144$\,MHz according to the LoTSS image (Fig.\,\ref{fig:J0813_lowres}) --, J0813$+$3508 can be resolved into two components. However, the low noise level of the third-epoch VLASS data allowed the detection of the $3$-GHz radio emission at the position of region E as well. In the FIRST catalogue, the flux densities given for the two components are $S_\mathrm{F2}^\mathrm{J0813}= (11.5 \pm 0.2)$\,mJy for the northern, and $S_\mathrm{F1}^\mathrm{J0813}= (37.5\pm 0.2)$\,mJy for the southern one (see Fig.\,\ref{fig:J0813_maps}). In case of the LoTSS data, the image can be described by several slightly different two-component fits, depending on the starting values of the components. Therefore, we fixed the position of the southern component at the approximate location of the optical galaxy (Fig.\,\ref{fig:J0813_maps}).

Using these low-resolution survey data, we attempted to estimate the spectral indices of the emission regions. Region N can be straightforwardly identified across the three surveys. In the case of the LoTSS image, its flux density, $(157.5\pm 15.8)$\,mJy, is slightly larger than the one derived from the international LOFAR data, $(136.4 \pm 0.7)$\,mJy, indicating that some large-scale emission is resolved out in the higher-resolution measurement. Using the data from the arcsec-scale resolution surveys, we obtained $\alpha^\mathrm{J0813}_\mathrm{N}=-1.1 \pm 0.1$ (shown by the blue solid line in Fig.~\ref{fig:J0813_spectrum}).

The flux density of component C in the LoTSS image, $\sim 139$\,mJy (Table\,\ref{tab:low-res}), is below the summed flux density of regions C and E, $(170.9\pm 1.0)$\,mJy, in the international LOFAR data. To obtain the spectral index of region C, we only used the LoTSS, and the third-epoch VLASS data, where it can be clearly resolved from region E. The two-point spectral index is $\alpha^\mathrm{J0813}_\mathrm{C}=-0.80 \pm 0.03$ (shown by the dashed turquoise line in Fig.\,\ref{fig:J0813_spectrum}). By extrapolation, this would imply a $1.6$-GHz flux density of $\sim 20$\,mJy and a $5$-GHz flux density of $\sim 8.1$\,mJy for region C, which are in rough agreement with the flux densities obtained from the mas-scale VLBI observations. On the other hand, the implied $1.4$-GHz flux density, $\sim 22.5$\,mJy is significantly lower than the flux density of F1, indicating that the FIRST feature is the blended emission of regions C and E. 

To approximate the spectral index of region E, the flux density derived from international LOFAR data and from the third-epoch VLASS observation can be used. We get $\alpha^\mathrm{J0813}_\mathrm{E}=-1.0 \pm 0.03$. Thus, all three regions have steep radio spectra, with region C having the highest spectral index value, i.e. the least steep radio spectrum.

The integrated spectrum of the whole source, derived from low-resolution measurements (Table~\ref{tab:radioflux}) and summed FIRST and VLASS flux densities, gives $\alpha^\mathrm{J0813}=-0.82 \pm 0.03$ (black line in Fig.~\ref{fig:J0813_spectrum}). 
However, the $150$-MHz TGSS \citep{tgss} and the LoTSS \citep{lotssdr2} flux densities taken at close frequencies differ significantly. 
This cannot be explained by resolution effect, since the higher-resolution LoTSS flux density is more than twice the TGSS value. The LoTSS observation of J0813$+$3508 was performed on 2016 September 27, while the TGSS data were taken between 2010 April and 2012 March \citep{tgss}. If the flux density difference is real, it implies substantial variability on months-to-year timescales in the rest frame of J0813$+$3508. The $74$-MHz VLSSr measurement \citep{VLSSr} also falls below the extrapolation of the steep spectrum, suggesting either a spectral peak below $150$\,MHz observed (i.e. $\sim 888$\,MHz rest-frame) frequency, or variability. In the latter case, J0813$+$3508 must have been significantly fainter during the VLSS and TGSS epochs compared to the later LOFAR observation.

Radio flux density variability can also be caused by source-extrinsic, propagation effects in the ionized interstellar medium of the Milky Way \citep{Rickett1990}. Specifically, refractive interstellar scintillation can cause slow, broad-band variability \citep[e.g.][and references therein]{Walker1998}. At the galactic position of J0813$+$3508, at a Galactic latitude of $\sim 30^\circ$, the expected transition frequency is $\sim 10$\,GHz, implying a refractive time scale of $\sim 2.4$\,yr at $150$\,MHz for a point source, a source with a size below the scattering disc size. This is $\sim 30$\,mas at the position of J0813$+$3508 at $150$\,MHz \citep{Walker1998}. The $144$-MHz structure and flux density of J0813$+$3508 is extended, however it contains a bright, dominant feature in region C (see Fig.\,\ref{fig:J0813_maps}) with a deconvolved source size of $\sim 180$\,mas according to our image fitting. Thus, refractive scintillation can cause flux density variability with characteristic time scale as long as $14$\,yr according to the scaling relations given in \cite{Walker1998}. On the other hand, the modulation index of the variability, the rms fractional variation, can be at most $10$\%, which is reduced severely with increasing source size. Thus, while the variability time scale can be reconciled with refractive interstellar scintillation, the amount of flux density variability is much larger than expected from this effect. We note that extreme scattering events (ESE) are known to be responsible for large dips in radio light curves of pulsars and quasars. The longest known ESEs, $(3-4)$\,yr, were observed towards two pulsars and they caused a fading of the flux density by $\sim 60$\% \citep{ese_pulsar}. 

Moderate variability is also seen at $1.4$\,GHz when comparing FIRST \citep{first}, NVSS \citep{nvss}, and RACS-mid \citep{racs_1.37} data of J0813$+$3508. The NVSS ($\sim 36$\,mJy) and RACS-mid ($\sim 42$\,mJy) flux densities are both lower than the FIRST value ($\sim 49$\,mJy). Since both NVSS and RACS-mid have larger restoring beam than the FIRST, these differences cannot be caused by resolution issues. The rest-frame time separations between these surveys are $\sim 28$\,days (NVSS–FIRST) and $\sim 5$\,years (FIRST–RACS). Such modest variability within such time range is plausible, but the much larger difference implied by the TGSS and LOFAR data (on rest-frame timescales of a few months) remains puzzling.

\subsection{J1548$+$3335}

\subsubsection{Comprehensive radio morphology}
\begin{figure*}
   \begin{minipage}[t]{\columnwidth}
   \centering
       \includegraphics[width=\columnwidth, bb = 80 0 500 310, clip]{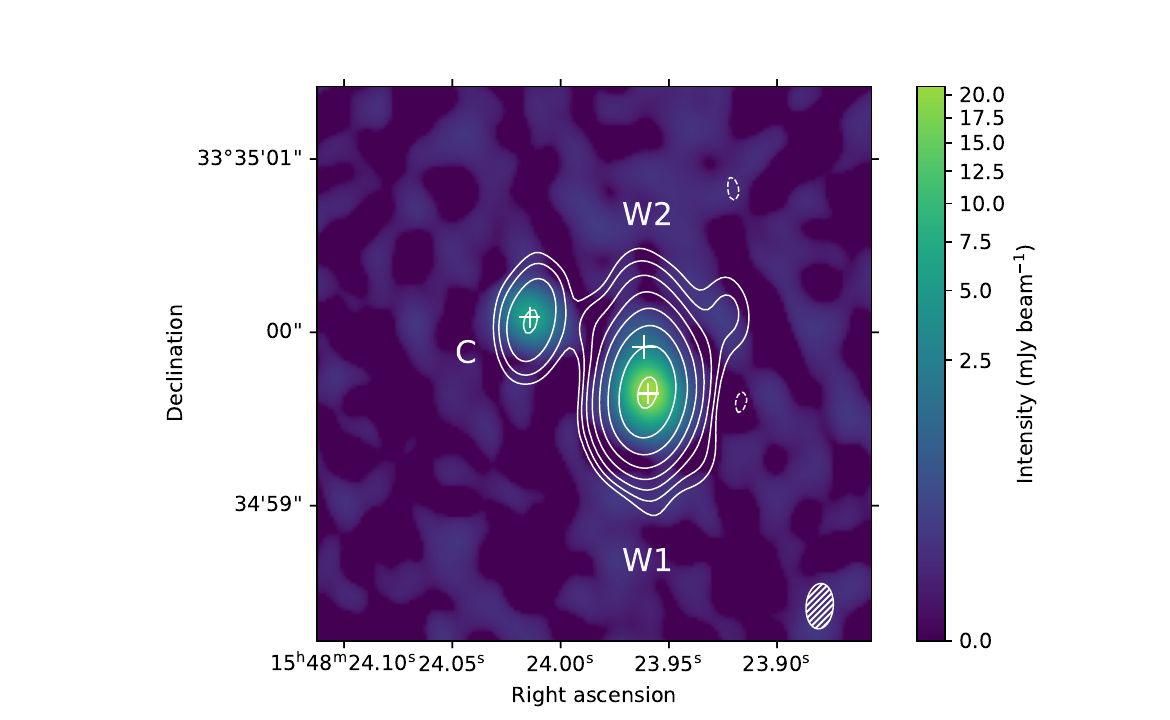}
      \caption{High-resolution LOFAR (white contours) and $1.5$-GHz e-MERLIN (colour scale) images of J1548$+$3335. The peak intensity in the LOFAR image is $56.0$\,mJy\,beam$^{-1}$. Contours start at $\pm 0.38$\,mJy\,beam$^{-1}$ ($\pm 5\sigma$ of the off-source image noise) and increase by a factor of $2$. The LOFAR restoring beam, shown as a hatched ellipse in the lower right corner, is $309\,\mathrm{mas} \times 185\,\mathrm{mas}$ with $\mathrm{PA}=4\fdg3$. The peak intensity in the e-MERLIN image is $23.2$\,mJy\,beam$^{-1}$. White crosses mark the positions of Gaussian features fitted to the e-MERLIN visibility data. In the case of W2, the size of the symbol represents the standard deviation of the coordinates of the fitted features in different spectral windows. For the other components, these differences are much smaller than the symbol size.
              }
         \label{fig:J1548_lofar_emerlin}
   \end{minipage}
   \hfill
   \begin{minipage}[t]{\columnwidth}
   \centering
       \includegraphics[width=\columnwidth, bb = 50 0 490 310, clip]{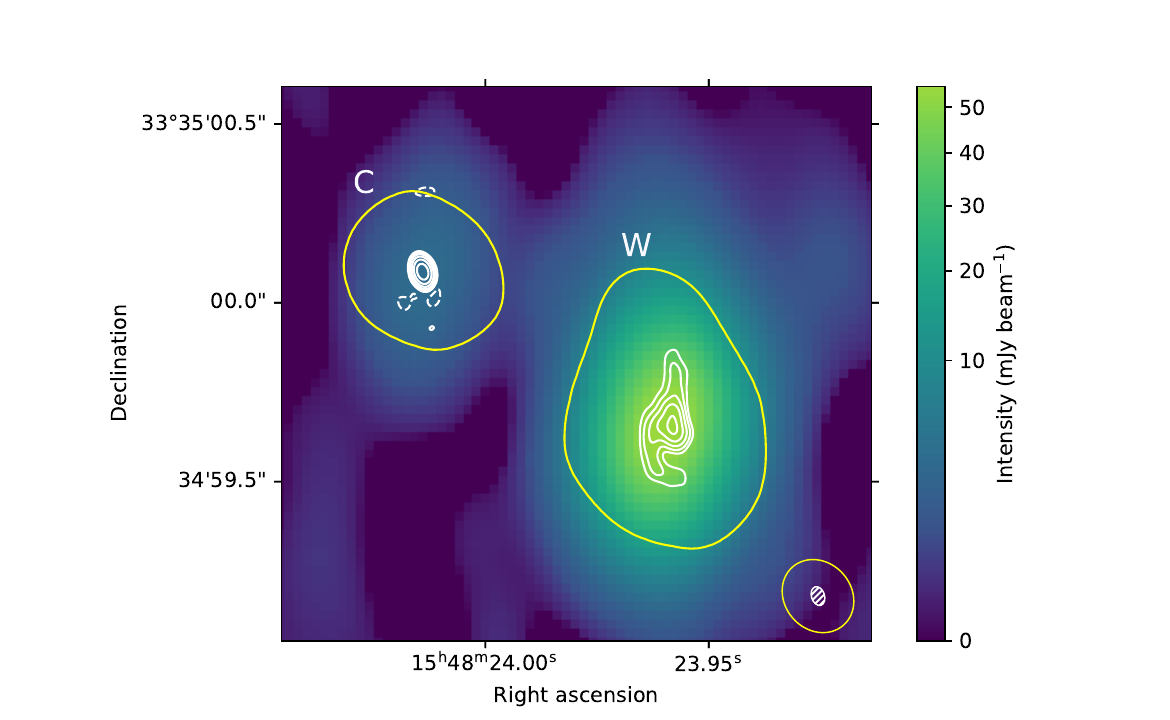}
      \caption{High-resolution LOFAR (colour scale) and $5$-GHz e-MERLIN (white contours) images of J1548$+$3335. The peak intensity in the e-MERLIN image is $4.2$\,mJy\,beam$^{-1}$. Contours start at $\pm 0.07$\,mJy\,beam$^{-1}$ ($\pm 4 \sigma$) and increase by a factor of $2$. The e-MERLIN $5$-GHz restoring beam, shown as a white hatched ellipse in the lower right corner, is $60\,\mathrm{mas} \times 40\,\mathrm{mas}$ with $\mathrm{PA}=16\degr$. For comparison, a single contour from the $1.5$-GHz e-MERLIN image is overlaid with thick yellow line at $0.3$\,mJy\,beam$^{-1}$ ($8\sigma$). The yellow ellipse indicates the $1.5$-GHz e-MERLIN restoring beam, $216\,\mathrm{mas} \times 187\,\mathrm{mas}$ with $\mathrm{PA}=41\degr$.
              }
         \label{fig:J1548_lofar_emerlinC}
   \end{minipage}
   \end{figure*}

The $1.5$- and $5$-GHz e-MERLIN images together with the $144$-MHz high-resolution LOFAR map of J1548$+$3335 are shown in Figs.~\ref{fig:J1548_lofar_emerlin} and \ref{fig:J1548_lofar_emerlinC}. 
All datasets reveal two emitting regions separated by $\sim 1\arcsec$, corresponding to a projected linear size of $\sim 6.7$\,kpc, arranged along a common axis. 
The western feature (W) lies at a position angle of $\mathrm{PA} \approx -125\degr$.  

Region C can be consistently fitted with a single circular Gaussian model in both the $1.5$- and $5$-GHz e-MERLIN observations, across all spectral windows. In contrast, region W requires two Gaussian components at $1.5$\,GHz: a brighter elliptical (W1) and a fainter circular one (W2). Stable models with the same number of free parameters were obtained in all but one spectral window; at $1.606$\,GHz, the position of the fainter component had to be fixed to achieve convergence. At $5$\,GHz, region W appears to be more resolved (Fig.\,\ref{fig:J1548_lofar_emerlinC}). The visibilities could be fitted with several model components of Gaussian brightness distribution, however, the individual components do not relate to physically separate emitting features. Their summed flux densities agree with the total flux densities of the \textsc{clean} components describing region W. Therefore, in Table~\ref{tab:J1548_comp}, where the flux densities of regions C and W are given, we listed the sum of \textsc{clean} components at $5$\,GHz. The high-resolution LOFAR image is also well described by two Gaussian components, with flux densities reported in Table~\ref{tab:lofar}.

\subsubsection{Radio spectrum}

The radio spectrum shown in Fig.~\ref{fig:J1548_spectrum} is based on flux densities from Table~\ref{tab:radioflux}. J1548$+$3335 has a broad-band spectral index of $\alpha^\mathrm{J1548}=-0.67 \pm 0.01$. 
It is undetected in the VLSSr at $74$\,MHz, with an upper limit of $\sim 213$\,mJy, whereas the extrapolated spectrum predicts a flux density of $\sim 272$\,mJy at $74$\,MHz. This discrepancy suggests a spectral turnover between $144$ and $74$\,MHz (rest-frame $\sim 0.8$–$0.4$\,GHz). The broad-band spectral index also predicts $\sim 16$\,mJy at $5$\,GHz, only $\sim 2.3$\,mJy higher than the combined e-MERLIN flux densities.

A single power-law cannot describe the  e-MERLIN and international LOFAR flux densities of components C and W. Fitting the e-MERLIN data only yields spectral indices $\alpha^\mathrm{J1548}_\mathrm{C}=-0.29 \pm 0.01$ and $\alpha^\mathrm{J1548}_\mathrm{W}=-1.13 \pm 0.03$, shown as solid blue and turquoise lines in Fig.~\ref{fig:J1548_spectrum}. However, extrapolation from these fits overpredicts the LOFAR flux densities of the corresponding features, suggesting a spectral flattening between $\sim 1.3$\,GHz and $144$\,MHz. Indeed, power-law fits including the LOFAR points give flatter slopes: $\alpha^{\mathrm{J1548}}_\mathrm{C}=-0.11 \pm 0.05$ and $\alpha^{\mathrm{J1548}}_\mathrm{W}=-0.61 \pm 0.01$ (dashed lines in Fig.~\ref{fig:J1548_spectrum}).

\section{Discussion}

Our international LOFAR, e-MERLIN, and EVN observations provide new insights into the nature of three $z>4$ radio-loud quasars. In all cases, we detected resolved radio structures on scales from a few hundred mas to several arcseconds, corresponding to projected linear sizes of several to tens of kpc. 

\subsection{J1231$+$3816}

The high-resolution LOFAR image of J1231$+$3816 reveals a complex, U-shaped morphology with three bright features (C, S, and W), as well as a possible faint bridge connecting them. The central, brightest feature (C) has the least steep radio spectrum, and contains the mas-scale compact radio core detected with VLBI \citep{Krezinger2022}. The non-Doppler-boosted nature of the mas-scale feature agrees with the arcsecond-scale radio morphology seen in the high-resolution LOFAR data (Fig.~\ref{fig:J1231_big}), both implying large jet viewing angle with respect to the line of sight.

Components S and W have steeper radio spectra and can be classified as radio lobes: in addition to the diffuse emission, they host hotspots compact at the scale of the international LOFAR resolution. The southern lobe (S) shows a hotspot farther away from the central region, while the hotspot in the western lobe (W) is located towards the central region (C), so it does not show an edge-brightened morphology. This hybrid character suggests a complex evolutionary history and development within a non-uniform environment.

At the highest resolution, in the $5$-GHz EVN data, only one compact feature could be detected \citep{Krezinger2022}. Its flux density, $(6.8\pm0.7)$\,mJy, roughly agrees with the value of $\sim 9$\,mJy implied by the radio spectrum of component C. Since $5$\,GHz observing frequency corresponds to $\sim 26$\,GHz in the rest frame of the source, it is not surprising that radio emission from the AGN centre can be detected.

\begin{figure}
    \centering
    \includegraphics[width=\hsize]{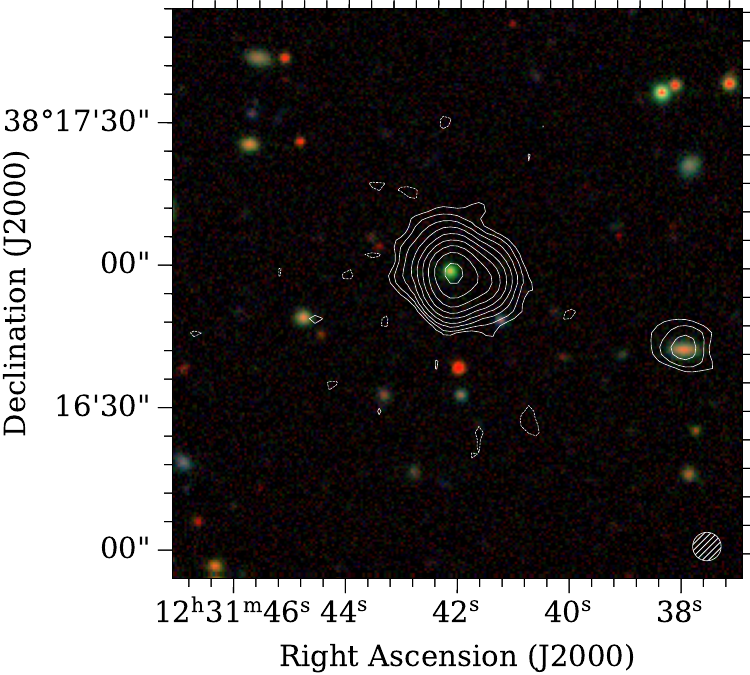}
    \caption{The low-resolution LoTSS-DR2 image (white contours) overlaid on the DESI \citep{desi_2019} three-color image of J1231$+$3816. The peak of the LOFAR image is $74.6$\,mJy\,beam$^{-1}$, the lowest contours are drawn at $\pm 0.25$\,mJy\,beam$^{-1}$ (corresponding to $\pm 2.5\sigma$ off-source image noise level), further contour levels increase by a factor of $2$. The restoring beam, shown in the lower right corner, has a size of $6\arcsec$.}
    \label{fig:J1231_lowres}
\end{figure}

On $\sim 10\arcsec$ scales, in the lower-resolution LoTSS-DR2 image, J1231$+$3816 exhibits a west--southwest elongation. In addition, a further emission feature is seen in the same direction at a separation of $\sim 1\arcmin$ (Fig.~\ref{fig:J1231_lowres}). This feature is not recovered in the high-resolution LOFAR data, likely due to its highly resolved structure at $144$\,MHz. If it were physically associated with J1231$+$3816, the angular separation would imply a projected linear size of $\sim 425$\,kpc. However, an extended optical source positionally coincident with this radio feature appears in SDSS DR12 \citep{sdss_dr12} and is classified as a galaxy. It was also detected with the Dark Energy Survey Instrument (see Fig.\,\ref{fig:J1231_lowres}) with an associated photometric redshift of $0.31\pm0.06$ \citep[DESI,][]{desi_2019}. These suggest that the emission more likely originates from a foreground galaxy rather than from J1231$+$3816 itself. Thus, in the following, we regard the radio structure recovered by the high-resolution LOFAR data as the full known extent of J1231$+$3816. Although its bent, U-shaped appearance makes the source appear relatively compact, the full extent of the emission on either side of the nucleus reaches a largest (projected) linear size of $\sim 85$\,kpc. 

J1231$+$3816 shows a modest arm-length asymmetry, with an arm-length ratio (ALR) of $\sim 1.4$. Such a value would already indicate only mild orientation effects in a straight, symmetric system. However, in the case of a strongly bent structure, it is far more likely that the jet dynamics are shaped primarily by environmental influences rather than by beaming alone. This kind of morphology could arise from interactions with neighboring galaxies, jet--intracluster medium interactions, or tidal forces within a galaxy group \citep[e.g., ][]{Wing2011, sasmal_2022}.

\subsection{J0813$+$3508}

\begin{figure}
    \centering
    \includegraphics[width=\hsize]{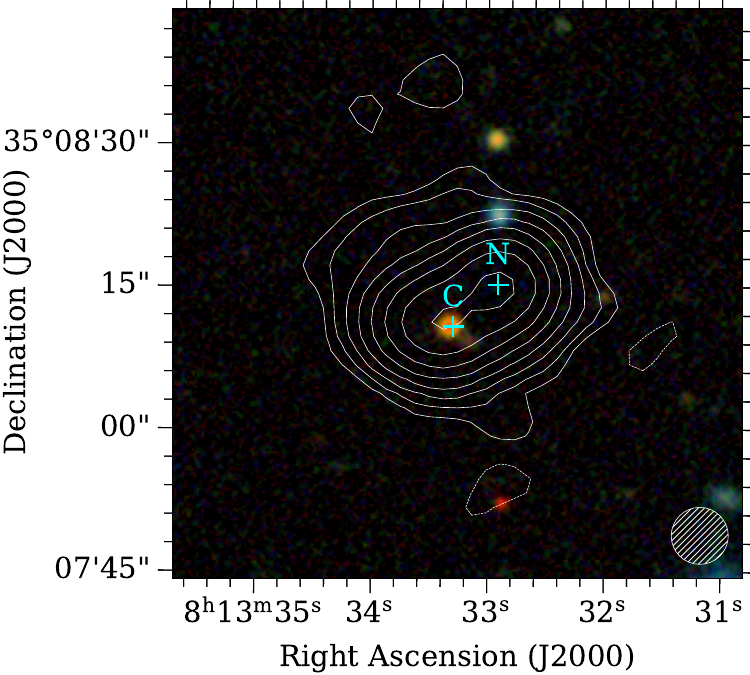}
    \caption{The low-resolution LoTSS-DR2 image (white contours) overlaid on the DESI \citep{desi_2019} three-color image of J0813$+$3508. The peak intensity of the LOFAR image is $119.6$\,mJy\,beam$^{-1}$, the lowest contour is drawn at $1.4$\,mJy\,beam$^{-1}$ (corresponding to $4\sigma$ off-source image noise level) and further contour levels increase by a factor of $2$. The restoring beam is shown in the lower right corner, its size is $6\arcsec$. The two fitted Gaussian components are marked with cyan crosses and labels.}
    \label{fig:J0813_lowres}
\end{figure}

The high-resolution LOFAR image of J0813$+$3508 reveals a complex morphology with three bright features (C, N, and E).
The brightest structure, C -- resolved into two subcomponents with $1.6$-GHz VLBI observation -- is elongated toward component N, with several faint radio features located between them (Fig.~\ref{fig:J0813_maps}). All three regions exhibit steep radio spectra, with region N showing the steepest slope. In optical, no Gaia or SDSS counterparts are found at the positions of the N and E components, thus they can be classified as radio lobes with FR~II-like morphology. 

The VLBI observations did not show Doppler boosting and the position of the central engine at that time could not be ascertained by \cite{Frey2010}. Now, however, the VLBI radio and the comparably accurate Gaia \citep{Gaia} optical positions can be compared. The $5$-GHz peak position determined by \citet{Frey2010} is consistent, within errors, with the Gaia Data Release 3 \citep[DR3,][]{gaia_dr3} optical position. Thus, the most compact, brightest radio feature, located in region C coincides with the quasar’s nucleus. The overlaid LoTSS radio and DESI DR9 optical images are shown in  Fig.\,\ref{fig:J0813_lowres}.

Similar to J1231$+$3816, this source also shows an arm-length asymmetry. The largest linear size, defined as the total extent of both arms, is $\sim 75$\,kpc, while the ALR is $\sim 1.5$. This value suggests only mild orientation effects. The fact that this asymmetry is already visible at low frequencies -- i.e., in the oldest plasma -- indicates that strong environmental effects have influenced the evolution of the source since its earliest stages.

\subsection{J1548$+$3335}

The high-resolution LOFAR and e-MERLIN images of J1548$+$3335 reveal a compact core–jet morphology (Fig.~\ref{fig:J1548_lofar_emerlin}). At mas scales, \cite{Coppejans2016} detected a compact feature within region C both at $1.7$\,GHz and $5$\,GHz. Its radio position agrees with the SDSS optical position of the quasar \citep{Coppejans2016}. The flux density of C measured in the $1.67$-GHz spectral window of the e-MERLIN data, $(6.68 \pm 0.09)$\,mJy, is close to the EVN value at $1.7$\,GHz, $(7.7 \pm 0.4)$\,mJy. At $5$\,GHz, e-MERLIN measures a flux density of $(5.01 \pm 0.05)$\,mJy, slightly higher than the EVN value, $(3.7 \pm 0.2)$\,mJy. Because the $5$-GHz EVN data were not self-calibrated, a $\sim 10$\% flux density loss is expected \citep{Coppejans2016}. A correction for this coherence loss reduces the flux density difference to $<1$\,mJy. These results indicate that component C is compact at GHz frequencies, and there is no sign of extended emission at $\sim 100$-mas scales sampled by e-MERLIN. The flat radio spectrum of component C further strengthens its association with the AGN core.  

The western component (W), located $\sim 1\arcsec$ away ($\sim 6.7$\,kpc) from C, shows a steep spectrum and extended structure, consistent with downstream jet emission. No counter-jet is detected on the opposite side of C either in the $1.6$-GHz e-MERLIN data (down to $0.46$\,mJy\,beam$^{-1}$) or in the $144$-MHz LOFAR image (down to $0.4$\,mJy\,beam$^{-1}$). This, together with the asymmetric structure, suggests that orientation effects play a role in determining the appearance of the radio source. 
While the one-sidedness of the radio structure is indicative of relativistic effects, \cite{Coppejans2016} did not detect relativistic boosting of the core emission. 
This is further supported by the moderate core dominance of J1548. Defining the core dominance parameter as
\begin{equation}
\label{eq:logR}
\log R_\mathrm{core} =\log \left( \frac{S_{\rm core}}{S_{\rm total}-S_{\rm core}}\right), 
\end{equation}
we obtain $\log\,R_\mathrm{core}=-1.2$, $-0.7$, and $-0.2$ at $144$\,MHz, $1.5$\,GHz, and $5$\,GHz, respectively. The increase with observing frequency is naturally explained by the flatter spectrum of the compact core relative to the steep-spectrum jet emission. The GHz-frequency values are comparable to those reported for compact steep-spectrum (CSS) quasars in the literature \citep[e.g.][]{Saikia2001,Fan2003}, indicating that the source is only moderately core-dominated. To our knowledge, the frequency dependence of the core dominance parameter has not yet been systematically investigated at LOFAR frequencies.

The jet-to-counter-jet flux density ratio is at least $348$ based on the $144$-MHz flux density and upper limit values. If orientation effect renders the counter-jet undetectable, one can estimate the possible ranges of intrinsic jet speeds, $\beta$ (measured in units of the speed of light) and orientation angles, $\theta$ as \citep[e.g.,][]{counterjet}:
\begin{equation}
       R =\left(\frac{1+\beta\cos \theta}{1-\beta \cos \theta}\right)^{2-\alpha},
\end{equation}
where $R$ is the jet-to-counter-jet flux density ratio and $\alpha$ is the spectral index. Using $\alpha_\mathrm{W}^\mathrm{J1548}=-0.61$, and the lower limit on $R$, one obtains $\beta \cos \theta \gtrsim 0.808$. Assuming $\beta=0.99$ (corresponding to a Lorentz factor value of $\sim 7.1$) and to remain consistent with the absence of strong Doppler boosting, the inclination angle should be between $30\degr$ and $35\degr$. This would imply a deprojected arm-length of $\sim(11-13)$\,pc. Principally, a lower $\beta$ value would allow for even smaller inclination angle, however, then relativistic Doppler boosting of the radiation would be expected. For example, the lowest possible $\beta$ value of $0.81$ (corresponding to a Lorentz factor of $\sim1.7$) that can result in such an asymmetric flux density difference would imply an inclination angle of $\theta\sim 4\degr$, thus would lead to detectable Doppler boosting.

The above approximation assumes that the surrounding environment is similar on either side of the nucleus. The complex shape of component W, being extended almost perpendicular to the presumed jet direction, indicates significant interaction between the jet and the environment, which questions the initial assumption. Large brightness asymmetries between the opposite sides of CSS sources have previously been reported for both powerful and weaker CSS samples \citep{Saikia2001,Magda2010}, and are generally interpreted as evidence for asymmetric environments in addition to orientation effects.

At much closer redshift, at $z=0.367$, the CSS source, 3C\,48 displays a somewhat similar morphology with an extended, highly-distorted jet seen on the northern side of the compact core shaped by jet–ISM interaction \citep{An_3C48}. However, it differs from J1548$+$3335 in showing evidence for a weak counter-jet \citep{Feng2005} and superluminal motion in the inner jet \citep{An_3C48}. Another example is 3C\,287, a bright quasar with a steep radio spectrum located at $z=1.055$. Recently, \cite{asym_Frey2024} revealed the mas-scale position of the radio core in the system enabling the interpretation of the much brighter (already known) extended feature as a jet-related radio emission. According to the VLBI observations of \cite{asym_Frey2024}, 3C\,287 shows an asymmetric radio lobe south of the central engine.

Within feature W, \cite{Coppejans2016} detected a compact component only at $1.7$\,GHz with the EVN with a flux density of $\sim2.3$\,mJy. Its position agrees with that of component W1 fitted to the $1.5$-GHz e-MERLIN visibilities. The EVN-recovered flux density is significantly below the $\sim 30$\,mJy seen in the $1.67$-GHz e-MERLIN data. Thus, emission from W is almost completely resolved out in the EVN observations, further supporting its interpretation as jet-related emission. 

\begin{figure}[t]
    \centering
    \includegraphics[width=\hsize]{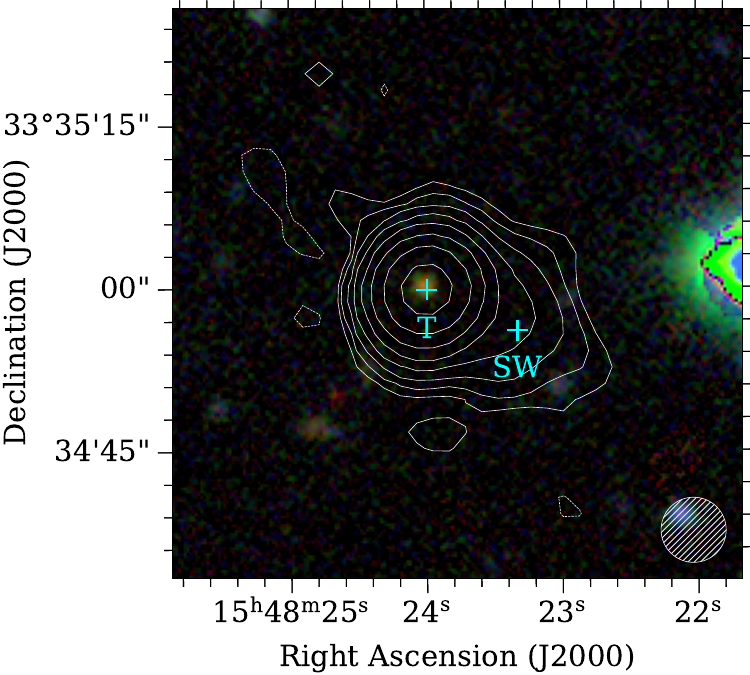}
    \caption{The low-resolution LoTSS-DR2 image (white contours) overlaid on the DESI three-color image \citep{desi_2019} of J1548$+$3335. The peak of the LOFAR image is $138.8$\,mJy\,beam$^{-1}$, the lowest contour is drawn at $\pm 0.8$\,mJy\,beam$^{-1}$ (corresponding to $\pm 4\sigma$ off-source image noise level) and further contour levels increase by a factor of $2$. The two fitted Gaussian components are marked with cyan crosses and labels. The restoring beam, shown in the lower right corner, has a size of $6\arcsec$.}
    \label{fig:J1548_lowres}
\end{figure}

At low resolution, J1548$+$3335 generally appears as a single radio source. An exception is the LoTSS DR2 image \citep{lotssdr2}, where a southwestern extension is visible, aligned with the direction of component W at a position angle of $\mathrm{PA}\approx -114\degr$ (Fig.\,\ref{fig:J1548_lofar_compare}). We modeled the LoTSS DR2 image with two Gaussian components (marked by cyan crosses and labels in Fig.~\ref{fig:J1548_lowres}), with flux densities of $S_\mathrm{J1548}^\mathrm{T}\sim 164$\,mJy for the central component and $S_\mathrm{J1548}^\mathrm{SW}\sim 13$\,mJy for the southwestern extension (see also Table\,\ref{tab:low-res}). Their summed flux density is consistent, within uncertainties, with the total flux density reported in the LoTSS DR2 catalogue \citep{lotssdr2}. The separation between the two components is $9\farcs18 \pm 0\farcs05$, corresponding to a projected linear distance of $\sim 61.6$\,kpc. The southwestern feature is not detected in the e-MERLIN data down to $0.1$\,mJy\,beam$^{-1}$, nor in the VLASS images down to $0.3$\,mJy\,beam$^{-1}$, and is likely resolved out in the high-resolution LOFAR imaging due to its low surface brightness and large angular size.

If this southwestern feature were physically associated with J1548$+$3335, the implied projected size of $\sim 61.6$\,kpc would be comparable to that of the recently reported “monster radio jet” at $z=4.9$ \citep{Gloudemans2025}. However, since such an association cannot be firmly established, we base our size estimate on the radio structure revealed by the e-MERLIN and international LOFAR observations, showing C and W emitting regions with a separation of $\sim 6.7$\,kpc. Assuming a counter-jet of similar extent to the detected jet, we estimate a largest linear size of $\sim 13$\,kpc for J1548$+$3335.

\subsection{Radio powers and multi-wavelength properties}

 \begin{figure*}
   \centering
   \includegraphics[scale=0.55]{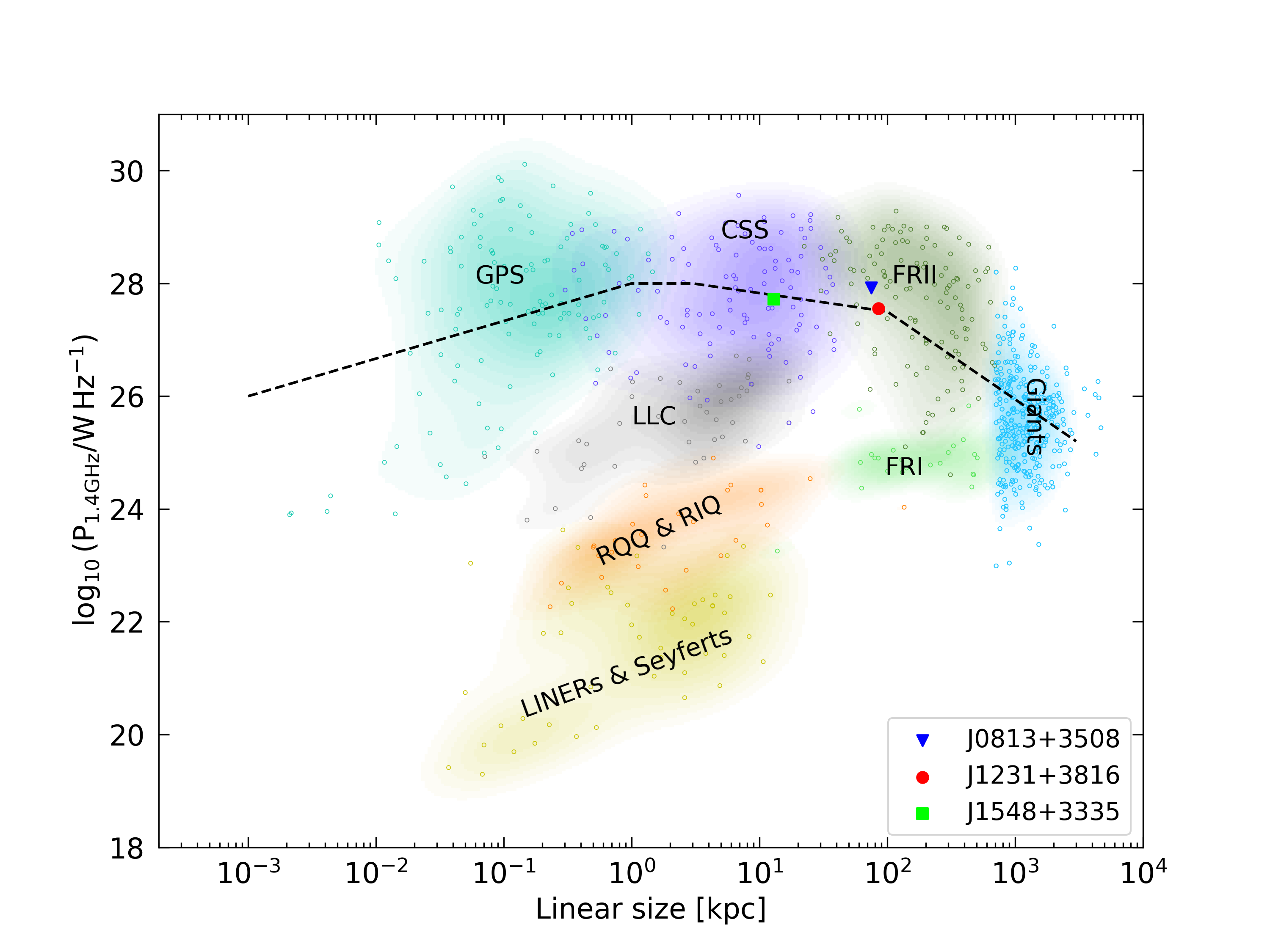}
      \caption{The radio power vs. linear size (P--D) diagram reproduced from \cite{Magda2025}, including various types of radio-selected AGN: gigahertz-peaked spectrum (GPS) sources, compact steep-spectrum (CSS) sources, low-luminosity compact (LLC) sources, Fanaroff--Riley type I
      (FR~I) or type II (FR~II) objects, radio-intermediate and radio-quiet quasars (RIQ/RQQ), low-ionization nuclear emission-line region (LINER) and Seyfert galaxies, and giant radio galaxies. The black dashed line shows a possible evolutionary track for high power objects as described in \cite{AnBaan}. The three large symbols indicate the $z>4$ quasars discussed in this article. 
              }
         \label{fig:sizepower}
   \end{figure*}

\begin{table*}
    \caption{Radio, X-ray, and (rest-frame) optical parameters of the three high-$z$ sources.}
    \label{tab:power}
    \centering
    \begin{tabular}{cccccccc}
    \hline
    \hline
    \noalign{\smallskip}
    Name & $P_{1.4\mathrm{\,GHz}}$ & $L_\mathrm{jet}$ & $L_\mathrm{rad}$ & $L_\mathrm{X}$ & $\alpha_\mathrm{ox}$ & $L_\mathrm{bol}$ & $\lambda_\mathrm{Edd}$ \\ 
     & ($10^{27}$\,W\,Hz$^{-1}$) & ($10^{38}$\,W) & ($10^{37}$\,W) &
     ($10^{38}$\,W) & & ($10^{38}$\,W)
     \\
     \noalign{\smallskip}
    \hline
    \noalign{\smallskip}
    J0813$+$3508 & $8.4 \pm 0.3$ & $43.0 \pm 1.5$ & $3.7-11.0$ & $4.8^{+1.9}_{-1.4}$ & $1.86\pm0.08$ & $145$ & $0.5$\\
    J1231$+$3816 & $3.6\pm 0.2$ & $20.7 \pm 1.2$ & $1.2-4.6$ & $2.5$ & $1.63$ & $74$ & 1.0\\ 
    J1548$+$3335 & $5.3 \pm 0.2$ & $20.9 \pm 0.8$ & $2.0-9.4$ & $23.2^{+7.5}_{-6.1}$ & $1.23^{+0.08}_{-0.07}$ & $26$ & $0.1$\\
   \noalign{\smallskip}
   \hline
    \end{tabular}
    \tablefoot{The sources names, $1.4$-GHz monochromatic radio power, the jet kinetic power, the radio luminosity are given in the first four columns. In the fifth and sixth columns, the X-ray luminosity in the rest-frame $(2-10)$\,keV interval and the $2$\,keV X-ray-to-optical luminosity ratio are given, respectively, from \cite{Snios2020} and from \cite{Zhu2019} (for J1231$+$3816). (We use the opposite sign definition for $\alpha_\mathrm{ox}$, so the values of \cite{Snios2020} and \cite{Zhu2019} are multiplied by $-1$.) In the last two columns, the bolometric luminosities and Eddington ratios from \cite{Rakshit2020} are listed. For the reliability of the bolometric luminosities and Eddington ratios, see the text for details.}
\end{table*}

The rest-frame $1.4$-GHz monochromatic radio powers, $P_{1.4\mathrm{\,GHz}}$ derived as,

\begin{equation}
  P_{1.4\mathrm{\,GHz}} = 4 \pi S D_\mathrm{L}^2 (1+z)^{-1-\alpha} ,
\end{equation}
where for the flux density, $S$, and for the spectral index, $\alpha$, we used the most recent RACS data, and the integrated radio spectral indices, respectively. The luminosity distance, $D_\mathrm{L}$ was calculated using the cosmology calculator of \citet{Wright_2006}. The radio power values are of the order of a few times $10^{27}$\,W\,Hz$^{-1}$ for all targets (Table\,\ref{tab:power}). When combined with the largest linear sizes measured from the high-resolution LOFAR images -- $\sim 70$\,kpc, $\sim 85$\,kpc, and $\sim 13$\,kpc for J0813$+$3508, J1231$+$3816, and J1548$+$3335, respectively -- J0813$+$3508 and J1231$+$3816 fall in the region typically populated by FR\,II-type radio galaxies, lying along the evolutionary track of high-power sources in the radio power--linear size (P--D) diagram (Fig.\,\ref{fig:sizepower}). In contrast, J1548$+$3335 has a smaller linear size, placing it closer to the locus of CSS sources in the P--D plane \citep{Magda2025}. This indicates that at relatively early stages of the Universe, at $\lesssim 1.6$\,Gyr age, already such objects with extended radio structures existed.

Recently, \cite{recent_radiogal} analysed a large sample of low-redshift radio galaxies with $z\lesssim 0.54$. The median $1.4$-GHz radio luminosities for the wide-angle tail (WAT), narrow-angle tail (NAT), and FR~II-type radio galaxy samples are three orders of magnitude below the values obtained for our targets, and their projected linear sizes are $\sim 4$ times larger than for J0813$+$3508 and J1231$+$3816. Thus, our targets represent high-power, possibly young and/or confined versions of these low-redshift radio galaxies.

Following \cite{Rusinek2017}, the jet kinetic power, $L_\mathrm{jet}$, can be estimated from $P_{1.4\mathrm{\,GHz}}$ as
\begin{equation}
    L_\mathrm{jet} [\mathrm{W}] = 5.0 \times 10^{15} \left(\frac{f}{10}\right)^{3/2} \left(P_{1.4\mathrm{\,GHz}} [\mathrm{W\,Hz}^{-1}] \right)^{6/7},
\end{equation}
where the value of $f$ accounts for the uncertainties of the modeling of the radio lobes, and it can be regarded as $\sim10$ in the case of FR~II galaxies. \cite{Wolowska2021} used the same formula to estimate the jet kinetic power also in younger radio sources. Even though we cannot detect the counter-jet in J1548$+$3335, its $1.4$-GHz radio power and assumed size are similar to those values of CSS objects, therefore, one can approximate the jet power similarly to the other sources. J0813$+$3508 has the largest $L_\mathrm{jet}$, while the other two sources have very similar values, half of that of J0813$+$3508 (see Table\,\ref{tab:power}). 

\cite{Rakshit2020} estimated the bolometric luminosities, $L_\mathrm{bol}$, from the $1350$-\AA \,monochromatic luminosity and the Eddington ratios, $\lambda_\mathrm{Edd}$, derived from the single-epoch black hole mass estimates obtained from the \ion{C}{IV} line. The uncertainties of the bolometric luminosities can be as high as $50$\% for single objects due to the uncertainties of the bolometric correction factors, according to \citet{Rakshit2020}. Additionally, the black hole mass estimate of J1548$+$3335 was flagged in \cite{Rakshit2020}, which can further increase the uncertainty of the $\lambda_\mathrm{Edd}$ value. In the $L_\mathrm{jet}/L_\mathrm{bol} - \lambda_\mathrm{Edd}$ diagram, the three sources fall into the region occupied mostly by the radiatively efficient FR~II-type quasars \citep{Wolowska2021}. Their $\lambda_\mathrm{Edd}$ is consistent with the findings that at $z\sim 6$, the radiatively efficient accretion mode dominates in quasars \citep[e.g., ][]{Trakhtenbrot2017}. The jet-production efficiency can be approximated in the following if one assumes $10$\% radiative efficiency \citep[e.g.][]{Wojtowicz2020}:

\begin{equation}
 \eta_\mathrm{jet}=\frac{L_\mathrm{jet}}{10L_\mathrm{bol}}   
\end{equation}

The jet production efficiency is $\sim 0.03$ for J0813$+$3508 and J1231$+$3816, while it is much larger, $\sim 0.08$ for J1548$+$3335. \cite{Wojtowicz2020} found that $\eta_\mathrm{jet}$ decreases with increasing $\lambda_\mathrm{Edd}$ for  young radio galaxies with $\lambda_\mathrm{Edd} > 0.05$. Our three sources roughly agree with this picture, as J1548$+$3335 with the lowest $\lambda_\mathrm{Edd}$ has the highest $\eta_\mathrm{jet}$.

We also estimated the integrated radio luminosity, $L_\mathrm{rad}$. It was obtained by integrating the best-fitting power-law spectrum over frequency. The lower values given in Table\,\ref{tab:power} are obtained by integrating over the frequency interval directly covered by the radio measurements, from $150$\,MHz to $3$\,GHz observed frequency, except for J0813$+$3508 where the $74$-MHz point is included; the upper values assume that the same power law extends over $10$\,MHz–$100$\,GHz. We therefore quote $L_\mathrm{rad}$ as a range rather than a single value.

These ranges imply radio luminosities a few times $10^{37}$\,W and are similar to the ones derived for the most energetic head-tail radio galaxies by \cite{sasmal_2022}. Head-tail radio galaxies are characterized by jets seen bent on the same side of the AGN. The high-resolution LOFAR images of J0813$+$3508 and J1231$+$3806 are reminiscent of such sources, the former having an apparently larger than $\sim 90\degr$ bending, thus similar to a WAT radio galaxy, while the latter with a smaller angle is similar to a NAT radio galaxy. Such bent radio sources can generally be found in rich galaxy clusters \citep[e.g., ][]{Wing2011, sasmal_2022}, their radio morphologies are thought to be the results of interaction of the jets with the intracluster medium (ICM). In the case of WAT galaxies, which are often found in merging galaxy clusters, the jet distortions can also be caused by the high-velocity, long-lived, large-scale sloshing motion of the ICM \citep{wat_review}. We are not aware of any galaxy clusters or groups in the vicinity of J1231$+$3816 or J0813$+$3508. Nevertheless, high-redshift radio galaxies are often found to be surrounded by galaxies of forming proto-clusters, and thus are thought to be in the process of forming the giant elliptical galaxies found in the centres of rich galaxy clusters in the local Universe \citep[e.g.,][and references therein]{Sax_TNJ1338}

All three sources were detected in X-rays with the Chandra satellite \citep{Zhu2019, Snios2020} as single point sources, with J1548$+$3335 having the largest X-ray luminosity of the three (Table~\ref{tab:power}). It was proposed that the increasing energy density of the CMB radiation at high redshifts results in extended X-ray-bright jets via the inverse-Compton scattering \citep{Fabian_CMB, Ghisellini_CMB}. While all three objects are detected in X-rays with Chandra, none of them show extended X-ray jets \citep{Snios2020, Zhu2019}. Specifically, the X-ray emission of J0813$+$3508 was only detected from region C, and according to \cite{Zhu2019}, the X-ray emission of region N must be at least $10$ times dimmer than that of region C.

The same effect, the inverse-Compton scattering on the CMB photons, is expected to dim the radio emission of extended jets. Nevertheless, \cite{Ghisellini_CMB} showed that the hotspot regions can still be efficient radio emitters if the magnetic field energy density can dominate above the CMB energy density. They conclude that the contrast between the compact magnetized regions and the more extended lobes should increase with increasing redshifts. In accordance with this, we found that the radio emission around hotspot regions, and at the central part of the AGN dominates, while extended, narrow jets are hardly seen. The longest jet emission is shown by the lowest-redshift object in our sample, J1231$+$3816. However, other effects can also hinder the detection of extended radio emission, i.e., the brightness decrease of the emission and the cosmological redshift due to extreme distances of these objects. The latter is related to the fact that the flux density of the optically thin syncrothron emission of the  extended features decreases with increasing frequency, but due to the high redshifts, relatively low observing frequencies correspond to high rest-frame values. In that context, the $144$-MHz LOFAR observations were crucial to reveal the rest-frame $(770-890)$-MHz emission. 

The X-ray-to-optical luminosity ratio, $\alpha_\mathrm{ox}$, is commonly used to differentiate between relativistically beamed blazars and radio-loud AGN with jets oriented at larger angles to the line of sight \citep[e.g.][]{Donato2001, Shemmer2006}. If the X-ray emission is dominated by the relativistic jet, $\alpha_\mathrm{ox}\lesssim1.5$ is expected. The $\alpha_\mathrm{ox}$ values of our targets are listed in Table\,\ref{tab:power} taken from \cite{Snios2020} and \cite{Zhu2019}.\footnote{We note that \cite{Snios2020} and \cite{Zhu2019} used this value with the opposite sign convention.} J0813$+$3508 and J1231$+$3816 are not blazars according to this classification, their $\alpha_\mathrm{ox}$ values are in excess of $1.5$. The $\alpha_\mathrm{ox}$ of J1548$+$3335, however, is below this limit. \cite{Ighina2019} showed that in the case of high-redshift AGN the ratio, $\tilde{\alpha}_\mathrm{ox}$ calculated from the higher-energy, $10$\,keV X-ray luminosity provides a more accurate estimator. Using the X-ray spectral index, $\tilde{\alpha}_\mathrm{ox}$ can be calculated from $\alpha_\mathrm{ox}$ \citep{Ighina2019}. The photon indices of the three objects given in \cite{Snios2020} and \cite{Zhu2019} are used to obtain the X-ray spectral indices. We found that $\tilde{\alpha}_\mathrm{ox}$ of J0813$+$3508 and J1231$+$3816 exceed $1.355$, so they are not blazars according to the finding of \cite{Ighina2019}. J1548$+$3335, however, has $\tilde{\alpha}_\mathrm{ox}=1.2$, thus the X-ray-to-optical ratio of this source, irrespectively whether the lower or higher energy X-ray luminosity is used, indicates a blazar nature. In the radio domain, blazars are often identified by containing bright compact features with brightness temperatures exceeding the equipartition limit \citep[$5\times10^{10}$\,K, ][]{Readhead}, or the slightly lower, empirically-obtained value for the intrinsic brightness temperature of $4 \times 10^{10}$\,K \citep{Homan2021}. The brightness temperature values measured with VLBI observations conducted at $5$\,GHz observing frequency of all three sources are a factor of ten below these limits \citep{Frey2010, Krezinger2022, Coppejans2016}. Thus, they do not indicate relativistic beaming. Interestingly, J1548$+$3335 has the highest brightness temperature, $\sim 2\times 10^{9}$\,K \citep[][if $10$\% coherence loss is assumed]{Coppejans2016} among the three studied sources. \cite{Krezinger2026} compared the X-ray and radio methods to identify the blazars in a sample of high-redshift AGN. They found a general agreement of the two classification methods but they also recovered a few objects with seemingly contradictory classification results close to the borders of the classification area. Similarly, the $\alpha_\mathrm{ox}$ and $\tilde{\alpha}_\mathrm{ox}$ values of J1548$+$3335 are rather close to the limiting values between blazars and misaligned sources.

The X-ray photon indices of J0813$+$3508 and J1231$+$3816 are very similar, $1.35^{+0.44}_{-0.17}$ and $1.37^{+0.44}_{-0.21}$, respectively \citep{Zhu2019}, while J1548$+$3335 has a larger value of $2.17^{+0.13}_{-0.11}$ \citep{Snios2020}. The lower photon index of radio-loud quasars compared to radio-quiet quasars can be caused by the contribution of the jet emission and/or denser intrinsic column densities \citep[e.g.,][]{Snios2020}. \cite{Snios2020} speculate that the relatively high level of the soft X-ray emission in J1548$+$3335 may be due to the presence of the $6.4$\,keV Fe K$\alpha$ line, observable at $1.13$\,keV at the redshift of the source. 

Recently \cite{Tullia_blazarcat} published a living catalog of $z>4$ blazar sources based on literature search and including the most recent extended ROentgen Survey with an Imaging Telescope Array \citep[eROSITA, ][]{erosita} data. Both J0813$+$3508 and J1231$+$3816 are listed in the catalog, and J0813$+$3508 was also detected by eROSITA. However, for both objects, \cite{Tullia_blazarcat} noted their extended radio structures shown by VLASS and FIRST. According to our results, these radio quasars are more similar to radio galaxies than blazars. This is not the first time that X-ray and radio emission do not appear consistent, with X-ray spectrum and broad-band SED pointing toward a blazar classification, while high-resolution radio images clearly showing misaligned extended jets or large-scale structures \citep[e.g.,][]{Cao_blazarcan, three_radiogal, J1420}. Such a diversity of features can be explained by jet reorientation. Changes in the position angles of the inner mas-scale jets of close-by ($z\lesssim 0.1$) blazars are reported on decadal time scales by e.g. \citet{Lister2021}. The more frequent changes in the direction of powerful relativistic jets at very high redshift are suggested by an earlier study of \cite{sbarrato} discussing the possibility that fast accretion expected in these systems may cause more frequent jet bending. Whether this is due to intrinsic reorientation of the jet base or, instead, jet bending on larger scales is currently unclear.

\section{Conclusions and future work}

We investigated the radio properties of three high-redshift ($z>4$) AGN using radio observations of various angular resolutions. Specifically, we analysed the $144$-MHz international LOFAR data and complemented them with e-MERLIN and archival EVN data taken at GHz frequencies. Additionally, we collected the flux densities of our targets measured in low-resolution radio surveys. 

Previous VLBI observations at GHz frequencies revealed compact features in all three objects, however, with relatively low brightness temperatures, a few times $10^9$\,K, indicating the absence of relativistic Doppler boosting. Nevertheless, from VLBI data alone, the exact nature of these objects could not be ascertained as large fraction of the total flux density was resolved out in those observations. The new intermediate-resolution data revealed extended radio features at kpc scales with complex morphologies in all three of our target sources. 

Two of the studied sources (J1231$+$3816 and J0813$+$3508) exhibit steep-spectrum, lobe-dominated morphologies, similar to FR~II-type radio galaxies, with sizes typical for radio sources at a mature age. In the case of J0813$+$3508, the FIRST data indicated two radio-emitting features at kpc scales, however, their relation to the mas-scale VLBI features could not be unambiguously ascertained in the past. The jet direction revealed by the international LOFAR observations confirmed their physical connection. Both sources show distorted, bent kpc-scale jets possibly due to the interactions with the surrounding dense material. Their radio luminosities are similar to those of the most luminous known head-tail radio galaxies. 

The third target, J1548$+$3335, contains two mas-scale compact features according to its previous 1.7-GHz EVN observation \citep{Coppejans2016}. Our e-MERLIN and international LOFAR data revealed that the fainter EVN-detected component is a hotspot embedded in an extended, steep-spectrum jet-related region. No radio emission is detected on the opposite side of the AGN centre in either the e-MERLIN or LOFAR data. 
Based on the radio structure unambiguously associated with the source, the derived largest linear size places J1548$+$3335 in the region of CSS sources in the radio power--linear size (P--D) diagram. 
The presence of a CSS quasar at $z>4$ suggests that radio-mode AGN activity was already established within the first $\sim 1.3$\,Gyr of cosmic history, producing jets and lobes on kpc scales despite the hostile high-redshift environment.

According to their multi-wavelength data, all three sources accrete in a radiatively efficient way. They are bright X-ray sources, however, they do not exhibit extended X-ray emission.

Our results highlight the importance of low-frequency, high-resolution observations to reveal the kpc-scale extended radio structures of such objects. Further observations are needed to refine these classifications. Broad-band radio spectra, particularly at higher frequencies ($>10$\,GHz), would help to constrain possible spectral breaks and aging effects. Deeper LOFAR or upgraded GMRT imaging could test for faint extended emission beyond the scales recovered here. Polarization studies would provide insights into the magnetic field configurations and jet interactions with the intergalactic medium. Multi-wavelength data (X-ray and infrared) could constrain the accretion properties of the central engines and test the connection between jet activity and black hole growth at high redshift. Additionally, sensitive spectral-line observations with the James Webb Space Telescope can provide more details on the jet--environment interactions around the radio lobes.

\begin{acknowledgements}
We thank the anonymous referee whose comments helped to improve the paper.

LOFAR is the Low Frequency Array designed and constructed by ASTRON. It has observing, data processing, and data storage facilities in several countries, which are owned by various parties (each with their own funding sources), and which are collectively operated by the LOFAR ERIC under a joint scientific policy. The LOFAR resources have benefited from the following recent major funding sources: CNRS-INSU, Observatoire de Paris and Université d'Orléans, France; BMFTR, MKW-NRW, MPG, Germany; Science Foundation Ireland (SFI), Department of Business, Enterprise and Innovation (DBEI), Ireland; NWO, The Netherlands; The Science and Technology Facilities Council, UK; Ministry of Science and Higher Education, Poland; The Istituto Nazionale di Astrofisica (INAF), Italy.

This research made use of the Dutch national e-infrastructure with support of the SURF Cooperative (e-infra 180169) and the LOFAR e-infra group. The Jülich LOFAR Long Term Archive and the German LOFAR network are both coordinated and operated by the Jülich Supercomputing Centre (JSC), and computing resources on the supercomputer JUWELS at JSC were provided by the Gauss Centre for Supercomputing e.V. (grant CHTB00) through the John von Neumann Institute for Computing (NIC).

This research made use of the University of Hertfordshire high-performance computing facility and the LOFAR-UK computing facility located at the University of Hertfordshire and supported by STFC [ST/P000096/1], and of the Italian LOFAR-IT computing infrastructure supported and operated by INAF, including the resources within the PLEIADI special "LOFAR" project by USC-C of INAF, and by the Physics Department of Turin University (under an agreement with Consorzio Interuniversitario per la Fisica Spaziale) at the C3S Supercomputing Centre, Italy.

This research is part of the project LOFAR Data Valorization (LDV) [project numbers 2020.031, 2022.033, and 2024.047] of the research programme Computing Time on National Computer Facilities using SPIDER that is (co-)funded by the Dutch Research Council (NWO), hosted by SURF through the call for proposals of Computing Time on National Computer Facilities. 

The e-MERLIN is a National Facility operated by the University of Manchester at Jodrell Bank Observatory on behalf of STFC. Scientific results presented in this publication are partly derived from the e-MERLIN project code: CY5211. The European VLBI Network is a joint facility of independent European, African, Asian, and North American radio astronomy institutes. Scientific results from data presented in this publication are derived from the following EVN project codes: EF021, EG102.

The Legacy Surveys consist of three individual and complementary projects: the Dark Energy Camera Legacy Survey (DECaLS; Proposal ID \#2014B-0404; PIs: David Schlegel and Arjun Dey), the Beijing-Arizona Sky Survey (BASS; NOAO Prop. ID \#2015A-0801; PIs: Zhou Xu and Xiaohui Fan), and the Mayall z-band Legacy Survey (MzLS; Prop. ID \#2016A-0453; PI: Arjun Dey). DECaLS, BASS and MzLS together include data obtained, respectively, at the Blanco telescope, Cerro Tololo Inter-American Observatory, NSF’s NOIRLab; the Bok telescope, Steward Observatory, University of Arizona; and the Mayall telescope, Kitt Peak National Observatory, NOIRLab. Pipeline processing and analyses of the data were supported by NOIRLab and the Lawrence Berkeley National Laboratory (LBNL). The Legacy Surveys project is honored to be permitted to conduct astronomical research on Iolkam Du’ag (Kitt Peak), a mountain with particular significance to the Tohono O’odham Nation.

NOIRLab is operated by the Association of Universities for Research in Astronomy (AURA) under a cooperative agreement with the National Science Foundation. LBNL is managed by the Regents of the University of California under contract to the U.S. Department of Energy.

This project used data obtained with the Dark Energy Camera (DECam), which was constructed by the Dark Energy Survey (DES) collaboration. Funding for the DES Projects has been provided by the U.S. Department of Energy, the U.S. National Science Foundation, the Ministry of Science and Education of Spain, the Science and Technology Facilities Council of the United Kingdom, the Higher Education Funding Council for England, the National Center for Supercomputing Applications at the University of Illinois at Urbana Champaign, the Kavli Institute of Cosmological Physics at the University of Chicago, Center for Cosmology and Astro-Particle Physics at the Ohio State University, the Mitchell Institute for Fundamental Physics and Astronomy at Texas A\&M University, Financiadora de Estudos e Projetos, Fundacao Carlos Chagas Filho de Amparo, Financiadora de Estudos e Projetos, Fundacao Carlos Chagas Filho de Amparo a Pesquisa do Estado do Rio de Janeiro, Conselho Nacional de Desenvolvimento Cientifico e Tecnologico and the Ministerio da Ciencia, Tecnologia e Inovacao, the Deutsche Forschungsgemeinschaft and the Collaborating Institutions in the Dark Energy Survey. The Collaborating Institutions are Argonne National Laboratory, the University of California at Santa Cruz, the University of Cambridge, Centro de Investigaciones Energeticas, Medioambientales y Tecnologicas-Madrid, the University of Chicago, University College London, the DES-Brazil Consortium, the University of Edinburgh, the Eidgenossische Technische Hochschule (ETH) Zurich, Fermi National Accelerator Laboratory, the University of Illinois at Urbana-Champaign, the Institut de Ciencies de l’Espai (IEEC/CSIC), the Institut de Fisica d’Altes Energies, Lawrence Berkeley National Laboratory, the Ludwig Maximilians Universitat Munchen and the associated Excellence Cluster Universe, the University of Michigan, NSF’s NOIRLab, the University of Nottingham, the Ohio State University, the University of Pennsylvania, the University of Portsmouth, SLAC National Accelerator Laboratory, Stanford University, the University of Sussex, and Texas A\&M University.

BASS is a key project of the Telescope Access Program (TAP), which has been funded by the National Astronomical Observatories of China, the Chinese Academy of Sciences (the Strategic Priority Research Program “The Emergence of Cosmological Structures” Grant \# XDB09000000), and the Special Fund for Astronomy from the Ministry of Finance. The BASS is also supported by the External Cooperation Program of Chinese Academy of Sciences (Grant \# 114A11KYSB20160057), and Chinese National Natural Science Foundation (Grant \# 12120101003, \# 11433005).

The Legacy Survey team makes use of data products from the Near-Earth Object Wide-field Infrared Survey Explorer (NEOWISE), which is a project of the Jet Propulsion Laboratory/California Institute of Technology. NEOWISE is funded by the National Aeronautics and Space Administration.

The Legacy Surveys imaging of the DESI footprint is supported by the Director, Office of Science, Office of High Energy Physics of the U.S. Department of Energy under Contract No. DE-AC02-05CH1123, by the National Energy Research Scientific Computing Center, a DOE Office of Science User Facility under the same contract; and by the U.S. National Science Foundation, Division of Astronomical Sciences under Contract No. AST-0950945 to NOAO.

LIG gratefully acknowledges support
by the PIFI International Distinguished Scholars Project 2026PD0183 of the Chinese Academy of Sciences.

H.M.C. acknowledges support from the Henan Natural Science Foundation of China (grant No. 252300420344).

This research was supported by HUN-REN and the NKFIH excellence grant TKP2021-NKTA-64. We thank for the usage of the HUN-REN Cloud.
\end{acknowledgements}

\bibliographystyle{aa}
\bibliography{ref}

@ARTICLE{Kravchenko2025,
       author = {{Kravchenko}, E.~V. and {Pashchenko}, I.~N. and {Homan}, D.~C. and {Kovalev}, Y.~Y. and {Lister}, M.~L. and {Pushkarev}, A.~B. and {Ros}, E. and {Savolainen}, T.},
        title = "{MOJAVE - XXII. Brightness temperature distributions and geometric profiles along parsec-scale active galactic nucleus jets}",
      journal = {\mnras},
         year = 2025,
        month = apr,
       volume = {538},
       number = {3},
        pages = {2008-2030},
          doi = {10.1093/mnras/staf343},
archivePrefix = {arXiv},
       eprint = {2502.14516},
 primaryClass = {astro-ph.HE},
       adsurl = {https://ui.adsabs.harvard.edu/abs/2025MNRAS.538.2008K}
}

@ARTICLE{Kharb2010,
       author = {{Kharb}, P. and {Lister}, M.~L. and {Cooper}, N.~J.},
        title = "{Extended Radio Emission in MOJAVE Blazars: Challenges to Unification}",
      journal = {\apj},
         year = 2010,
        month = feb,
       volume = {710},
       number = {1},
        pages = {764-782},
          doi = {10.1088/0004-637X/710/1/764},
archivePrefix = {arXiv},
       eprint = {1001.0731},
 primaryClass = {astro-ph.CO},
       adsurl = {https://ui.adsabs.harvard.edu/abs/2010ApJ...710..764K}
}

@ARTICLE{Saikia2001,
       author = {{Saikia}, D.~J. and {Jeyakumar}, S. and {Salter}, C.~J. and {Thomasson}, P. and {Spencer}, R.~E. and {Mantovani}, F.},
        title = "{Compact steep-spectrum sources from the S4 sample}",
      journal = {\mnras},
         year = 2001,
        month = feb,
       volume = {321},
       number = {1},
        pages = {37-43},
          doi = {10.1046/j.1365-8711.2001.04017.x},
archivePrefix = {arXiv},
       eprint = {astro-ph/0009175},
 primaryClass = {astro-ph},
       adsurl = {https://ui.adsabs.harvard.edu/abs/2001MNRAS.321...37S}
}

@ARTICLE{Gloudemans2025,
       author = {{Gloudemans}, Anniek J. and {Sweijen}, Frits and {Morabito}, Leah K. and {Farina}, Emanuele Paolo and {Duncan}, Kenneth J. and {Harikane}, Yuichi and {R{\"o}ttgering}, Huub J.~A. and {Saxena}, Aayush and {Schindler}, Jan-Torge},
        title = "{Monster Radio Jet (>66 kpc) Observed in Quasar at z {\ensuremath{\sim}} 5}",
      journal = {\apjl},
         year = 2025,
        month = feb,
       volume = {980},
       number = {1},
          eid = {L8},
        pages = {L8},
          doi = {10.3847/2041-8213/ad9609},
archivePrefix = {arXiv},
       eprint = {2411.16838},
 primaryClass = {astro-ph.GA},
       adsurl = {https://ui.adsabs.harvard.edu/abs/2025ApJ...980L...8G}
}

@ARTICLE{Parijskij2014,
       author = {{Parijskij}, Yu. N. and {Thomasson}, P. and {Kopylov}, A.~I. and {Zhelenkova}, O.~P. and {Muxlow}, T.~W.~B. and {Beswick}, R. and {Soboleva}, N.~S. and {Temirova}, A.~V. and {Verkhodanov}, O.~V.},
        title = "{Observations of the z = 4.514 radio galaxy RC J0311+0507}",
      journal = {\mnras},
         year = 2014,
        month = apr,
       volume = {439},
       number = {3},
        pages = {2314-2322},
          doi = {10.1093/mnras/stu047},
       adsurl = {https://ui.adsabs.harvard.edu/abs/2014MNRAS.439.2314P}
}

@ARTICLE{Volonteri2011,
       author = {{Volonteri}, M. and {Haardt}, F. and {Ghisellini}, G. and {Della Ceca}, R.},
        title = "{Blazars in the early Universe}",
      journal = {\mnras},
         year = 2011,
        month = sep,
       volume = {416},
       number = {1},
        pages = {216-224},
          doi = {10.1111/j.1365-2966.2011.19024.x},
archivePrefix = {arXiv},
       eprint = {1103.5565},
 primaryClass = {astro-ph.HE},
       adsurl = {https://ui.adsabs.harvard.edu/abs/2011MNRAS.416..216V}
}

@ARTICLE{Krezinger2022,
       author = {{Krezinger}, M{\'a}t{\'e} and {Perger}, Krisztina and {Gab{\'a}nyi}, Krisztina {\'E}va and {Frey}, S{\'a}ndor and {Gurvits}, Leonid I. and {Paragi}, Zsolt and {An}, Tao and {Zhang}, Yingkang and {Cao}, Hongmin and {Sbarrato}, Tullia},
        title = "{Radio-loud Quasars above Redshift 4: Very Long Baseline Interferometry (VLBI) Imaging of an Extended Sample}",
      journal = {\apjs},
         year = 2022,
        month = jun,
       volume = {260},
       number = {2},
          eid = {49},
        pages = {49},
          doi = {10.3847/1538-4365/ac63b8},
archivePrefix = {arXiv},
       eprint = {2204.02114},
 primaryClass = {astro-ph.GA},
       adsurl = {https://ui.adsabs.harvard.edu/abs/2022ApJS..260...49K}
}

@INPROCEEDINGS{casa,
   author = {{McMullin}, J.~P. and {Waters}, B. and {Schiebel}, D. and {Young}, W. and 
	{Golap}, K.},
    title = "{CASA Architecture and Applications}",
booktitle = {Astronomical Data Analysis Software and Systems XVI},
     year = 2007,
   series = {Astronomical Society of the Pacific Conference Series},
   volume = 376,
   editor = {{Shaw}, R.~A. and {Hill}, F. and {Bell}, D.~J.},
    pages = {127},
   adsurl = {http://adsabs.harvard.edu/abs/2007ASPC..376..127M}
}

@INPROCEEDINGS{difmap,
       author = {{Shepherd}, M.~C.},
        title = "{Difmap: an Interactive Program for Synthesis Imaging}",
    booktitle = {Astronomical Data Analysis Software and Systems VI},
         year = 1997,
       editor = {{Hunt}, Gareth and {Payne}, Harry},
       series = {Astronomical Society of the Pacific Conference Series},
       volume = {125},
        pages = {77},
       adsurl = {https://ui.adsabs.harvard.edu/abs/1997ASPC..125...77S}
}

@ARTICLE{aoflagger,
  title = "A morphological algorithm for improved radio-frequency interference detection",
  author = {A. R. Offringa and J. J. van de Gronde and J. B. T. M. Roerdink},
  journal = "A\&A",
  year = 2012,
  volume = 539,
  month = mar,
  issue = "A95"
}

@INPROCEEDINGS{p-ref,
       author = {{Beasley}, A.~J. and {Conway}, J.~E.},
        title = "{VLBI Phase-Referencing}",
    booktitle = {Very Long Baseline Interferometry and the VLBA},
         year = 1995,
       editor = {{Zensus}, J.~A. and {Diamond}, P.~J. and {Napier}, P.~J.},
       series = {Astronomical Society of the Pacific Conference Series},
       volume = {82},
        month = jan,
        pages = {327},
       adsurl = {https://ui.adsabs.harvard.edu/abs/1995ASPC...82..327B}
}

@ARTICLE{Frey2010,
       author = {{Frey}, S. and {Paragi}, Z. and {Gurvits}, L.~I. and {Cseh}, D. and {Gab{\'a}nyi}, K. {\'E}.},
        title = "{High-resolution images of five radio quasars at early cosmological epochs}",
      journal = {\aap},
         year = 2010,
        month = dec,
       volume = {524},
          eid = {A83},
        pages = {A83},
          doi = {10.1051/0004-6361/201015554},
archivePrefix = {arXiv},
       eprint = {1009.5023},
 primaryClass = {astro-ph.CO},
       adsurl = {https://ui.adsabs.harvard.edu/abs/2010A&A...524A..83F}
}

@ARTICLE{lotssdr2,
       author = {{Shimwell}, T.~W. and {Hardcastle}, M.~J. and {Tasse}, C. and {Best}, P.~N. and {R{\"o}ttgering}, H.~J.~A. and {Williams}, W.~L. and {Botteon}, A. and {Drabent}, A. and {Mechev}, A. and {Shulevski}, A. and {van Weeren}, R.~J. and {Bester}, L. and {Br{\"u}ggen}, M. and {Brunetti}, G. and {Callingham}, J.~R. and {Chy{\.z}y}, K.~T. and {Conway}, J.~E. and {Dijkema}, T.~J. and {Duncan}, K. and {de Gasperin}, F. and {Hale}, C.~L. and {Haverkorn}, M. and {Hugo}, B. and {Jackson}, N. and {Mevius}, M. and {Miley}, G.~K. and {Morabito}, L.~K. and {Morganti}, R. and {Offringa}, A. and {Oonk}, J.~B.~R. and {Rafferty}, D. and {Sabater}, J. and {Smith}, D.~J.~B. and {Schwarz}, D.~J. and {Smirnov}, O. and {O'Sullivan}, S.~P. and {Vedantham}, H. and {White}, G.~J. and {Albert}, J.~G. and {Alegre}, L. and {Asabere}, B. and {Bacon}, D.~J. and {Bonafede}, A. and {Bonnassieux}, E. and {Brienza}, M. and {Bilicki}, M. and {Bonato}, M. and {Calistro Rivera}, G. and {Cassano}, R. and {Cochrane}, R. and {Croston}, J.~H. and {Cuciti}, V. and {Dallacasa}, D. and {Danezi}, A. and {Dettmar}, R.~J. and {Di Gennaro}, G. and {Edler}, H.~W. and {En{\ss}lin}, T.~A. and {Emig}, K.~L. and {Franzen}, T.~M.~O. and {Garc{\'\i}a-Vergara}, C. and {Grange}, Y.~G. and {G{\"u}rkan}, G. and {Hajduk}, M. and {Heald}, G. and {Heesen}, V. and {Hoang}, D.~N. and {Hoeft}, M. and {Horellou}, C. and {Iacobelli}, M. and {Jamrozy}, M. and {Jeli{\'c}}, V. and {Kondapally}, R. and {Kukreti}, P. and {Kunert-Bajraszewska}, M. and {Magliocchetti}, M. and {Mahatma}, V. and {Ma{\l}ek}, K. and {Mandal}, S. and {Massaro}, F. and {Meyer-Zhao}, Z. and {Mingo}, B. and {Mostert}, R.~I.~J. and {Nair}, D.~G. and {Nakoneczny}, S.~J. and {Nikiel-Wroczy{\'n}ski}, B. and {Orr{\'u}}, E. and {Pajdosz-{\'S}mierciak}, U. and {Pasini}, T. and {Prandoni}, I. and {van Piggelen}, H.~E. and {Rajpurohit}, K. and {Retana-Montenegro}, E. and {Riseley}, C.~J. and {Rowlinson}, A. and {Saxena}, A. and {Schrijvers}, C. and {Sweijen}, F. and {Siewert}, T.~M. and {Timmerman}, R. and {Vaccari}, M. and {Vink}, J. and {West}, J.~L. and {Wo{\l}owska}, A. and {Zhang}, X. and {Zheng}, J.},
        title = "{The LOFAR Two-metre Sky Survey. V. Second data release}",
      journal = {\aap},
         year = 2022,
        month = mar,
       volume = {659},
          eid = {A1},
        pages = {A1},
          doi = {10.1051/0004-6361/202142484},
archivePrefix = {arXiv},
       eprint = {2202.11733},
 primaryClass = {astro-ph.GA},
       adsurl = {https://ui.adsabs.harvard.edu/abs/2022A&A...659A...1S}
}

@ARTICLE{wenss,
       author = {{Rengelink}, R.~B. and {Tang}, Y. and {de Bruyn}, A.~G. and {Miley}, G.~K. and {Bremer}, M.~N. and {Roettgering}, H.~J.~A. and {Bremer}, M.~A.~R.},
        title = "{The Westerbork Northern Sky Survey (WENSS), I. A 570 square degree Mini-Survey around the North Ecliptic Pole}",
      journal = {\aaps},
         year = 1997,
        month = aug,
       volume = {124},
        pages = {259-280},
          doi = {10.1051/aas:1997358},
       adsurl = {https://ui.adsabs.harvard.edu/abs/1997A&AS..124..259R}
}

@ARTICLE{racs_1.37,
       author = {{Duchesne}, S.~W. and {Grundy}, J.~A. and {Heald}, George H. and {Lenc}, Emil and {Leung}, James K. and {McConnell}, David and {Murphy}, Tara and {Pritchard}, Joshua and {Rose}, Kovi and {Thomson}, Alec J.~M. and {Wang}, Yuanming and {Wang}, Ziteng and {Whiting}, Matthew T.},
        title = "{The Rapid ASKAP Continuum Survey V: Cataloguing the sky at 1 367.5 MHz and the second data release of RACS-mid}",
      journal = {\pasa},
         year = 2024,
        month = jan,
       volume = {41},
          eid = {e003},
        pages = {e003},
          doi = {10.1017/pasa.2023.60},
archivePrefix = {arXiv},
       eprint = {2311.12369},
 primaryClass = {astro-ph.GA},
       adsurl = {https://ui.adsabs.harvard.edu/abs/2024PASA...41....3D}
}

@ARTICLE{nvss,
       author = {{Condon}, J.~J. and {Cotton}, W.~D. and {Greisen}, E.~W. and {Yin}, Q.~F. and {Perley}, R.~A. and {Taylor}, G.~B. and {Broderick}, J.~J.},
        title = "{The NRAO VLA Sky Survey}",
      journal = {\aj},
         year = 1998,
        month = may,
       volume = {115},
       number = {5},
        pages = {1693-1716},
          doi = {10.1086/300337},
       adsurl = {https://ui.adsabs.harvard.edu/abs/1998AJ....115.1693C}
}

@ARTICLE{first,
       author = {{Helfand}, David J. and {White}, Richard L. and {Becker}, Robert H.},
        title = "{The Last of FIRST: The Final Catalog and Source Identifications}",
      journal = {\apj},
         year = 2015,
        month = mar,
       volume = {801},
       number = {1},
          eid = {26},
        pages = {26},
          doi = {10.1088/0004-637X/801/1/26},
archivePrefix = {arXiv},
       eprint = {1501.01555},
 primaryClass = {astro-ph.GA},
       adsurl = {https://ui.adsabs.harvard.edu/abs/2015ApJ...801...26H}
}

@ARTICLE{lacy_vlass,
       author = {{Lacy}, M. and {Baum}, S.~A. and {Chandler}, C.~J. and {Chatterjee}, S. and {Clarke}, T.~E. and {Deustua}, S. and {English}, J. and {Farnes}, J. and {Gaensler}, B.~M. and {Gugliucci}, N. and {Hallinan}, G. and {Kent}, B.~R. and {Kimball}, A. and {Law}, C.~J. and {Lazio}, T.~J.~W. and {Marvil}, J. and {Mao}, S.~A. and {Medlin}, D. and {Mooley}, K. and {Murphy}, E.~J. and {Myers}, S. and {Osten}, R. and {Richards}, G.~T. and {Rosolowsky}, E. and {Rudnick}, L. and {Schinzel}, F. and {Sivakoff}, G.~R. and {Sjouwerman}, L.~O. and {Taylor}, R. and {White}, R.~L. and {Wrobel}, J. and {Andernach}, H. and {Beasley}, A.~J. and {Berger}, E. and {Bhatnager}, S. and {Birkinshaw}, M. and {Bower}, G.~C. and {Brandt}, W.~N. and {Brown}, S. and {Burke-Spolaor}, S. and {Butler}, B.~J. and {Comerford}, J. and {Demorest}, P.~B. and {Fu}, H. and {Giacintucci}, S. and {Golap}, K. and {G{\"u}th}, T. and {Hales}, C.~A. and {Hiriart}, R. and {Hodge}, J. and {Horesh}, A. and {Ivezi{\'c}}, {\v{Z}}. and {Jarvis}, M.~J. and {Kamble}, A. and {Kassim}, N. and {Liu}, X. and {Loinard}, L. and {Lyons}, D.~K. and {Masters}, J. and {Mezcua}, M. and {Moellenbrock}, G.~A. and {Mroczkowski}, T. and {Nyland}, K. and {O'Dea}, C.~P. and {O'Sullivan}, S.~P. and {Peters}, W.~M. and {Radford}, K. and {Rao}, U. and {Robnett}, J. and {Salcido}, J. and {Shen}, Y. and {Sobotka}, A. and {Witz}, S. and {Vaccari}, M. and {van Weeren}, R.~J. and {Vargas}, A. and {Williams}, P.~K.~G. and {Yoon}, I.},
        title = "{The Karl G. Jansky Very Large Array Sky Survey (VLASS). Science Case and Survey Design}",
      journal = {\pasp},
         year = 2020,
        month = mar,
       volume = {132},
       number = {1009},
          eid = {035001},
        pages = {035001},
          doi = {10.1088/1538-3873/ab63eb},
archivePrefix = {arXiv},
       eprint = {1907.01981},
 primaryClass = {astro-ph.IM},
       adsurl = {https://ui.adsabs.harvard.edu/abs/2020PASP..132c5001L}
}

@ARTICLE{sdss_dr12,
       author = {{Alam}, Shadab and {Albareti}, Franco D. and {Allende Prieto}, Carlos and {Anders}, F. and {Anderson}, Scott F. and {Anderton}, Timothy and {Andrews}, Brett H. and {Armengaud}, Eric and {Aubourg}, {\'E}ric and {Bailey}, Stephen and {Basu}, Sarbani and {Bautista}, Julian E. and {Beaton}, Rachael L. and {Beers}, Timothy C. and {Bender}, Chad F. and {Berlind}, Andreas A. and {Beutler}, Florian and {Bhardwaj}, Vaishali and {Bird}, Jonathan C. and {Bizyaev}, Dmitry and {Blake}, Cullen H. and {Blanton}, Michael R. and {Blomqvist}, Michael and {Bochanski}, John J. and {Bolton}, Adam S. and {Bovy}, Jo and {Shelden Bradley}, A. and {Brandt}, W.~N. and {Brauer}, D.~E. and {Brinkmann}, J. and {Brown}, Peter J. and {Brownstein}, Joel R. and {Burden}, Angela and {Burtin}, Etienne and {Busca}, Nicol{\'a}s G. and {Cai}, Zheng and {Capozzi}, Diego and {Carnero Rosell}, Aurelio and {Carr}, Michael A. and {Carrera}, Ricardo and {Chambers}, K.~C. and {Chaplin}, William James and {Chen}, Yen-Chi and {Chiappini}, Cristina and {Chojnowski}, S. Drew and {Chuang}, Chia-Hsun and {Clerc}, Nicolas and {Comparat}, Johan and {Covey}, Kevin and {Croft}, Rupert A.~C. and {Cuesta}, Antonio J. and {Cunha}, Katia and {da Costa}, Luiz N. and {Da Rio}, Nicola and {Davenport}, James R.~A. and {Dawson}, Kyle S. and {De Lee}, Nathan and {Delubac}, Timoth{\'e}e and {Deshpande}, Rohit and {Dhital}, Saurav and {Dutra-Ferreira}, Let{\'\i}cia and {Dwelly}, Tom and {Ealet}, Anne and {Ebelke}, Garrett L. and {Edmondson}, Edward M. and {Eisenstein}, Daniel J. and {Ellsworth}, Tristan and {Elsworth}, Yvonne and {Epstein}, Courtney R. and {Eracleous}, Michael and {Escoffier}, Stephanie and {Esposito}, Massimiliano and {Evans}, Michael L. and {Fan}, Xiaohui and {Fern{\'a}ndez-Alvar}, Emma and {Feuillet}, Diane and {Filiz Ak}, Nurten and {Finley}, Hayley and {Finoguenov}, Alexis and {Flaherty}, Kevin and {Fleming}, Scott W. and {Font-Ribera}, Andreu and {Foster}, Jonathan and {Frinchaboy}, Peter M. and {Galbraith-Frew}, J.~G. and {Garc{\'\i}a}, Rafael A. and {Garc{\'\i}a-Hern{\'a}ndez}, D.~A. and {Garc{\'\i}a P{\'e}rez}, Ana E. and {Gaulme}, Patrick and {Ge}, Jian and {G{\'e}nova-Santos}, R. and {Georgakakis}, A. and {Ghezzi}, Luan and {Gillespie}, Bruce A. and {Girardi}, L{\'e}o and {Goddard}, Daniel and {Gontcho}, Satya Gontcho A. and {Gonz{\'a}lez Hern{\'a}ndez}, Jonay I. and {Grebel}, Eva K. and {Green}, Paul J. and {Grieb}, Jan Niklas and {Grieves}, Nolan and {Gunn}, James E. and {Guo}, Hong and {Harding}, Paul and {Hasselquist}, Sten and {Hawley}, Suzanne L. and {Hayden}, Michael and {Hearty}, Fred R. and {Hekker}, Saskia and {Ho}, Shirley and {Hogg}, David W. and {Holley-Bockelmann}, Kelly and {Holtzman}, Jon A. and {Honscheid}, Klaus and {Huber}, Daniel and {Huehnerhoff}, Joseph and {Ivans}, Inese I. and {Jiang}, Linhua and {Johnson}, Jennifer A. and {Kinemuchi}, Karen and {Kirkby}, David and {Kitaura}, Francisco and {Klaene}, Mark A. and {Knapp}, Gillian R. and {Kneib}, Jean-Paul and {Koenig}, Xavier P. and {Lam}, Charles R. and {Lan}, Ting-Wen and {Lang}, Dustin and {Laurent}, Pierre and {Le Goff}, Jean-Marc and {Leauthaud}, Alexie and {Lee}, Khee-Gan and {Lee}, Young Sun and {Licquia}, Timothy C. and {Liu}, Jian and {Long}, Daniel C. and {L{\'o}pez-Corredoira}, Mart{\'\i}n and {Lorenzo-Oliveira}, Diego and {Lucatello}, Sara and {Lundgren}, Britt and {Lupton}, Robert H. and {Mack}, III, Claude E. and {Mahadevan}, Suvrath and {Maia}, Marcio A.~G. and {Majewski}, Steven R. and {Malanushenko}, Elena and {Malanushenko}, Viktor and {Manchado}, A. and {Manera}, Marc and {Mao}, Qingqing and {Maraston}, Claudia and {Marchwinski}, Robert C. and {Margala}, Daniel and {Martell}, Sarah L. and {Martig}, Marie and {Masters}, Karen L. and {Mathur}, Savita and {McBride}, Cameron K. and {McGehee}, Peregrine M. and {McGreer}, Ian D. and {McMahon}, Richard G. and {M{\'e}nard}, Brice and {Menzel}, Marie-Luise and {Merloni}, Andrea and {M{\'e}sz{\'a}ros}, Szabolcs and {Miller}, Adam A. and {Miralda-Escud{\'e}}, Jordi and {Miyatake}, Hironao and {Montero-Dorta}, Antonio D. and {More}, Surhud and {Morganson}, Eric and {Morice-Atkinson}, Xan and {Morrison}, Heather L. and {Mosser}, Ben{\^o}it and {Muna}, Demitri and {Myers}, Adam D. and {Nandra}, Kirpal and {Newman}, Jeffrey A. and {Neyrinck}, Mark and {Nguyen}, Duy Cuong and {Nichol}, Robert C. and {Nidever}, David L. and {Noterdaeme}, Pasquier and {Nuza}, Sebasti{\'a}n E. and {O'Connell}, Julia E. and {O'Connell}, Robert W. and {O'Connell}, Ross and {Ogando}, Ricardo L.~C. and {Olmstead}, Matthew D. and {Oravetz}, Audrey E. and {Oravetz}, Daniel J. and {Osumi}, Keisuke and {Owen}, Russell and {Padgett}, Deborah L. and {Padmanabhan}, Nikhil and {Paegert}, Martin and {Palanque-Delabrouille}, Nathalie and {Pan}, Kaike},
        title = "{The Eleventh and Twelfth Data Releases of the Sloan Digital Sky Survey: Final Data from SDSS-III}",
      journal = {\apjs},
         year = 2015,
        month = jul,
       volume = {219},
       number = {1},
          eid = {12},
        pages = {12},
          doi = {10.1088/0067-0049/219/1/12},
archivePrefix = {arXiv},
       eprint = {1501.00963},
 primaryClass = {astro-ph.IM},
       adsurl = {https://ui.adsabs.harvard.edu/abs/2015ApJS..219...12A}
}

@ARTICLE{tgss,
       author = {{Intema}, H.~T. and {Jagannathan}, P. and {Mooley}, K.~P. and {Frail}, D.~A.},
        title = "{The GMRT 150 MHz all-sky radio survey. First alternative data release TGSS ADR1}",
      journal = {\aap},
         year = 2017,
        month = feb,
       volume = {598},
          eid = {A78},
        pages = {A78},
          doi = {10.1051/0004-6361/201628536},
archivePrefix = {arXiv},
       eprint = {1603.04368},
 primaryClass = {astro-ph.CO},
       adsurl = {https://ui.adsabs.harvard.edu/abs/2017A&A...598A..78I}
}

@ARTICLE{LOFAR-VLBI_pipeline,
       author = {{Morabito}, L.~K. and {Jackson}, N.~J. and {Mooney}, S. and {Sweijen}, F. and {Badole}, S. and {Kukreti}, P. and {Venkattu}, D. and {Groeneveld}, C. and {Kappes}, A. and {Bonnassieux}, E. and {Drabent}, A. and {Iacobelli}, M. and {Croston}, J.~H. and {Best}, P.~N. and {Bondi}, M. and {Callingham}, J.~R. and {Conway}, J.~E. and {Deller}, A.~T. and {Hardcastle}, M.~J. and {McKean}, J.~P. and {Miley}, G.~K. and {Moldon}, J. and {R{\"o}ttgering}, H.~J.~A. and {Tasse}, C. and {Shimwell}, T.~W. and {van Weeren}, R.~J. and {Anderson}, J.~M. and {Asgekar}, A. and {Avruch}, I.~M. and {van Bemmel}, I.~M. and {Bentum}, M.~J. and {Bonafede}, A. and {Brouw}, W.~N. and {Butcher}, H.~R. and {Ciardi}, B. and {Corstanje}, A. and {Coolen}, A. and {Damstra}, S. and {de Gasperin}, F. and {Duscha}, S. and {Eisl{\"o}ffel}, J. and {Engels}, D. and {Falcke}, H. and {Garrett}, M.~A. and {Griessmeier}, J. and {Gunst}, A.~W. and {van Haarlem}, M.~P. and {Hoeft}, M. and {van der Horst}, A.~J. and {J{\"u}tte}, E. and {Kadler}, M. and {Koopmans}, L.~V.~E. and {Krankowski}, A. and {Mann}, G. and {Nelles}, A. and {Oonk}, J.~B.~R. and {Orru}, E. and {Paas}, H. and {Pandey}, V.~N. and {Pizzo}, R.~F. and {Pandey-Pommier}, M. and {Reich}, W. and {Rothkaehl}, H. and {Ruiter}, M. and {Schwarz}, D.~J. and {Shulevski}, A. and {Soida}, M. and {Tagger}, M. and {Vocks}, C. and {Wijers}, R.~A.~M.~J. and {Wijnholds}, S.~J. and {Wucknitz}, O. and {Zarka}, P. and {Zucca}, P.},
        title = "{Sub-arcsecond imaging with the International LOFAR Telescope. I. Foundational calibration strategy and pipeline}",
      journal = {\aap},
         year = 2022,
        month = feb,
       volume = {658},
          eid = {A1},
        pages = {A1},
          doi = {10.1051/0004-6361/202140649},
archivePrefix = {arXiv},
       eprint = {2108.07283},
 primaryClass = {astro-ph.IM},
       adsurl = {https://ui.adsabs.harvard.edu/abs/2022A&A...658A...1M}
}

@ARTICLE{lofar_facet_selfcal,
       author = {{van Weeren}, R.~J. and {Shimwell}, T.~W. and {Botteon}, A. and {Brunetti}, G. and {Br{\"u}ggen}, M. and {Boxelaar}, J.~M. and {Cassano}, R. and {Di Gennaro}, G. and {Andrade-Santos}, F. and {Bonnassieux}, E. and {Bonafede}, A. and {Cuciti}, V. and {Dallacasa}, D. and {de Gasperin}, F. and {Gastaldello}, F. and {Hardcastle}, M.~J. and {Hoeft}, M. and {Kraft}, R.~P. and {Mandal}, S. and {Rossetti}, M. and {R{\"o}ttgering}, H.~J.~A. and {Tasse}, C. and {Wilber}, A.~G.},
        title = "{LOFAR observations of galaxy clusters in HETDEX. Extraction and self-calibration of individual LOFAR targets}",
      journal = {\aap},
         year = 2021,
        month = jul,
       volume = {651},
          eid = {A115},
        pages = {A115},
          doi = {10.1051/0004-6361/202039826},
archivePrefix = {arXiv},
       eprint = {2011.02387},
 primaryClass = {astro-ph.CO},
       adsurl = {https://ui.adsabs.harvard.edu/abs/2021A&A...651A.115V}
}

@ARTICLE{racs_1.66,
       author = {{Duchesne}, S.~W. and {Ross}, K. and {Thomson}, A.~J.~M. and {Lenc}, E. and {Murphy}, Tara and {Galvin}, T.~J. and {Hotan}, A.~W. and {Moss}, V. and {Whiting}, Matthew T.},
        title = "{The Rapid ASKAP Continuum Survey (RACS) VI: The RACS-high 1 655.5 MHz images and catalogue}",
      journal = {\pasa},
         year = 2025,
        month = jan,
       volume = {42},
          eid = {e038},
        pages = {e038},
          doi = {10.1017/pasa.2025.2},
archivePrefix = {arXiv},
       eprint = {2501.04978},
 primaryClass = {astro-ph.GA},
       adsurl = {https://ui.adsabs.harvard.edu/abs/2025PASA...42...38D}
}

@ARTICLE{Coppejans2016,
       author = {{Coppejans}, Rocco and {Frey}, S{\'a}ndor and {Cseh}, D{\'a}vid and {M{\"u}ller}, Cornelia and {Paragi}, Zsolt and {Falcke}, Heino and {Gab{\'a}nyi}, Krisztina {\'E}. and {Gurvits}, Leonid I. and {An}, Tao and {Titov}, Oleg},
        title = "{On the nature of bright compact radio sources at z > 4.5}",
      journal = {\mnras},
         year = 2016,
        month = dec,
       volume = {463},
       number = {3},
        pages = {3260-3275},
          doi = {10.1093/mnras/stw2236},
archivePrefix = {arXiv},
       eprint = {1609.00575},
 primaryClass = {astro-ph.GA},
       adsurl = {https://ui.adsabs.harvard.edu/abs/2016MNRAS.463.3260C}
}

@ARTICLE{coherence_loss1,
       author = {{Mosoni}, L. and {Frey}, S. and {Gurvits}, L.~I. and {Garrett}, M.~A. and {Garrington}, S.~T. and {Tsvetanov}, Z.~I.},
        title = "{Deep Extragalactic VLBI-Optical Survey (DEVOS). I. Pilot MERLIN and VLBI observations}",
      journal = {\aap},
         year = 2006,
        month = jan,
       volume = {445},
       number = {2},
        pages = {413-422},
          doi = {10.1051/0004-6361:20053473},
archivePrefix = {arXiv},
       eprint = {astro-ph/0508579},
 primaryClass = {astro-ph},
       adsurl = {https://ui.adsabs.harvard.edu/abs/2006A&A...445..413M}
}

@ARTICLE{coherence_loss2,
       author = {{Gab{\'a}nyi}, K. {\'E}. and {Frey}, S. and {Satyapal}, S. and {Constantin}, A. and {Pfeifle}, R.~W.},
        title = "{Very long baseline interferometry observation of the triple AGN candidate J0849+1114}",
      journal = {\aap},
         year = 2019,
        month = oct,
       volume = {630},
          eid = {L5},
        pages = {L5},
          doi = {10.1051/0004-6361/201936519},
archivePrefix = {arXiv},
       eprint = {1909.03259},
 primaryClass = {astro-ph.GA},
       adsurl = {https://ui.adsabs.harvard.edu/abs/2019A&A...630L...5G}
}

@ARTICLE{VLSSr,
       author = {{Lane}, W.~M. and {Cotton}, W.~D. and {van Velzen}, S. and {Clarke}, T.~E. and {Kassim}, N.~E. and {Helmboldt}, J.~F. and {Lazio}, T.~J.~W. and {Cohen}, A.~S.},
        title = "{The Very Large Array Low-frequency Sky Survey Redux (VLSSr)}",
      journal = {\mnras},
         year = 2014,
        month = may,
       volume = {440},
       number = {1},
        pages = {327-338},
          doi = {10.1093/mnras/stu256},
archivePrefix = {arXiv},
       eprint = {1404.0694},
 primaryClass = {astro-ph.IM},
       adsurl = {https://ui.adsabs.harvard.edu/abs/2014MNRAS.440..327L}
}

@INPROCEEDINGS{aips,
       author = {{Greisen}, E.~W.},
        title = "{AIPS, the VLA, and the VLBA}",
    booktitle = {Information Handling in Astronomy - Historical Vistas},
         year = 2003,
       editor = {{Heck}, Andr{\'e}},
       series = {Astrophysics and Space Science Library},
       volume = {285},
        month = mar,
        pages = {109},
          doi = {10.1007/0-306-48080-8_7},
       adsurl = {https://ui.adsabs.harvard.edu/abs/2003ASSL..285..109G}
}

@ARTICLE{Rakshit2020,
       author = {{Rakshit}, Suvendu and {Stalin}, C.~S. and {Kotilainen}, Jari},
        title = "{Spectral Properties of Quasars from Sloan Digital Sky Survey Data Release 14: The Catalog}",
      journal = {\apjs},
         year = 2020,
        month = jul,
       volume = {249},
       number = {1},
          eid = {17},
        pages = {17},
          doi = {10.3847/1538-4365/ab99c5},
archivePrefix = {arXiv},
       eprint = {1910.10395},
 primaryClass = {astro-ph.GA},
       adsurl = {https://ui.adsabs.harvard.edu/abs/2020ApJS..249...17R}
}

@ARTICLE{Snios2020,
       author = {{Snios}, Bradford and {Siemiginowska}, Aneta and {Sobolewska}, Ma{\l}gosia and {Cheung}, C.~C. and {Kashyap}, Vinay and {Migliori}, Giulia and {Schwartz}, Daniel A. and {Stawarz}, {\L}ukasz and {Worrall}, Diana M.},
        title = "{X-Ray Properties of Young Radio Quasars at z > 4.5}",
      journal = {\apj},
         year = 2020,
        month = aug,
       volume = {899},
       number = {2},
          eid = {127},
        pages = {127},
          doi = {10.3847/1538-4357/aba2ca},
archivePrefix = {arXiv},
       eprint = {2007.01342},
 primaryClass = {astro-ph.HE},
       adsurl = {https://ui.adsabs.harvard.edu/abs/2020ApJ...899..127S}
}

@ARTICLE{Rusinek2017,
       author = {{Rusinek}, Katarzyna and {Sikora}, Marek and {Kozie{\l}-Wierzbowska}, Dorota and {Godfrey}, Leith},
        title = "{On the efficiency of jet production in FR II radio galaxies and quasars}",
      journal = {\mnras},
         year = 2017,
        month = apr,
       volume = {466},
       number = {2},
        pages = {2294-2301},
          doi = {10.1093/mnras/stw3330},
archivePrefix = {arXiv},
       eprint = {1612.07392},
 primaryClass = {astro-ph.GA},
       adsurl = {https://ui.adsabs.harvard.edu/abs/2017MNRAS.466.2294R}
}

@ARTICLE{Wolowska2021,
       author = {{Wo{\l}owska}, Aleksandra and {Kunert-Bajraszewska}, Magdalena and {Mooley}, Kunal P. and {Siemiginowska}, Aneta and {Kharb}, Preeti and {Ishwara-Chandra}, C.~H. and {Hallinan}, Gregg and {Gromadzki}, Mariusz and {Kozie{\l}-Wierzbowska}, Dorota},
        title = "{Caltech-NRAO Stripe 82 Survey (CNSS). V. AGNs That Transitioned to Radio-loud State}",
      journal = {\apj},
         year = 2021,
        month = jun,
       volume = {914},
       number = {1},
          eid = {22},
        pages = {22},
          doi = {10.3847/1538-4357/abe62d},
archivePrefix = {arXiv},
       eprint = {2103.08422},
 primaryClass = {astro-ph.GA},
       adsurl = {https://ui.adsabs.harvard.edu/abs/2021ApJ...914...22W}
}

@ARTICLE{Coppejans2017,
       author = {{Coppejans}, Rocco and {van Velzen}, Sjoert and {Intema}, Huib T. and {M{\"u}ller}, Cornelia and {Frey}, S{\'a}ndor and {Coppejans}, Deanne L. and {Cseh}, D{\'a}vid and {Williams}, Wendy L. and {Falcke}, Heino and {K{\"o}rding}, Elmar G. and {Orr{\'u}}, Emanuela and {Paragi}, Zsolt and {Gab{\'a}nyi}, Krisztina {\'E}.},
        title = "{Radio spectra of bright compact sources at z > 4.5}",
      journal = {\mnras},
         year = 2017,
        month = may,
       volume = {467},
       number = {2},
        pages = {2039-2060},
          doi = {10.1093/mnras/stx215},
archivePrefix = {arXiv},
       eprint = {1701.06622},
 primaryClass = {astro-ph.GA},
       adsurl = {https://ui.adsabs.harvard.edu/abs/2017MNRAS.467.2039C}
}

@ARTICLE{Magda2010,
       author = {{Kunert-Bajraszewska}, M. and {Gawro{\'n}ski}, M.~P. and {Labiano}, A. and {Siemiginowska}, A.},
        title = "{A survey of low-luminosity compact sources and its implication for the evolution of radio-loud active galactic nuclei - I. Radio data}",
      journal = {\mnras},
         year = 2010,
        month = nov,
       volume = {408},
       number = {4},
        pages = {2261-2278},
          doi = {10.1111/j.1365-2966.2010.17271.x},
archivePrefix = {arXiv},
       eprint = {1009.5235},
 primaryClass = {astro-ph.CO},
       adsurl = {https://ui.adsabs.harvard.edu/abs/2010MNRAS.408.2261K}
}

@ARTICLE{Fan2003,
       author = {{Fan}, J.~H. and {Zhang}, J.~S.},
        title = "{The core dominance parameter of extragalactic radio sources}",
      journal = {\aap},
         year = 2003,
        month = sep,
       volume = {407},
        pages = {899-904},
          doi = {10.1051/0004-6361:20030896},
       adsurl = {https://ui.adsabs.harvard.edu/abs/2003A&A...407..899F}
}

@ARTICLE{Magda2025,
       author = {{Kunert-Bajraszewska}, Magdalena and {Krauze}, Aleksandra and {Kimball}, Amy E. and {Stawarz}, {\L}ukasz and {Kharb}, Preeti and {Stern}, Daniel and {Mooley}, Kunal and {Nyland}, Kristina and {Kozie{\l}-Wierzbowska}, Dorota},
        title = "{VLASS-based Survey of Transition State Galaxies and Their Relationship to Compact Peaked-spectrum Radio Sources}",
      journal = {\apjs},
         year = 2025,
        month = apr,
       volume = {277},
       number = {2},
          eid = {50},
        pages = {50},
          doi = {10.3847/1538-4365/ada68f},
archivePrefix = {arXiv},
       eprint = {2412.07702},
 primaryClass = {astro-ph.GA},
       adsurl = {https://ui.adsabs.harvard.edu/abs/2025ApJS..277...50K}
}

@ARTICLE{AnBaan,
       author = {{An}, Tao and {Baan}, Willem A.},
        title = "{The Dynamic Evolution of Young Extragalactic Radio Sources}",
      journal = {\apj},
         year = 2012,
        month = nov,
       volume = {760},
       number = {1},
          eid = {77},
        pages = {77},
          doi = {10.1088/0004-637X/760/1/77},
archivePrefix = {arXiv},
       eprint = {1211.1760},
 primaryClass = {astro-ph.CO},
       adsurl = {https://ui.adsabs.harvard.edu/abs/2012ApJ...760...77A}
}

@ARTICLE{gaia_dr3,
       author = {{Gaia Collaboration} and {Vallenari}, A. and {Brown}, A.~G.~A. and {Prusti}, T. and {de Bruijne}, J.~H.~J. and {Arenou}, F. and {Babusiaux}, C. and {Biermann}, M. and {Creevey}, O.~L. and {Ducourant}, C. and {Evans}, D.~W. and {Eyer}, L. and {Guerra}, R. and {Hutton}, A. and {Jordi}, C. and {Klioner}, S.~A. and {Lammers}, U.~L. and {Lindegren}, L. and {Luri}, X. and {Mignard}, F. and {Panem}, C. and {Pourbaix}, D. and {Randich}, S. and {Sartoretti}, P. and {Soubiran}, C. and {Tanga}, P. and {Walton}, N.~A. and {Bailer-Jones}, C.~A.~L. and {Bastian}, U. and {Drimmel}, R. and {Jansen}, F. and {Katz}, D. and {Lattanzi}, M.~G. and {van Leeuwen}, F. and {Bakker}, J. and {Cacciari}, C. and {Casta{\~n}eda}, J. and {De Angeli}, F. and {Fabricius}, C. and {Fouesneau}, M. and {Fr{\'e}mat}, Y. and {Galluccio}, L. and {Guerrier}, A. and {Heiter}, U. and {Masana}, E. and {Messineo}, R. and {Mowlavi}, N. and {Nicolas}, C. and {Nienartowicz}, K. and {Pailler}, F. and {Panuzzo}, P. and {Riclet}, F. and {Roux}, W. and {Seabroke}, G.~M. and {Sordo}, R. and {Th{\'e}venin}, F. and {Gracia-Abril}, G. and {Portell}, J. and {Teyssier}, D. and {Altmann}, M. and {Andrae}, R. and {Audard}, M. and {Bellas-Velidis}, I. and {Benson}, K. and {Berthier}, J. and {Blomme}, R. and {Burgess}, P.~W. and {Busonero}, D. and {Busso}, G. and {C{\'a}novas}, H. and {Carry}, B. and {Cellino}, A. and {Cheek}, N. and {Clementini}, G. and {Damerdji}, Y. and {Davidson}, M. and {de Teodoro}, P. and {Nu{\~n}ez Campos}, M. and {Delchambre}, L. and {Dell'Oro}, A. and {Esquej}, P. and {Fern{\'a}ndez-Hern{\'a}ndez}, J. and {Fraile}, E. and {Garabato}, D. and {Garc{\'\i}a-Lario}, P. and {Gosset}, E. and {Haigron}, R. and {Halbwachs}, J. -L. and {Hambly}, N.~C. and {Harrison}, D.~L. and {Hern{\'a}ndez}, J. and {Hestroffer}, D. and {Hodgkin}, S.~T. and {Holl}, B. and {Jan{\ss}en}, K. and {Jevardat de Fombelle}, G. and {Jordan}, S. and {Krone-Martins}, A. and {Lanzafame}, A.~C. and {L{\"o}ffler}, W. and {Marchal}, O. and {Marrese}, P.~M. and {Moitinho}, A. and {Muinonen}, K. and {Osborne}, P. and {Pancino}, E. and {Pauwels}, T. and {Recio-Blanco}, A. and {Reyl{\'e}}, C. and {Riello}, M. and {Rimoldini}, L. and {Roegiers}, T. and {Rybizki}, J. and {Sarro}, L.~M. and {Siopis}, C. and {Smith}, M. and {Sozzetti}, A. and {Utrilla}, E. and {van Leeuwen}, M. and {Abbas}, U. and {{\'A}brah{\'a}m}, P. and {Abreu Aramburu}, A. and {Aerts}, C. and {Aguado}, J.~J. and {Ajaj}, M. and {Aldea-Montero}, F. and {Altavilla}, G. and {{\'A}lvarez}, M.~A. and {Alves}, J. and {Anders}, F. and {Anderson}, R.~I. and {Anglada Varela}, E. and {Antoja}, T. and {Baines}, D. and {Baker}, S.~G. and {Balaguer-N{\'u}{\~n}ez}, L. and {Balbinot}, E. and {Balog}, Z. and {Barache}, C. and {Barbato}, D. and {Barros}, M. and {Barstow}, M.~A. and {Bartolom{\'e}}, S. and {Bassilana}, J. -L. and {Bauchet}, N. and {Becciani}, U. and {Bellazzini}, M. and {Berihuete}, A. and {Bernet}, M. and {Bertone}, S. and {Bianchi}, L. and {Binnenfeld}, A. and {Blanco-Cuaresma}, S. and {Blazere}, A. and {Boch}, T. and {Bombrun}, A. and {Bossini}, D. and {Bouquillon}, S. and {Bragaglia}, A. and {Bramante}, L. and {Breedt}, E. and {Bressan}, A. and {Brouillet}, N. and {Brugaletta}, E. and {Bucciarelli}, B. and {Burlacu}, A. and {Butkevich}, A.~G. and {Buzzi}, R. and {Caffau}, E. and {Cancelliere}, R. and {Cantat-Gaudin}, T. and {Carballo}, R. and {Carlucci}, T. and {Carnerero}, M.~I. and {Carrasco}, J.~M. and {Casamiquela}, L. and {Castellani}, M. and {Castro-Ginard}, A. and {Chaoul}, L. and {Charlot}, P. and {Chemin}, L. and {Chiaramida}, V. and {Chiavassa}, A. and {Chornay}, N. and {Comoretto}, G. and {Contursi}, G. and {Cooper}, W.~J. and {Cornez}, T. and {Cowell}, S. and {Crifo}, F. and {Cropper}, M. and {Crosta}, M. and {Crowley}, C. and {Dafonte}, C. and {Dapergolas}, A. and {David}, M. and {David}, P. and {de Laverny}, P. and {De Luise}, F. and {De March}, R.},
        title = "{Gaia Data Release 3. Summary of the content and survey properties}",
      journal = {\aap},
         year = 2023,
        month = jun,
       volume = {674},
          eid = {A1},
        pages = {A1},
          doi = {10.1051/0004-6361/202243940},
archivePrefix = {arXiv},
       eprint = {2208.00211},
 primaryClass = {astro-ph.GA},
       adsurl = {https://ui.adsabs.harvard.edu/abs/2023A&A...674A...1G}
}

@ARTICLE{Hogbom_clean,
       author = {{H{\"o}gbom}, J.~A.},
        title = "{Aperture Synthesis with a Non-Regular Distribution of Interferometer Baselines}",
      journal = {\aaps},
         year = 1974,
        month = jun,
       volume = {15},
        pages = {417},
       adsurl = {https://ui.adsabs.harvard.edu/abs/1974A&AS...15..417H}
}

@ARTICLE{selfcal,
       author = {{Alef}, W. and {Porcas}, R.~W.},
        title = "{VLBI fringe-fitting with antenna-based residuals.}",
      journal = {\aap},
         year = 1986,
        month = nov,
       volume = {168},
        pages = {365-368},
       adsurl = {https://ui.adsabs.harvard.edu/abs/1986A&A...168..365A}
}

@ARTICLE{wat_review,
       author = {{O'Dea}, Christopher P. and {Baum}, Stefi A.},
        title = "{Wide-Angle-Tail (WAT) Radio Sources}",
      journal = {Galaxies},
         year = 2023,
        month = may,
       volume = {11},
       number = {3},
          eid = {67},
        pages = {67},
          doi = {10.3390/galaxies11030067},
       adsurl = {https://ui.adsabs.harvard.edu/abs/2023Galax..11...67O}
}

@ARTICLE{sasmal_2022,
       author = {{Sasmal}, Tapan K. and {Bera}, Soumen and {Pal}, Sabyasachi and {Mondal}, Soumen},
        title = "{A New Catalog of Head-Tail Radio Galaxies from the VLA FIRST Survey}",
      journal = {\apjs},
         year = 2022,
        month = apr,
       volume = {259},
       number = {2},
          eid = {31},
        pages = {31},
          doi = {10.3847/1538-4365/ac4473},
       adsurl = {https://ui.adsabs.harvard.edu/abs/2022ApJS..259...31S}
}

@ARTICLE{Wing2011,
       author = {{Wing}, Joshua D. and {Blanton}, Elizabeth L.},
        title = "{Galaxy Cluster Environments of Radio Sources}",
      journal = {\aj},
         year = 2011,
        month = mar,
       volume = {141},
       number = {3},
          eid = {88},
        pages = {88},
          doi = {10.1088/0004-6256/141/3/88},
archivePrefix = {arXiv},
       eprint = {1008.1099},
 primaryClass = {astro-ph.CO},
       adsurl = {https://ui.adsabs.harvard.edu/abs/2011AJ....141...88W}
}

@ARTICLE{Zhu2019,
       author = {{Zhu}, S.~F. and {Brandt}, W.~N. and {Wu}, Jianfeng and {Garmire}, G.~P. and {Miller}, B.~P.},
        title = "{Investigating the X-ray enhancements of highly radio-loud quasars at z > 4}",
      journal = {\mnras},
         year = 2019,
        month = jan,
       volume = {482},
       number = {2},
        pages = {2016-2038},
          doi = {10.1093/mnras/sty2832},
archivePrefix = {arXiv},
       eprint = {1810.06572},
 primaryClass = {astro-ph.HE},
       adsurl = {https://ui.adsabs.harvard.edu/abs/2019MNRAS.482.2016Z}
}

@ARTICLE{Ghisellini_CMB,
       author = {{Ghisellini}, G. and {Celotti}, A. and {Tavecchio}, F. and {Haardt}, F. and {Sbarrato}, T.},
        title = "{Radio-loud active galactic nuclei at high redshifts and the cosmic microwave background}",
      journal = {\mnras},
         year = 2014,
        month = mar,
       volume = {438},
       number = {3},
        pages = {2694-2700},
          doi = {10.1093/mnras/stt2394},
archivePrefix = {arXiv},
       eprint = {1311.7147},
 primaryClass = {astro-ph.CO},
       adsurl = {https://ui.adsabs.harvard.edu/abs/2014MNRAS.438.2694G}
}

@ARTICLE{Fabian_CMB,
       author = {{Fabian}, A.~C. and {Walker}, S.~A. and {Celotti}, A. and {Ghisellini}, G. and {Mocz}, P. and {Blundell}, K.~M. and {McMahon}, R.~G.},
        title = "{Do high-redshift quasars have powerful jets ?}",
      journal = {\mnras},
         year = 2014,
        month = jul,
       volume = {442},
        pages = {L81-L84},
          doi = {10.1093/mnrasl/slu065},
archivePrefix = {arXiv},
       eprint = {1404.7367},
 primaryClass = {astro-ph.HE},
       adsurl = {https://ui.adsabs.harvard.edu/abs/2014MNRAS.442L..81F}
}

@ARTICLE{Wright_2006,
       author = {{Wright}, E.~L.},
        title = "{A Cosmology Calculator for the World Wide Web}",
      journal = {\pasp},
         year = 2006,
        month = dec,
       volume = {118},
       number = {850},
        pages = {1711-1715},
          doi = {10.1086/510102},
archivePrefix = {arXiv},
       eprint = {astro-ph/0609593},
 primaryClass = {astro-ph},
       adsurl = {https://ui.adsabs.harvard.edu/abs/2006PASP..118.1711W}
}

@ARTICLE{Trakhtenbrot2017,
       author = {{Trakhtenbrot}, Benny and {Volonteri}, Marta and {Natarajan}, Priyamvada},
        title = "{On the Accretion Rates and Radiative Efficiencies of the Highest-redshift Quasars}",
      journal = {\apjl},
         year = 2017,
        month = feb,
       volume = {836},
       number = {1},
          eid = {L1},
        pages = {L1},
          doi = {10.3847/2041-8213/836/1/L1},
archivePrefix = {arXiv},
       eprint = {1611.00772},
 primaryClass = {astro-ph.GA},
       adsurl = {https://ui.adsabs.harvard.edu/abs/2017ApJ...836L...1T}
}

@ARTICLE{SDSS_DR16,
       author = {{Lyke}, Brad W. and {Higley}, Alexandra N. and {McLane}, J.~N. and {Schurhammer}, Danielle P. and {Myers}, Adam D. and {Ross}, Ashley J. and {Dawson}, Kyle and {Chabanier}, Sol{\`e}ne and {Martini}, Paul and {Busca}, Nicol{\'a}s G. and {Mas des Bourboux}, H{\'e}lion du and {Salvato}, Mara and {Streblyanska}, Alina and {Zarrouk}, Pauline and {Burtin}, Etienne and {Anderson}, Scott F. and {Bautista}, Julian and {Bizyaev}, Dmitry and {Brandt}, W.~N. and {Brinkmann}, Jonathan and {Brownstein}, Joel R. and {Comparat}, Johan and {Green}, Paul and {de la Macorra}, Axel and {Mu{\~n}oz Guti{\'e}rrez}, Andrea and {Hou}, Jiamin and {Newman}, Jeffrey A. and {Palanque-Delabrouille}, Nathalie and {P{\^a}ris}, Isabelle and {Percival}, Will J. and {Petitjean}, Patrick and {Rich}, James and {Rossi}, Graziano and {Schneider}, Donald P. and {Smith}, Alexander and {Vivek}, M. and {Weaver}, Benjamin Alan},
        title = "{The Sloan Digital Sky Survey Quasar Catalog: Sixteenth Data Release}",
      journal = {\apjs},
         year = 2020,
        month = sep,
       volume = {250},
       number = {1},
          eid = {8},
        pages = {8},
          doi = {10.3847/1538-4365/aba623},
archivePrefix = {arXiv},
       eprint = {2007.09001},
 primaryClass = {astro-ph.GA},
       adsurl = {https://ui.adsabs.harvard.edu/abs/2020ApJS..250....8L}
}

@ARTICLE{J1420,
       author = {{Gab{\'a}nyi}, Krisztina {\'E}. and {Frey}, S{\'a}ndor and {An}, Tao and {Cao}, Hongmin and {Paragi}, Zsolt and {Gurvits}, Leonid I. and {Zhang}, Yingkang and {Sbarrato}, Tullia and {Krezinger}, M{\'a}t{\'e} and {Perger}, Krisztina and {Mez{\"o}}, Gy{\"o}rgy},
        title = "{A small radio galaxy at z = 4.026}",
      journal = {Astronomische Nachrichten},
         year = 2021,
        month = nov,
       volume = {342},
       number = {1092},
        pages = {1092-1096},
          doi = {10.1002/asna.20210057},
archivePrefix = {arXiv},
       eprint = {2110.10964},
 primaryClass = {astro-ph.GA},
       adsurl = {https://ui.adsabs.harvard.edu/abs/2021AN....342.1092G}
}

@INPROCEEDINGS{three_radiogal,
       author = {{Gab{\'a}nyi}, K. and {Frey}, S. and {Paragi}, Z. and {Cao}, H. and {An}, T. and {Gurvits}, L. and {Sbarrato}, T. and {Perger}, K. and {Rozgonyi}, K. and {Mez{\H{o}}}, G.},
        title = "{Three little radio galaxies in the early Universe}",
    booktitle = {14th European VLBI Network Symposium \& Users Meeting (EVN 2018)},
         year = 2018,
        month = nov,
          eid = {31},
        pages = {31},
          doi = {10.22323/1.344.0031},
archivePrefix = {arXiv},
       eprint = {1902.07272},
 primaryClass = {astro-ph.GA},
       adsurl = {https://ui.adsabs.harvard.edu/abs/2018evn..confE..31G}
}

@ARTICLE{recent_radiogal,
       author = {{Goyal}, Arti and {Misra}, Arpita and {Dey}, Subhrata and {Sureshkumar}, Unnikrishnan and {Soida}, Marian and {W{\'o}jtowicz}, Anna and {Stasi{\'n}ska}, Gra{\.z}yna and {Vale Asari}, Natalia and {Naqvi}, Syed},
        title = "{Size Measurements and Characterization of the Astrophysical Properties of Multiple-component Radio Active Galactic Nuclei in the ROGUE I Catalog}",
      journal = {\apjs},
         year = 2026,
        month = apr,
       volume = {283},
       number = {2},
          eid = {76},
        pages = {76},
          doi = {10.3847/1538-4365/ae5058},
archivePrefix = {arXiv},
       eprint = {2603.06215},
 primaryClass = {astro-ph.GA},
       adsurl = {https://ui.adsabs.harvard.edu/abs/2026ApJS..283...76G}
}

@ARTICLE{counterjet,
       author = {{Scheuer}, P.~A.~G. and {Readhead}, A.~C.~S.},
        title = "{Superluminally expanding radio sources and the radio-quiet QSOs}",
      journal = {\nat},
         year = 1979,
        month = jan,
       volume = {277},
        pages = {182-185},
          doi = {10.1038/277182a0},
       adsurl = {https://ui.adsabs.harvard.edu/abs/1979Natur.277..182S}
}

@ARTICLE{FR,
       author = {{Fanaroff}, B.~L. and {Riley}, J.~M.},
        title = "{The morphology of extragalactic radio sources of high and low luminosity}",
      journal = {\mnras},
         year = 1974,
        month = may,
       volume = {167},
        pages = {31P-36P},
          doi = {10.1093/mnras/167.1.31P},
       adsurl = {https://ui.adsabs.harvard.edu/abs/1974MNRAS.167P..31F}
}

@ARTICLE{Gaia,
       author = {{Gaia Collaboration} and {Prusti}, T. and {de Bruijne}, J.~H.~J. and {Brown}, A.~G.~A. and {Vallenari}, A. and {Babusiaux}, C. and {Bailer-Jones}, C.~A.~L. and {Bastian}, U. and {Biermann}, M. and {Evans}, D.~W. and {Eyer}, L. and {Jansen}, F. and {Jordi}, C. and {Klioner}, S.~A. and {Lammers}, U. and {Lindegren}, L. and {Luri}, X. and {Mignard}, F. and {Milligan}, D.~J. and {Panem}, C. and {Poinsignon}, V. and {Pourbaix}, D. and {Randich}, S. and {Sarri}, G. and {Sartoretti}, P. and {Siddiqui}, H.~I. and {Soubiran}, C. and {Valette}, V. and {van Leeuwen}, F. and {Walton}, N.~A. and {Aerts}, C. and {Arenou}, F. and {Cropper}, M. and {Drimmel}, R. and {H{\o}g}, E. and {Katz}, D. and {Lattanzi}, M.~G. and {O'Mullane}, W. and {Grebel}, E.~K. and {Holland}, A.~D. and {Huc}, C. and {Passot}, X. and {Bramante}, L. and {Cacciari}, C. and {Casta{\~n}eda}, J. and {Chaoul}, L. and {Cheek}, N. and {De Angeli}, F. and {Fabricius}, C. and {Guerra}, R. and {Hern{\'a}ndez}, J. and {Jean-Antoine-Piccolo}, A. and {Masana}, E. and {Messineo}, R. and {Mowlavi}, N. and {Nienartowicz}, K. and {Ord{\'o}{\~n}ez-Blanco}, D. and {Panuzzo}, P. and {Portell}, J. and {Richards}, P.~J. and {Riello}, M. and {Seabroke}, G.~M. and {Tanga}, P. and {Th{\'e}venin}, F. and {Torra}, J. and {Els}, S.~G. and {Gracia-Abril}, G. and {Comoretto}, G. and {Garcia-Reinaldos}, M. and {Lock}, T. and {Mercier}, E. and {Altmann}, M. and {Andrae}, R. and {Astraatmadja}, T.~L. and {Bellas-Velidis}, I. and {Benson}, K. and {Berthier}, J. and {Blomme}, R. and {Busso}, G. and {Carry}, B. and {Cellino}, A. and {Clementini}, G. and {Cowell}, S. and {Creevey}, O. and {Cuypers}, J. and {Davidson}, M. and {De Ridder}, J. and {de Torres}, A. and {Delchambre}, L. and {Dell'Oro}, A. and {Ducourant}, C. and {Fr{\'e}mat}, Y. and {Garc{\'\i}a-Torres}, M. and {Gosset}, E. and {Halbwachs}, J.-L. and {Hambly}, N.~C. and {Harrison}, D.~L. and {Hauser}, M. and {Hestroffer}, D. and {Hodgkin}, S.~T. and {Huckle}, H.~E. and {Hutton}, A. and {Jasniewicz}, G. and {Jordan}, S. and {Kontizas}, M. and {Korn}, A.~J. and {Lanzafame}, A.~C. and {Manteiga}, M. and {Moitinho}, A. and {Muinonen}, K. and {Osinde}, J. and {Pancino}, E. and {Pauwels}, T. and {Petit}, J.-M. and {Recio-Blanco}, A. and {Robin}, A.~C. and {Sarro}, L.~M. and {Siopis}, C. and {Smith}, M. and {Smith}, K.~W. and {Sozzetti}, A. and {Thuillot}, W. and {van Reeven}, W. and {Viala}, Y. and {Abbas}, U. and {Abreu Aramburu}, A. and {Accart}, S. and {Aguado}, J.~J. and {Allan}, P.~M. and {Allasia}, W. and {Altavilla}, G. and {{\'A}lvarez}, M.~A. and {Alves}, J. and {Anderson}, R.~I. and {Andrei}, A.~H. and {Anglada Varela}, E. and {Antiche}, E. and {Antoja}, T. and {Ant{\'o}n}, S. and {Arcay}, B. and {Atzei}, A. and {Ayache}, L. and {Bach}, N. and {Baker}, S.~G. and {Balaguer-N{\'u}{\~n}ez}, L. and {Barache}, C. and {Barata}, C. and {Barbier}, A. and {Barblan}, F. and {Baroni}, M. and {Barrado y Navascu{\'e}s}, D. and {Barros}, M. and {Barstow}, M.~A. and {Becciani}, U. and {Bellazzini}, M. and {Bellei}, G. and {Bello Garc{\'\i}a}, A. and {Belokurov}, V. and {Bendjoya}, P. and {Berihuete}, A. and {Bianchi}, L. and {Bienaym{\'e}}, O. and {Billebaud}, F. and {Blagorodnova}, N. and {Blanco-Cuaresma}, S. and {Boch}, T. and {Bombrun}, A. and {Borrachero}, R. and {Bouquillon}, S. and {Bourda}, G. and {Bouy}, H. and {Bragaglia}, A. and {Breddels}, M.~A. and {Brouillet}, N. and {Br{\"u}semeister}, T. and {Bucciarelli}, B. and {Budnik}, F. and {Burgess}, P. and {Burgon}, R. and {Burlacu}, A. and {Busonero}, D. and {Buzzi}, R. and {Caffau}, E. and {Cambras}, J. and {Campbell}, H. and {Cancelliere}, R. and {Cantat-Gaudin}, T. and {Carlucci}, T. and {Carrasco}, J.~M. and {Castellani}, M. and {Charlot}, P. and {Charnas}, J. and {Charvet}, P. and {Chassat}, F. and {Chiavassa}, A. and {Clotet}, M. and {Cocozza}, G. and {Collins}, R.~S. and {Collins}, P. and {Costigan}, G.},
        title = "{The Gaia mission}",
      journal = {\aap},
         year = 2016,
        month = nov,
       volume = {595},
          eid = {A1},
        pages = {A1},
          doi = {10.1051/0004-6361/201629272},
archivePrefix = {arXiv},
       eprint = {1609.04153},
 primaryClass = {astro-ph.IM},
       adsurl = {https://ui.adsabs.harvard.edu/abs/2016A&A...595A...1G}
}

@ARTICLE{desi_2019,
       author = {{Dey}, Arjun and {Schlegel}, David J. and {Lang}, Dustin and {Blum}, Robert and {Burleigh}, Kaylan and {Fan}, Xiaohui and {Findlay}, Joseph R. and {Finkbeiner}, Doug and {Herrera}, David and {Juneau}, St{\'e}phanie and {Landriau}, Martin and {Levi}, Michael and {McGreer}, Ian and {Meisner}, Aaron and {Myers}, Adam D. and {Moustakas}, John and {Nugent}, Peter and {Patej}, Anna and {Schlafly}, Edward F. and {Walker}, Alistair R. and {Valdes}, Francisco and {Weaver}, Benjamin A. and {Y{\`e}che}, Christophe and {Zou}, Hu and {Zhou}, Xu and {Abareshi}, Behzad and {Abbott}, T.~M.~C. and {Abolfathi}, Bela and {Aguilera}, C. and {Alam}, Shadab and {Allen}, Lori and {Alvarez}, A. and {Annis}, James and {Ansarinejad}, Behzad and {Aubert}, Marie and {Beechert}, Jacqueline and {Bell}, Eric F. and {BenZvi}, Segev Y. and {Beutler}, Florian and {Bielby}, Richard M. and {Bolton}, Adam S. and {Brice{\~n}o}, C{\'e}sar and {Buckley-Geer}, Elizabeth J. and {Butler}, Karen and {Calamida}, Annalisa and {Carlberg}, Raymond G. and {Carter}, Paul and {Casas}, Ricard and {Castander}, Francisco J. and {Choi}, Yumi and {Comparat}, Johan and {Cukanovaite}, Elena and {Delubac}, Timoth{\'e}e and {DeVries}, Kaitlin and {Dey}, Sharmila and {Dhungana}, Govinda and {Dickinson}, Mark and {Ding}, Zhejie and {Donaldson}, John B. and {Duan}, Yutong and {Duckworth}, Christopher J. and {Eftekharzadeh}, Sarah and {Eisenstein}, Daniel J. and {Etourneau}, Thomas and {Fagrelius}, Parker A. and {Farihi}, Jay and {Fitzpatrick}, Mike and {Font-Ribera}, Andreu and {Fulmer}, Leah and {G{\"a}nsicke}, Boris T. and {Gaztanaga}, Enrique and {George}, Koshy and {Gerdes}, David W. and {Gontcho}, Satya Gontcho A. and {Gorgoni}, Claudio and {Green}, Gregory and {Guy}, Julien and {Harmer}, Diane and {Hernandez}, M. and {Honscheid}, Klaus and {Huang}, Lijuan Wendy and {James}, David J. and {Jannuzi}, Buell T. and {Jiang}, Linhua and {Joyce}, Richard and {Karcher}, Armin and {Karkar}, Sonia and {Kehoe}, Robert and {Kneib}, Jean-Paul and {Kueter-Young}, Andrea and {Lan}, Ting-Wen and {Lauer}, Tod R. and {Le Guillou}, Laurent and {Le Van Suu}, Auguste and {Lee}, Jae Hyeon and {Lesser}, Michael and {Perreault Levasseur}, Laurence and {Li}, Ting S. and {Mann}, Justin L. and {Marshall}, Robert and {Mart{\'\i}nez-V{\'a}zquez}, C.~E. and {Martini}, Paul and {du Mas des Bourboux}, H{\'e}lion and {McManus}, Sean and {Meier}, Tobias Gabriel and {M{\'e}nard}, Brice and {Metcalfe}, Nigel and {Mu{\~n}oz-Guti{\'e}rrez}, Andrea and {Najita}, Joan and {Napier}, Kevin and {Narayan}, Gautham and {Newman}, Jeffrey A. and {Nie}, Jundan and {Nord}, Brian and {Norman}, Dara J. and {Olsen}, Knut A.~G. and {Paat}, Anthony and {Palanque-Delabrouille}, Nathalie and {Peng}, Xiyan and {Poppett}, Claire L. and {Poremba}, Megan R. and {Prakash}, Abhishek and {Rabinowitz}, David and {Raichoor}, Anand and {Rezaie}, Mehdi and {Robertson}, A.~N. and {Roe}, Natalie A. and {Ross}, Ashley J. and {Ross}, Nicholas P. and {Rudnick}, Gregory and {Safonova}, Sasha and {Saha}, Abhijit and {S{\'a}nchez}, F. Javier and {Savary}, Elodie and {Schweiker}, Heidi and {Scott}, Adam and {Seo}, Hee-Jong and {Shan}, Huanyuan and {Silva}, David R. and {Slepian}, Zachary and {Soto}, Christian and {Sprayberry}, David and {Staten}, Ryan and {Stillman}, Coley M. and {Stupak}, Robert J. and {Summers}, David L. and {Sien Tie}, Suk and {Tirado}, H. and {Vargas-Maga{\~n}a}, Mariana and {Vivas}, A. Katherina and {Wechsler}, Risa H. and {Williams}, Doug and {Yang}, Jinyi and {Yang}, Qian and {Yapici}, Tolga and {Zaritsky}, Dennis and {Zenteno}, A. and {Zhang}, Kai and {Zhang}, Tianmeng and {Zhou}, Rongpu and {Zhou}, Zhimin},
        title = "{Overview of the DESI Legacy Imaging Surveys}",
      journal = {\aj},
         year = 2019,
        month = may,
       volume = {157},
       number = {5},
          eid = {168},
        pages = {168},
          doi = {10.3847/1538-3881/ab089d},
archivePrefix = {arXiv},
       eprint = {1804.08657},
 primaryClass = {astro-ph.IM},
       adsurl = {https://ui.adsabs.harvard.edu/abs/2019AJ....157..168D}
}

@ARTICLE{Sax_TNJ1338,
       author = {{Saxena}, Aayush and {Overzier}, Roderik A. and {Villar-Mart{\'\i}n}, Montserrat and {Heckman}, Tim and {Roy}, Namrata and {Duncan}, Kenneth J. and {R{\"o}ttgering}, Huub and {Miley}, George and {Aydar}, Catarina and {Best}, Philip and {Bosman}, Sarah E.~I. and {Cameron}, Alex J. and {Gab{\'a}nyi}, Krisztina {\'E}va and {Humphrey}, Andrew and {Morais}, Sandy and {Onoue}, Masafusa and {Pentericci}, Laura and {Reynaldi}, Victoria and {Venemans}, Bram},
        title = "{Widespread AGN feedback in a forming brightest cluster galaxy at z = 4.1, unveiled by JWST}",
      journal = {\mnras},
         year = 2024,
        month = jul,
       volume = {531},
       number = {4},
        pages = {4391-4407},
          doi = {10.1093/mnras/stae1406},
archivePrefix = {arXiv},
       eprint = {2401.12199},
 primaryClass = {astro-ph.GA},
       adsurl = {https://ui.adsabs.harvard.edu/abs/2024MNRAS.531.4391S}
}

@ARTICLE{J1430_largejet,
       author = {{Cheung}, C.~C. and {Stawarz}, {\L}. and {Siemiginowska}, A. and {Gobeille}, D. and {Wardle}, J.~F.~C. and {Harris}, D.~E. and {Schwartz}, D.~A.},
        title = "{Discovery of a Kiloparsec-scale X-Ray/Radio Jet in the z = 4.72 Quasar GB 1428+4217}",
      journal = {\apjl},
         year = 2012,
        month = sep,
       volume = {756},
       number = {1},
          eid = {L20},
        pages = {L20},
          doi = {10.1088/2041-8205/756/1/L20},
archivePrefix = {arXiv},
       eprint = {1208.0584},
 primaryClass = {astro-ph.HE},
       adsurl = {https://ui.adsabs.harvard.edu/abs/2012ApJ...756L..20C}
}

@ARTICLE{Tullia_blazarcat,
       author = {{Sbarrato}, Tullia and {Belladitta}, Silvia and {Wolf}, Julien and {Baldini}, Pietro and {Tub{\'\i}n-Arenas}, Dusan and {Salvato}, Mara and {Momjian}, Emmanuel and {H{\"a}mmerich}, Steven and {Merloni}, Andrea and {Collmar}, Werner and {Wilms}, Joern},
        title = "{BLAZ4R and the eROSITA view of z>4 blazars}",
      journal = {arXiv e-prints},
         year = 2026,
        month = mar,
          eid = {arXiv:2603.24659},
        pages = {arXiv:2603.24659},
          doi = {10.48550/arXiv.2603.24659},
archivePrefix = {arXiv},
       eprint = {2603.24659},
 primaryClass = {astro-ph.HE},
       adsurl = {https://ui.adsabs.harvard.edu/abs/2026arXiv260324659S}
}

@ARTICLE{erosita,
       author = {{Predehl}, P. and {Andritschke}, R. and {Arefiev}, V. and {Babyshkin}, V. and {Batanov}, O. and {Becker}, W. and {B{\"o}hringer}, H. and {Bogomolov}, A. and {Boller}, T. and {Borm}, K. and {Bornemann}, W. and {Br{\"a}uninger}, H. and {Br{\"u}ggen}, M. and {Brunner}, H. and {Brusa}, M. and {Bulbul}, E. and {Buntov}, M. and {Burwitz}, V. and {Burkert}, W. and {Clerc}, N. and {Churazov}, E. and {Coutinho}, D. and {Dauser}, T. and {Dennerl}, K. and {Doroshenko}, V. and {Eder}, J. and {Emberger}, V. and {Eraerds}, T. and {Finoguenov}, A. and {Freyberg}, M. and {Friedrich}, P. and {Friedrich}, S. and {F{\"u}rmetz}, M. and {Georgakakis}, A. and {Gilfanov}, M. and {Granato}, S. and {Grossberger}, C. and {Gueguen}, A. and {Gureev}, P. and {Haberl}, F. and {H{\"a}lker}, O. and {Hartner}, G. and {Hasinger}, G. and {Huber}, H. and {Ji}, L. and {Kienlin}, A. v. and {Kink}, W. and {Korotkov}, F. and {Kreykenbohm}, I. and {Lamer}, G. and {Lomakin}, I. and {Lapshov}, I. and {Liu}, T. and {Maitra}, C. and {Meidinger}, N. and {Menz}, B. and {Merloni}, A. and {Mernik}, T. and {Mican}, B. and {Mohr}, J. and {M{\"u}ller}, S. and {Nandra}, K. and {Nazarov}, V. and {Pacaud}, F. and {Pavlinsky}, M. and {Perinati}, E. and {Pfeffermann}, E. and {Pietschner}, D. and {Ramos-Ceja}, M.~E. and {Rau}, A. and {Reiffers}, J. and {Reiprich}, T.~H. and {Robrade}, J. and {Salvato}, M. and {Sanders}, J. and {Santangelo}, A. and {Sasaki}, M. and {Scheuerle}, H. and {Schmid}, C. and {Schmitt}, J. and {Schwope}, A. and {Shirshakov}, A. and {Steinmetz}, M. and {Stewart}, I. and {Str{\"u}der}, L. and {Sunyaev}, R. and {Tenzer}, C. and {Tiedemann}, L. and {Tr{\"u}mper}, J. and {Voron}, V. and {Weber}, P. and {Wilms}, J. and {Yaroshenko}, V.},
        title = "{The eROSITA X-ray telescope on SRG}",
      journal = {\aap},
         year = 2021,
        month = mar,
       volume = {647},
          eid = {A1},
        pages = {A1},
          doi = {10.1051/0004-6361/202039313},
archivePrefix = {arXiv},
       eprint = {2010.03477},
 primaryClass = {astro-ph.HE},
       adsurl = {https://ui.adsabs.harvard.edu/abs/2021A&A...647A...1P}
}

@ARTICLE{Cao_blazarcan,
       author = {{Cao}, H.-M. and {Frey}, S. and {Gab{\'a}nyi}, K. {\'E}. and {Paragi}, Z. and {Yang}, J. and {Cseh}, D. and {Hong}, X.-Y. and {An}, T.},
        title = "{VLBI observations of four radio quasars at z > 4: blazars or not?}",
      journal = {\mnras},
         year = 2017,
        month = may,
       volume = {467},
       number = {1},
        pages = {950-960},
          doi = {10.1093/mnras/stx160},
archivePrefix = {arXiv},
       eprint = {1701.04760},
 primaryClass = {astro-ph.GA},
       adsurl = {https://ui.adsabs.harvard.edu/abs/2017MNRAS.467..950C}
}

@ARTICLE{sbarrato,
       author = {{Sbarrato}, T. and {Ghisellini}, G. and {Tagliaferri}, G. and {Tavecchio}, F. and {Ghirlanda}, G. and {Costamante}, L.},
        title = "{Blazar nature of high-z radio-loud quasars}",
      journal = {\aap},
         year = 2022,
        month = jul,
       volume = {663},
          eid = {A147},
        pages = {A147},
          doi = {10.1051/0004-6361/202243569},
archivePrefix = {arXiv},
       eprint = {2203.09527},
 primaryClass = {astro-ph.HE},
       adsurl = {https://ui.adsabs.harvard.edu/abs/2022A&A...663A.147S}
}

@ARTICLE{drouart2020,
       author = {{Drouart}, Guillaume and {Seymour}, Nick and {Galvin}, Tim J. and {Afonso}, Jose and {Callingham}, Joseph R. and {De Breuck}, Carlos and {Johnston-Hollitt}, Melanie and {Kapi{\'n}ska}, Anna D. and {Lehnert}, Matthew D. and {Vernet}, Jo{\"e}l},
        title = "{The GLEAMing of the first supermassive black holes}",
      journal = {\pasa},
         year = 2020,
        month = jul,
       volume = {37},
          eid = {e026},
        pages = {e026},
          doi = {10.1017/pasa.2020.6},
archivePrefix = {arXiv},
       eprint = {2111.08104},
 primaryClass = {astro-ph.GA},
       adsurl = {https://ui.adsabs.harvard.edu/abs/2020PASA...37...26D}
}

@ARTICLE{vanBreugel1999,
       author = {{van Breugel}, Wil and {De Breuck}, Carlos and {Stanford}, S.~A. and {Stern}, Daniel and {R{\"o}ttgering}, Huub and {Miley}, George},
        title = "{A Radio Galaxy at z = 5.19}",
      journal = {\apjl},
         year = 1999,
        month = jun,
       volume = {518},
       number = {2},
        pages = {L61-L64},
          doi = {10.1086/312080},
archivePrefix = {arXiv},
       eprint = {astro-ph/9904272},
 primaryClass = {astro-ph},
       adsurl = {https://ui.adsabs.harvard.edu/abs/1999ApJ...518L..61V}
}

@ARTICLE{Walker1998,
       author = {{Walker}, Mark A.},
        title = "{Interstellar scintillation of compact extragalactic radio sources}",
      journal = {\mnras},
         year = 1998,
        month = feb,
       volume = {294},
        pages = {307-311},
          doi = {10.1046/j.1365-8711.1998.01238.x10.1111/j.1365-8711.1998.01238.x},
       adsurl = {https://ui.adsabs.harvard.edu/abs/1998MNRAS.294..307W}
}

@ARTICLE{Rickett1990,
       author = {{Rickett}, B.~J.},
        title = "{Radio propagation through the turbulent interstellar plasma.}",
      journal = {\araa},
         year = 1990,
        month = jan,
       volume = {28},
        pages = {561-605},
          doi = {10.1146/annurev.aa.28.090190.003021},
       adsurl = {https://ui.adsabs.harvard.edu/abs/1990ARA&A..28..561R}
}

@ARTICLE{ese_pulsar,
       author = {{Kerr}, M. and {Coles}, W.~A. and {Ward}, C.~A. and {Johnston}, S. and {Tuntsov}, A.~V. and {Shannon}, R.~M.},
        title = "{Extreme scattering events towards two young pulsars}",
      journal = {\mnras},
         year = 2018,
        month = mar,
       volume = {474},
       number = {4},
        pages = {4637-4647},
          doi = {10.1093/mnras/stx3101},
archivePrefix = {arXiv},
       eprint = {1712.00426},
 primaryClass = {astro-ph.HE},
       adsurl = {https://ui.adsabs.harvard.edu/abs/2018MNRAS.474.4637K}
}

@ARTICLE{Ighina2019,
       author = {{Ighina}, L. and {Caccianiga}, A. and {Moretti}, A. and {Belladitta}, S. and {Della Ceca}, R. and {Ballo}, L. and {Dallacasa}, D.},
        title = "{X-ray properties of z > 4 blazars}",
      journal = {\mnras},
         year = 2019,
        month = oct,
       volume = {489},
       number = {2},
        pages = {2732-2745},
          doi = {10.1093/mnras/stz2340},
archivePrefix = {arXiv},
       eprint = {1908.08084},
 primaryClass = {astro-ph.GA},
       adsurl = {https://ui.adsabs.harvard.edu/abs/2019MNRAS.489.2732I}
}

@ARTICLE{Krezinger2026,
       author = {{Krezinger}, M{\'a}t{\'e} and {Caccianiga}, Alessandro and {Dallacasa}, Daniele and {Ighina}, Luca and {Frey}, S{\'a}ndor and {Moretti}, Alberto and {Ant{\'o}n}, Sonia and {Belladitta}, Silvia and {Cicone}, Claudia and {Gab{\'a}nyi}, Krisztina {\'E}. and {March{\~a}}, M.~J.~M. and {Perger}, Krisztina},
        title = "{Milliarcsecond-resolution Radio-imaging Survey of Blazar Candidates at 4 < z < 5.4}",
      journal = {\apjs},
         year = 2026,
        month = jan,
       volume = {282},
       number = {1},
          eid = {21},
        pages = {21},
          doi = {10.3847/1538-4365/ae1ef0},
archivePrefix = {arXiv},
       eprint = {2601.05736},
 primaryClass = {astro-ph.GA},
       adsurl = {https://ui.adsabs.harvard.edu/abs/2026ApJS..282...21K}
}

@ARTICLE{Donato2001,
       author = {{Donato}, D. and {Ghisellini}, G. and {Tagliaferri}, G. and {Fossati}, G.},
        title = "{Hard X-ray properties of blazars}",
      journal = {\aap},
         year = 2001,
        month = sep,
       volume = {375},
        pages = {739-751},
          doi = {10.1051/0004-6361:20010675},
archivePrefix = {arXiv},
       eprint = {astro-ph/0105203},
 primaryClass = {astro-ph},
       adsurl = {https://ui.adsabs.harvard.edu/abs/2001A&A...375..739D}
}

@ARTICLE{Shemmer2006,
       author = {{Shemmer}, Ohad and {Brandt}, W.~N. and {Schneider}, Donald P. and {Fan}, Xiaohui and {Strauss}, Michael A. and {Diamond-Stanic}, Aleksandar M. and {Richards}, Gordon T. and {Anderson}, Scott F. and {Gunn}, James E. and {Brinkmann}, Jon},
        title = "{Chandra Observations of the Highest Redshift Quasars from the Sloan Digital Sky Survey}",
      journal = {\apj},
         year = 2006,
        month = jun,
       volume = {644},
       number = {1},
        pages = {86-99},
          doi = {10.1086/503543},
archivePrefix = {arXiv},
       eprint = {astro-ph/0602442},
 primaryClass = {astro-ph},
       adsurl = {https://ui.adsabs.harvard.edu/abs/2006ApJ...644...86S}
}

@ARTICLE{Readhead,
       author = {{Readhead}, Anthony C.~S.},
        title = "{Equipartition Brightness Temperature and the Inverse Compton Catastrophe}",
      journal = {\apj},
         year = 1994,
        month = may,
       volume = {426},
        pages = {51},
          doi = {10.1086/174038},
       adsurl = {https://ui.adsabs.harvard.edu/abs/1994ApJ...426...51R}
}

@ARTICLE{Homan2021,
       author = {{Homan}, D.~C. and {Cohen}, M.~H. and {Hovatta}, T. and {Kellermann}, K.~I. and {Kovalev}, Y.~Y. and {Lister}, M.~L. and {Popkov}, A.~V. and {Pushkarev}, A.~B. and {Ros}, E. and {Savolainen}, T.},
        title = "{MOJAVE. XIX. Brightness Temperatures and Intrinsic Properties of Blazar Jets}",
      journal = {\apj},
         year = 2021,
        month = dec,
       volume = {923},
       number = {1},
          eid = {67},
        pages = {67},
          doi = {10.3847/1538-4357/ac27af},
archivePrefix = {arXiv},
       eprint = {2109.04977},
 primaryClass = {astro-ph.HE},
       adsurl = {https://ui.adsabs.harvard.edu/abs/2021ApJ...923...67H}
}

@ARTICLE{Lister2021,
       author = {{Lister}, M.~L. and {Homan}, D.~C. and {Kellermann}, K.~I. and {Kovalev}, Y.~Y. and {Pushkarev}, A.~B. and {Ros}, E. and {Savolainen}, T.},
        title = "{Monitoring Of Jets in Active Galactic Nuclei with VLBA Experiments. XVIII. Kinematics and Inner Jet Evolution of Bright Radio-loud Active Galaxies}",
      journal = {\apj},
         year = 2021,
        month = dec,
       volume = {923},
       number = {1},
          eid = {30},
        pages = {30},
          doi = {10.3847/1538-4357/ac230f},
archivePrefix = {arXiv},
       eprint = {2108.13358},
 primaryClass = {astro-ph.HE},
       adsurl = {https://ui.adsabs.harvard.edu/abs/2021ApJ...923...30L}
}

@ARTICLE{Padovani,
       author = {{Padovani}, P. and {Alexander}, D.~M. and {Assef}, R.~J. and {De Marco}, B. and {Giommi}, P. and {Hickox}, R.~C. and {Richards}, G.~T. and {Smol{\v{c}}i{\'c}}, V. and {Hatziminaoglou}, E. and {Mainieri}, V. and {Salvato}, M.},
        title = "{Active galactic nuclei: what's in a name?}",
      journal = {\aapr},
         year = 2017,
        month = aug,
       volume = {25},
       number = {1},
          eid = {2},
        pages = {2},
          doi = {10.1007/s00159-017-0102-9},
archivePrefix = {arXiv},
       eprint = {1707.07134},
 primaryClass = {astro-ph.GA},
       adsurl = {https://ui.adsabs.harvard.edu/abs/2017A&ARv..25....2P}
}

@ARTICLE{Kellermann1989,
       author = {{Kellermann}, K.~I. and {Sramek}, R. and {Schmidt}, M. and {Shaffer}, D.~B. and {Green}, R.},
        title = "{VLA Observations of Objects in the Palomar Bright Quasar Survey}",
      journal = {\aj},
         year = 1989,
        month = oct,
       volume = {98},
        pages = {1195},
          doi = {10.1086/115207},
       adsurl = {https://ui.adsabs.harvard.edu/abs/1989AJ.....98.1195K}
}

@ARTICLE{Kellermann2016,
       author = {{Kellermann}, K.~I. and {Condon}, J.~J. and {Kimball}, A.~E. and {Perley}, R.~A. and {Ivezi{\'c}}, {\v{Z}}eljko},
        title = "{Radio-loud and Radio-quiet QSOs}",
      journal = {\apj},
         year = 2016,
        month = nov,
       volume = {831},
       number = {2},
          eid = {168},
        pages = {168},
          doi = {10.3847/0004-637X/831/2/168},
archivePrefix = {arXiv},
       eprint = {1608.04586},
 primaryClass = {astro-ph.GA},
       adsurl = {https://ui.adsabs.harvard.edu/abs/2016ApJ...831..168K}
}

@ARTICLE{Wojtowicz2020,
       author = {{W{\'o}jtowicz}, A. and {Stawarz}, {\l}. and {Cheung}, C.~C. and {Ostorero}, L. and {Kosmaczewski}, E. and {Siemiginowska}, A.},
        title = "{On the Jet Production Efficiency in a Sample of the Youngest Radio Galaxies}",
      journal = {\apj},
         year = 2020,
        month = apr,
       volume = {892},
       number = {2},
          eid = {116},
        pages = {116},
          doi = {10.3847/1538-4357/ab7930},
archivePrefix = {arXiv},
       eprint = {1911.01197},
 primaryClass = {astro-ph.HE},
       adsurl = {https://ui.adsabs.harvard.edu/abs/2020ApJ...892..116W}
}

@ARTICLE{An_3C48,
       author = {{An}, T. and {Hong}, X.~Y. and {Hardcastle}, M.~J. and {Worrall}, D.~M. and {Venturi}, T. and {Pearson}, T.~J. and {Shen}, Z.-Q. and {Zhao}, W. and {Feng}, W.~X.},
        title = "{Kinematics of the parsec-scale radio jet in 3C 48}",
      journal = {\mnras},
         year = 2010,
        month = feb,
       volume = {402},
       number = {1},
        pages = {87-104},
          doi = {10.1111/j.1365-2966.2009.15899.x},
archivePrefix = {arXiv},
       eprint = {0910.3782},
 primaryClass = {astro-ph.CO},
       adsurl = {https://ui.adsabs.harvard.edu/abs/2010MNRAS.402...87A}
}

@BOOK{VLASS-memo13,
       author = {{Lacy}, M. and {Myers}, S. T. and {Chandler}, C. and {Kapinska}, A. D. and {Kent}, B. and {Kimball}, A. and {Marvil}, J. and {Masters}, J. and {Medlin}, D. and {Radford}, K. and {Schinzel}, F. and {Sjouwerman}, L. and {Sobotka}, A. and {Vargas}, A.},
        title = "{VLASS Project Memo \#13: Pilot and Quick Look Data Release (v2)}",
        publisher ="{US National Radio Astronomy Observatory}",
        year = 2022,
        url = {https://library.nrao.edu/public/memos/vla/vlass/VLASS_013.pdf}
}

@INPROCEEDINGS{asym_Frey2024,
       author = {{Frey}, S. and {Titov}, O. and {Melnikov}, A. and {Lambert}, S.},
        title = "{A new look at old friends: EVN imaging of prominent radio-loud AGNs with large radio-optical positional offsets}",
    booktitle = {Proceedings of the 16th EVN Symposium},
         year = 2024,
       editor = {{Ros}, E. and {Benke}, P. and {Dzib}, S.~A. and {Rottmann}, I. and {Zensus}, J.~A.},
        month = sep,
        pages = {141-144},
          doi = {10.48550/arXiv.2501.06513},
archivePrefix = {arXiv},
       eprint = {2501.06513},
 primaryClass = {astro-ph.GA},
       adsurl = {https://ui.adsabs.harvard.edu/abs/2024evn..conf..141F}
}

@ARTICLE{Feng2005,
       author = {{Feng}, W.~X. and {An}, T. and {Hong}, X.~Y. and {Zhao}, Jun-Hui and {Venturi}, T. and {Shen}, Z.-Q. and {Wang}, W.~H.},
        title = "{The radio counter-jet of the QSO 3C 48}",
      journal = {\aap},
         year = 2005,
        month = apr,
       volume = {434},
       number = {1},
        pages = {101-107},
          doi = {10.1051/0004-6361:20042316},
archivePrefix = {arXiv},
       eprint = {astro-ph/0412507},
 primaryClass = {astro-ph},
       adsurl = {https://ui.adsabs.harvard.edu/abs/2005A&A...434..101F}
}

\begin{appendix}
\section{Flux density measurements}

\begin{table}[h]
    \centering
    \caption{Flux densities and peak intensities of the different emission regions revealed by the $144$\-MHz high-resolution LOFAR imaging.}
    \begin{tabular}{cccr}
    \hline
    \hline
    \noalign{\smallskip}
         Source& Comp. & Peak intensity     & \multicolumn{1}{c}{Flux density} \\
               &           & (mJy\,beam$^{-1}$) & \multicolumn{1}{c}{(mJy)} \\
    \noalign{\smallskip}
    \hline
    \noalign{\smallskip}
    J1231+3816 & C & $28.6 \pm 3.3$ & $84.6 \pm 1.5$\\
               & S &     $8.2\pm1.1$   & $25.8\pm0.2$ \\
               & W &    $4.9 \pm 1.0$    & $23.7 \pm0.4$ \\
    \noalign{\smallskip}
    \hline
    \noalign{\smallskip}
    J0813+3508 & C &     $44.2\pm9.6$   &$103.2 \pm 0.5$ \\
               & E &   $8.3\pm 2.0$     &$67.7 \pm 0.9$\\
               & N &   $32.6\pm5.5$     &$136.4 \pm 0.7$ \\
    \noalign{\smallskip}
    \hline
    \noalign{\smallskip}
    J1548+3335 & C &   $3.6\pm1.0$     &$8.4 \pm 0.4$ \\
               & W &   $59.2\pm 9.6$     &$139.3 \pm 0.4$\\
    \noalign{\smallskip}
    \hline
    \end{tabular}
    \tablefoot{The first two columns show the component designations, while the third and fourth columns list the peak intensity and the integrated flux density of each component, respectively. The flux densities were obtained by Gaussian fitting of the components in the image; further details are given in the text.}
    \label{tab:lofar}
\end{table}

\begin{table}[h]
    \centering
    \caption{Peak intensities and flux densities of the Gaussian brightness distributions fitted to the LoTSS images of J0813$+$3508 and J1548$+$3335, shown in Figs.\,\ref{fig:J0813_lowres} and \ref{fig:J1548_lowres}, respectively.}
    \begin{tabular}{cccr}
    \hline
    \hline
    \noalign{\smallskip}
Source & Comp. & Peak intensity     & \multicolumn{1}{c}{Flux density} \\
                        &  & (mJy\,beam$^{-1}$) & \multicolumn{1}{c}{(mJy)} \\
                        \noalign{\smallskip}
               \hline
               \noalign{\smallskip}
 J0813$+$3508 & C & $109.9 \pm 0.4$ & $138.6 \pm 13.9$ \\
 & N & $118.1 \pm 0.4$ &$157.5\pm15.8$  \\
 \noalign{\smallskip}
 \hline
 \noalign{\smallskip}
 J1548$+$3335 & T & $147.8 \pm 0.8 $ & $163.9\pm 16.4$\\
  & SW & $8.9 \pm 0.8$ & $12.9 \pm 1.3$ \\
  \noalign{\smallskip}
  \hline
    \end{tabular}
     \tablefoot{The columns are the same as in Table\,\ref{tab:lofar}}
    \label{tab:low-res}
\end{table}

\begin{table}[h]
\caption{Parameters of the Gaussian components fitted to the $1.7$-GHz e-MERLIN visibility data of J1231+3816.}             
\label{table:J1231_eMERLIN}      
\centering                          
\begin{tabular}{ c r r c c}        
\hline\hline                 
\noalign{\smallskip}
 Comp. & \multicolumn{1}{c}{Flux density} & \multicolumn{1}{c}{FWHM} & Axial ratio & PA \\ 
 & \multicolumn{1}{c}{(mJy)} & \multicolumn{1}{c}{(mas)} & & ($\degr$) \\
\noalign{\smallskip}
\hline                        
\noalign{\smallskip}
 C1 & $13.8 \pm 0.9 $ & $70.9\pm 0.7$ & $1.0$ & --\\   
 C2 & $5.6\pm 0.4$ & $207.7\pm 2.8$  & $0.26 \pm 0.01$ & $56$ \\
\noalign{\smallskip}
\hline  
\end{tabular}
 \tablefoot{The columns show the component designations, the integrated flux density, FWHM size, axial ratio and the position angle of the major axis of the fitted Gaussian.}
\end{table}

\begin{table}[h]
\caption{The parameters of the Gaussian components fitted to the EVN visibility data of J0813+3508.} \label{table:J0813_EVN}      
\centering                          
\begin{tabular}{c c r r c c}         
\hline\hline 
\noalign{\smallskip}
$\nu$ & Comp. & \multicolumn{1}{c}{Flux density} & \multicolumn{1}{c}{FWHM} & Axial ratio & PA \\ 
(GHz) & & \multicolumn{1}{c}{(mJy)} & \multicolumn{1}{c}{(mas)} & & ($\degr$) \\
\noalign{\smallskip}
\hline    
\noalign{\smallskip}
 $1.6$ & C1 & $14.5 \pm 0.9 $ & $8.9 \pm 0.1$ & $0.4 \pm 0.03$ & $-46$\\      
  $1.6$ & C2 & $3.5\pm 0.2$ & $37.1 \pm 2.1$  & $0.2 \pm 0.02$ & $-29$ \\
$5.0$ & C1 & $7.0 \pm 0.4$ & $2.6 \pm 0.1$ & $1.0$ & -- \\
\noalign{\smallskip}
\hline    
\end{tabular}
 \tablefoot{The first column lists the frequencies of the observations. The further columns are the same as in Table\,\ref{table:J1231_eMERLIN}.}
\end{table}

\begin{table}[h]
    \centering
    \caption{Flux densities of the different emitting regions of J1548$+$3335 observed by e-MERLIN. At $5$\,GHz, we list the sum of the \textsc{clean} components of region W (for details see text).} \label{tab:J1548_comp}
    \begin{tabular}{cccr}
    \hline
    \hline
    \noalign{\smallskip}
    $\nu$  & \multicolumn{3}{c}{Flux density of the component} \\
    (GHz)  & \multicolumn{3}{c}{(mJy)} \\
     & C & \multicolumn{1}{c}{W1} & \multicolumn{1}{c}{W2} \\
    \noalign{\smallskip}
    \hline
    \noalign{\smallskip}
    $1.286$ & $7.39\pm0.22$ & $35.2 \pm 0.3$ & $1.8\pm 0.2$ \\ 
    $1.350$ & $7.62\pm 0.12$ & $35.9 \pm 0.2$ & $1.4 \pm 1.2$\\
    $1.414$ & $7.23\pm0.09$ & $34.2 \pm 0.1$ & $1.4\pm 0.1$\\ 
    $1.478$ & $7.04\pm0.10$ & $32.1 \pm 0.1$ & $1.4 \pm 0.1$\\
    $1.542$ & $7.18\pm0.17$ & $32.0 \pm 0.2$ & $1.4 \pm 0.1$ \\
    $1.606$ & $6.92\pm0.26$ & $29.6\pm 0.3$ & $1.5 \pm 0.3$ \\  
    $1.670$ & $6.68\pm0.09$ & $28.6 \pm 0.1$ & $1.4 \pm 0.1$\\ 
    $1.734$ & $6.62\pm0.11$ & $28.3\pm0.2$ & $1.5 \pm 0.1$\\ 
    \noalign{\smallskip}
    \hline
    \noalign{\smallskip}
    $4.880$ & $5.08\pm0.05$ & \multicolumn{2}{c}{$8.95\pm0.12$}\\ 
    $5.008$ & $5.01\pm0.05$ & \multicolumn{2}{c}{$8.66\pm0.11$}\\ 
    $5.136$ & $4.97\pm0.05$ & \multicolumn{2}{c}{$8.05\pm0.12$} \\ 
    $5.264$ & $4.86\pm0.05$ & \multicolumn{2}{c}{$7.57 \pm 0.13$} \\ 
    \noalign{\smallskip}
    \hline
    \end{tabular}
\end{table}

\begin{table*}
\caption{\label{tab:radioflux}Flux density measurements of the target sources from various radio surveys.}
\centering
\begin{tabular}{lrcrccc}
\hline\hline
\noalign{\smallskip}
Survey & \multicolumn{1}{c}{$\nu$} & Beam size & \multicolumn{1}{c}{Reference} & \multicolumn{3}{c}{Flux density} \\
 & (MHz) & ($\arcsec$) & & \multicolumn{3}{c}{(mJy)} \\
\noalign{\smallskip}
\noalign{\smallskip}
 &       &           & &\multicolumn{1}{c}{J1231+3816} &\multicolumn{1}{c}{J0813+3508} & \multicolumn{1}{c}{J1548+3335} \\
 \noalign{\smallskip}
 \hline
 \noalign{\smallskip}
 VLSSr & $74$ & $75$ & \cite{VLSSr} & $<272$ & $420 \pm 74$ & $<213$ \\
TGSS & $150$ & $25$ & \cite{tgss} & $158 \pm 17$ & $151 \pm 20$ & $135 \pm 15$ \\ 
LoTSS DR2 & $144$ & $6$ & \cite{lotssdr2} &  $146.3 \pm 0.7$\hspace{0.5em} & $311.2 \pm 2.3$\hspace{0.5em} & \hspace{-0.5em}$176 \pm 1$\\ 
WENSS & $330$ & $54$ & \citet{wenss} & $77 \pm 3$ & $140 \pm 4$\hspace{0.5em} & \hspace{-0.5em}$103 \pm 4$ \\
GMRT610 & $610$ & $9.4 \times 4.0$ & \cite{Coppejans2017} & -- & -- & $77.6 \pm 7.8$ \\
RACS-mid & $1370$ & $11.2 \times 9.3$ & \cite{racs_1.37} & $27.7 \pm 1.7$ & $41.5 \pm 1.7$ & $42.8 \pm 2.6$ \\
RACS-high & $1660$ & $11.9 \times 8.1$ & \cite{racs_1.66} & $23.0 \pm 2.3$ & $34.2 \pm 2.3 $ & $37.3 \pm 3.7$ \\
NVSS & $1400$ & $45$ & \cite{nvss} & $25.7 \pm 0.9$ & $35.6 \pm 1.1$ & $38.2 \pm 1.2$\\
FIRST & $1400$ & $5.4$ & \cite{first} & $24.0 \pm 0.1$ & \hspace{1.4em}$49 \pm 0.3$\tablefootmark{a} & $37.8 \pm 0.1$\\
VLASS & $3000$ & $2.5$ & \cite{lacy_vlass} & \hspace{0.5em}$16.7\pm 0.5$\tablefootmark{b} & \hspace{0.5em}$21.8 \pm 0.7$\tablefootmark{c} & \hspace{0.2em} $25.6 \pm 0.5$ \\
\noalign{\smallskip}
\hline
\hline
\end{tabular}
\tablefoot{Col.~1 -- survey name, Col.~2 -- central frequency, Col.~3 -- median angular resolution, Col.~4 -- literature reference, Cols.~5--7 -- flux densities with their uncertainties (and upper limits, where applicable) for the three high-redshift sources.
\tablefoottext{a}{Sum of the flux densities of the two features given in the FIRST catalogue.}
\tablefoottext{b}{Sum of the flux densities of components fitted to the third-epoch VLASS image.}
\tablefoottext{c}{Sum of the flux densities of the three components fitted to the VLASS images.}
}
\end{table*}

\section{Image noise levels}

\begin{table*}[h]
    \centering
    \caption{Noise levels of the radio images presented in this paper. Empty cells indicate that the corresponding image is not shown here. References are provided where the image has been previously published. A horizontal line indicates that no radio image at the corresponding frequency is available to our knowledge. All values are given in units of mJy\,beam$^{-1}$.} \label{tab:image_noise}
    \begin{tabular}{ccccccccc}
    \hline  \hline
     \noalign{\smallskip}
ID & LoTSS DR2 & LOFAR & FIRST & VLASS -- 3rd epoch & \multicolumn{2}{c}{e-MERLIN} & \multicolumn{2}{c}{EVN} \\
 & & & & & $1.6$\,GHz & $5$\,GHz & $1.7$\,GHz & $5$\,GHz \\
\noalign{\smallskip}
\hline
\noalign{\smallskip}
      J1231$+$3816  & $0.10$ & $0.06$ & $0.18$ & $0.12$ & $0.12$\tablefootmark{a} & -- & \multicolumn{2}{c}{\cite{Krezinger2022}}\\
       J0813$+$3508 & $0.35$ & $0.20$ & ... & $0.10$ & -- & -- & $0.04$\tablefootmark{b} & $0.08$\tablefootmark{b} \\
       J1548$+$3335 & $0.20$ & $0.08$ & ... & ... & $0.04$ & $0.02$ & \multicolumn{2}{c}{\cite{Coppejans2016}} \\
       \noalign{\smallskip}
       \hline
       \hline
    \end{tabular}
    \tablefoot{\tablefoottext{a}{e-MERLIN was used as part of an EVN+e-MERLIN observation. The central observing frequency was $1.7$\,GHz.} \tablefoottext{b}{EVN array restricted to the European anntennas.}}
\end{table*}

The noise levels of the radio images presented in this paper are given in Table \,\ref{tab:image_noise}.

\section{Comparison of LoTSS and high-resolution LOFAR image of J1548$+$3335}

\begin{figure}
    \centering
    \includegraphics[width=\hsize, bb=80 0 500 310, clip]{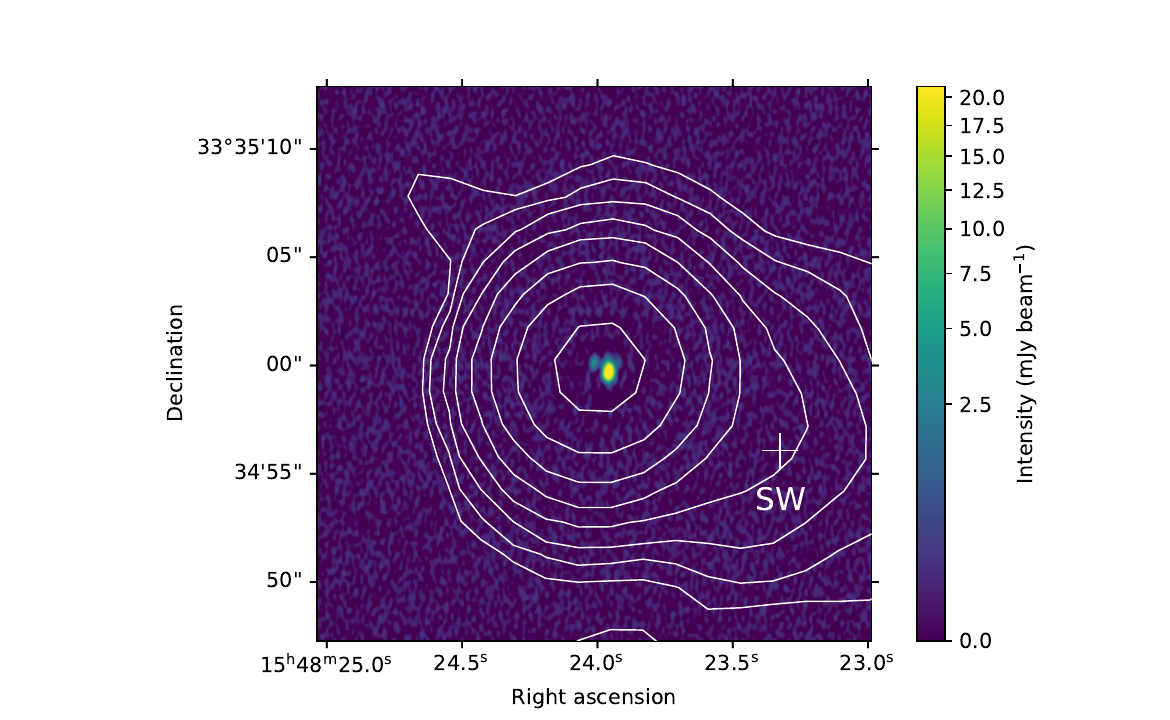}
    \caption{LOFAR images of J1548$+$3335. The white contours show the LoTSS-DR2 image; the contour levels are the same as in Fig.~\ref{fig:J1231_lowres}, except that negative contours are omitted. The colour scale shows the high-resolution LOFAR image. The white cross and label indicate the location of component SW.}
    \label{fig:J1548_lofar_compare}
\end{figure}

\section{Radio spectral plots}

\begin{figure}[h]
    \centering
    \includegraphics[bb= 30 40 720 560, clip, width=\hsize]{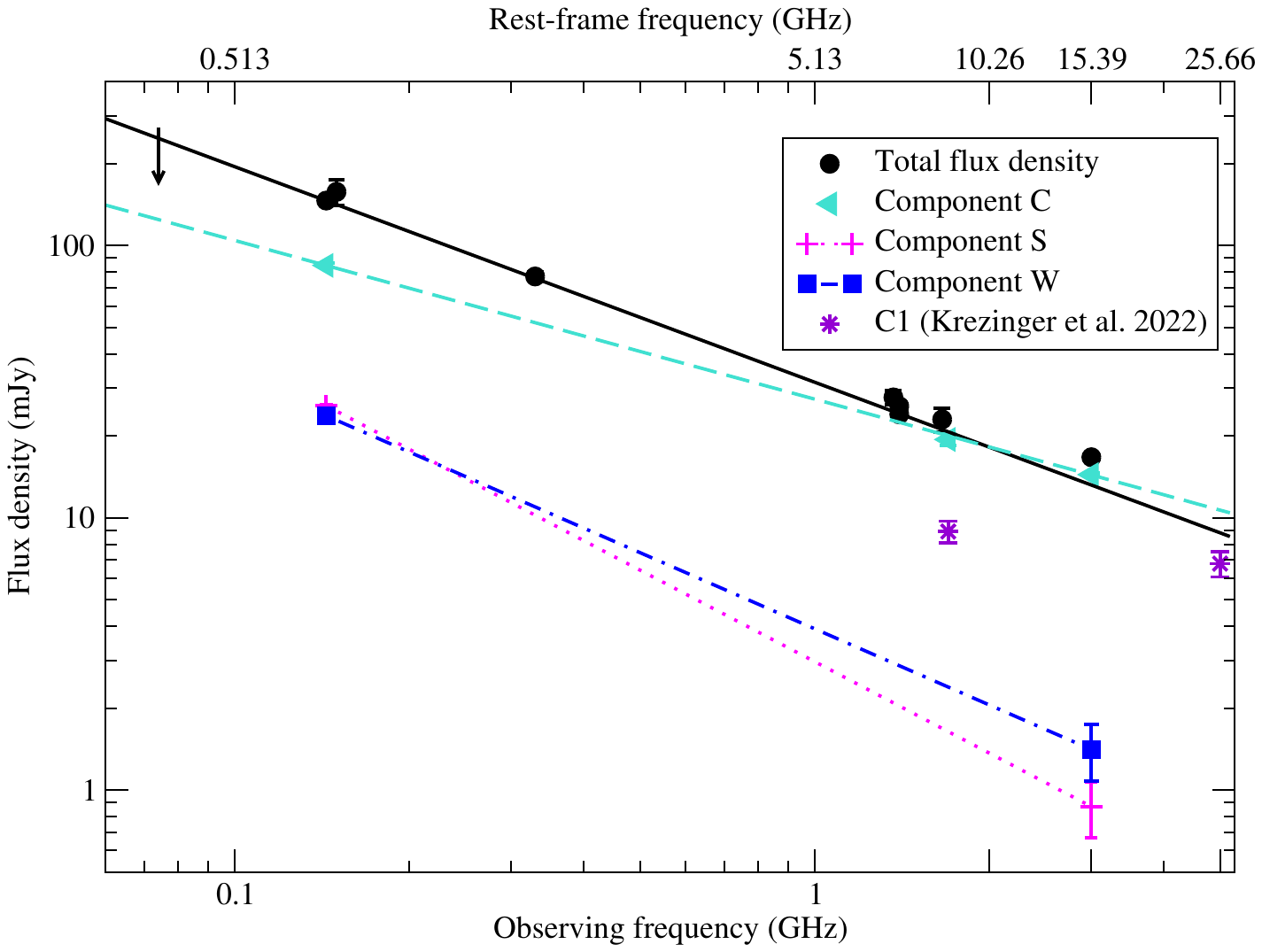}
    \caption{Radio spectrum of J1231$+$3816. Black dots show integrated flux densities from TGSS, LoTSS-DR2, WENSS, RACS, FIRST, NVSS, and VLASS (Table~\ref{tab:radioflux}). The solid black line is the corresponding power-law fit. The black downward arrow indicates the VLSSr upper limit. Filled turquoise triangles mark the flux densities of feature C, with the dashed turquoise line showing the power-law fit. Magenta symbols and the dotted line represent flux densities of region S. Blue squares and the dash-dotted line correspond to region W. Violet stars show the flux densities of the fitted mas-scale component from \cite{Krezinger2022}.}
    \label{fig:J1231_spectrum}
\end{figure}

\begin{figure}
    \centering
    \includegraphics[bb= 30 40 720 560, clip, width=\hsize]{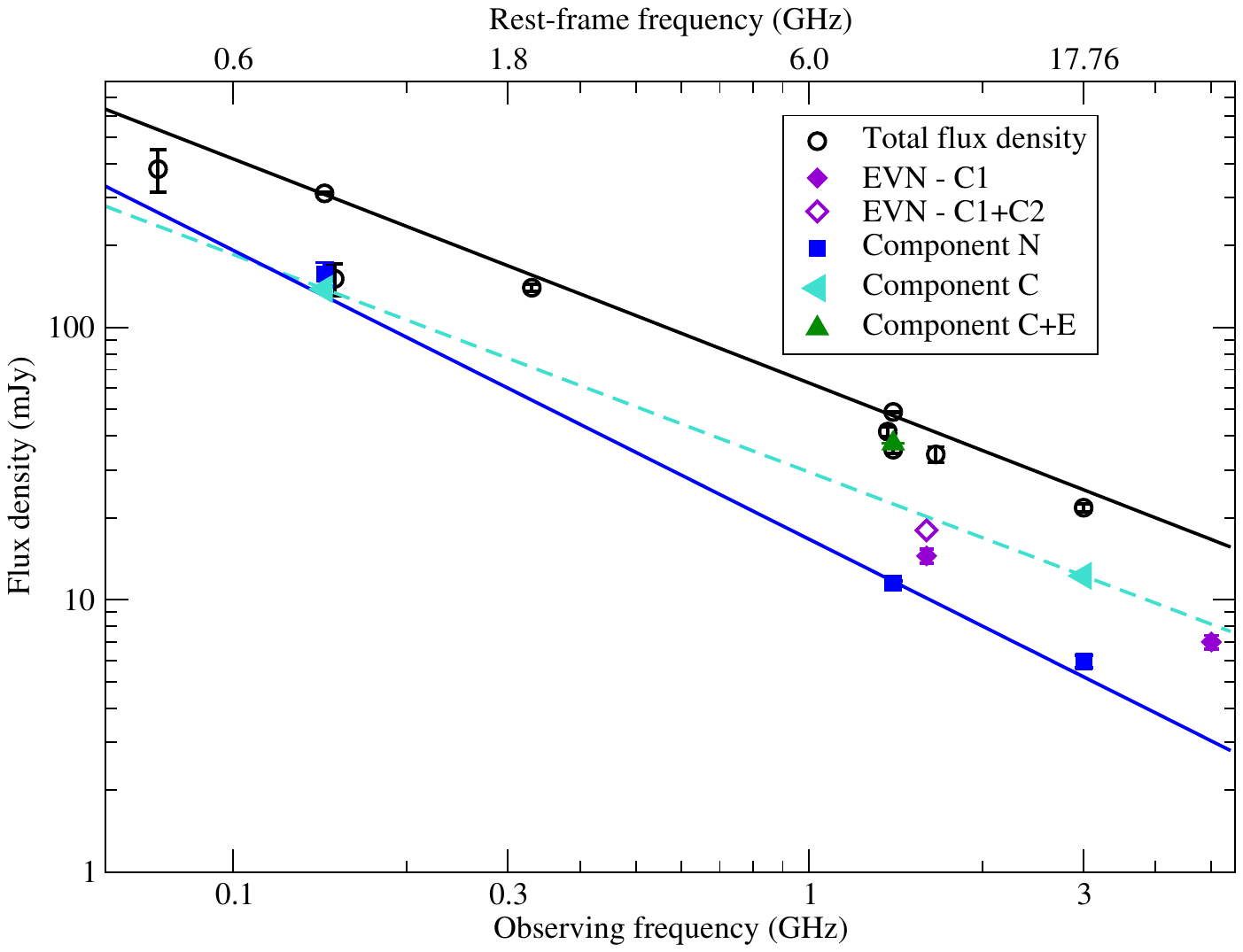}
    \caption{Radio spectrum of J0813$+$3508. Black open circles show flux densities from VLSSr, TGSS, LoTSS-DR2, WENSS, RACS, FIRST, NVSS, and VLASS (Table~\ref{tab:radioflux}). The black solid line represents a power-law fit to these data points. Blue squares and filled turquoise triangles denote the flux densities of components N and C, respectively; the corresponding blue and turquoise lines show their power-law fits. Green triangle shows the FIRST component which is a blend of C and E (for details see text). The open magenta diamond marks the summed flux density of components C1 and C2 from the $1.6$-GHz EVN observation. Filled magenta diamonds represent the flux densities of the EVN component C1 only.}
    \label{fig:J0813_spectrum}
\end{figure}

\begin{figure}
    \centering
    \includegraphics[bb= 40 50 730 570, clip, width=\hsize]{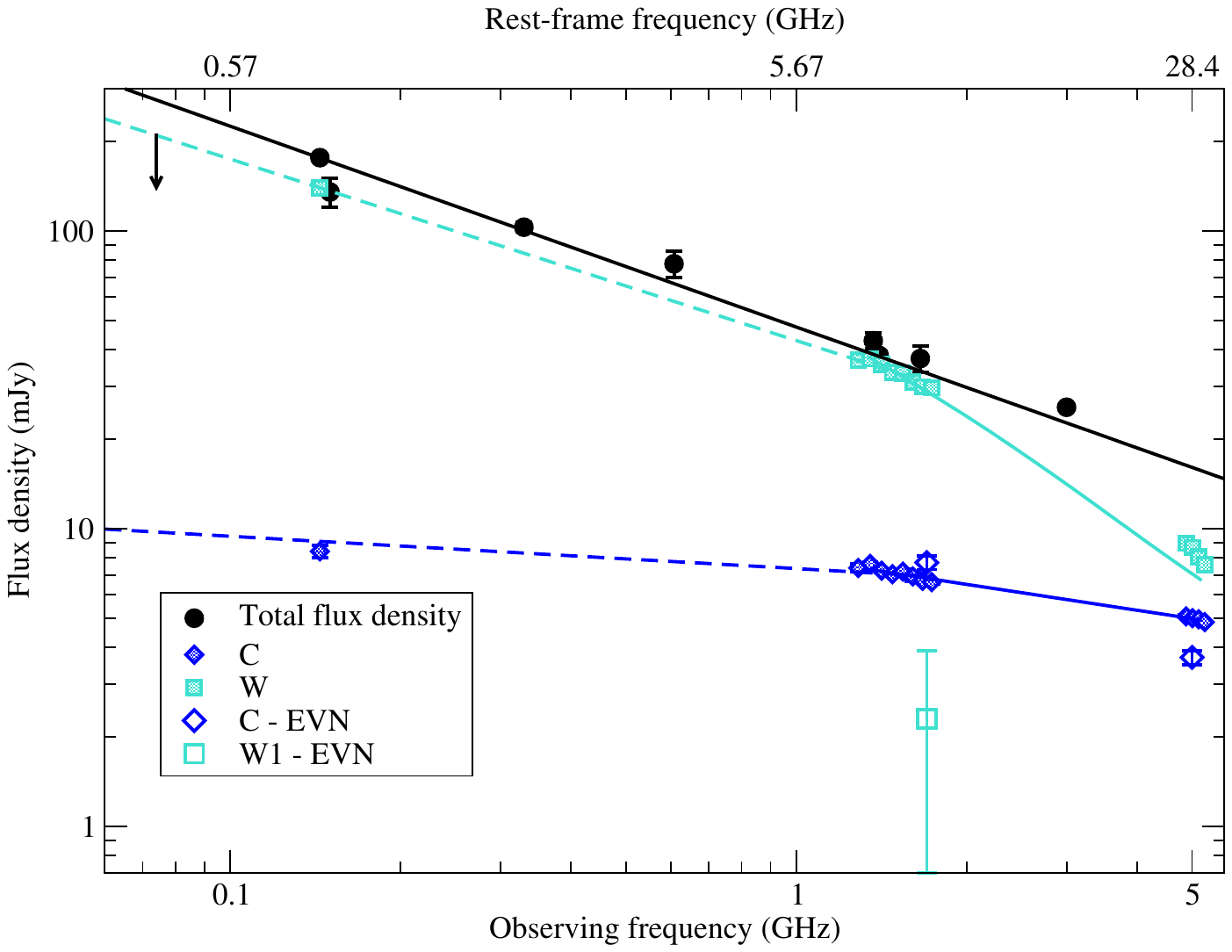}
    \caption{Radio spectrum of J1548$+$3335. Black dots show flux densities from TGSS, LoTSS-DR2, WENSS, RACS, FIRST, NVSS, and VLASS (Table\,\ref{tab:radioflux}), with the black line representing a power-law fit. The downward black arrow indicates the VLSSr upper limit \citep{VLSSr}. Blue filled diamonds and turquoise squares show the flux densities of components C and W measured with LOFAR and e-MERLIN. Solid blue and turquoise lines are power-law fits to the e-MERLIN measurements of C and W, respectively, while dashed lines show fits including the LOFAR and lower-frequency e-MERLIN data. Open blue diamonds and turquoise squares indicate the flux densities of C and W from the EVN observations of \citet{Coppejans2016}.
    }
    \label{fig:J1548_spectrum}
\end{figure}

\end{appendix}

\end{document}